\documentclass[floatfix,superscriptaddress,dvipsnames,nofootinbib,amsmath,amssymb,aps,prd,reprint,nobibnotes]{revtex4-2}

\usepackage[titletoc,toc,title]{appendix}

\usepackage{mathtools}
\usepackage{tensor}
\usepackage{amsmath, amsfonts,bm}
\usepackage{derivative}
\usepackage{bm}
\usepackage[utf8]{inputenc}
\usepackage{siunitx}
\usepackage{multirow}

\usepackage{suffix}

\usepackage{amsmath, amsfonts,bm}
\usepackage{xcolor}
\usepackage{graphicx}
\usepackage{float}
\usepackage[normalem]{ulem}
\usepackage{comment}
\usepackage[caption=false,justification=Justified]{subfig}
\usepackage[a4paper, left=20mm,right=20mm,top=20mm,bottom=20mm]{geometry}
\usepackage[colorlinks=true,hyperfootnotes=true, linkcolor=RedViolet, citecolor=BlueViolet, urlcolor=BlueViolet]{hyperref}
\usepackage{cleveref}

\newcommand\mathcomma{\,,}
\newcommand\mathperiod{\,.}

\DeclareMathAlphabet{\mathup}{OT1}{\familydefault}{m}{n}

\newcommand{\be}{\begin{equation}} 
\newcommand{\ee}{\end{equation}}

\usepackage{array}
\newcommand{\PreserveBackslash}[1]{\let\temp=\\#1\let\\=\temp}
\newcolumntype{C}[1]{>{\PreserveBackslash\centering}p{#1}}
\newcolumntype{R}[1]{>{\PreserveBackslash\raggedleft}p{#1}}
\newcolumntype{L}[1]{>{\PreserveBackslash\raggedright}p{#1}}

\begin{document}

\title{Forecasts on interacting dark energy with standard sirens}
\author{Elsa M. Teixeira}
\author{Richard Daniel}
\affiliation{Consortium for Fundamental Physics, School of Mathematics and Statistics, University of Sheffield, Hounsfield Road, Sheffield S3 7RH, United Kingdom}
\author{Noemi Frusciante}
\affiliation{Dipartimento di Fisica ``E. Pancini", Universit\`a degli Studi  di Napoli  ``Federico II", Compl. Univ. di Monte S. Angelo, Edificio G, Via Cinthia, I-80126, Napoli, Italy}
\author{Carsten van de Bruck}
\affiliation{Consortium for Fundamental Physics, School of Mathematics and Statistics, University of Sheffield, Hounsfield Road, Sheffield S3 7RH, United Kingdom}

\date{\today}

\begin{abstract}
We present the predictions with standard sirens at Gravitational Waves detectors, such as  the Laser Interferometer Space Antenna (LISA) and the Einstein Telescope (ET), for interacting dark energy theories. We focus on four models characterised by couplings between the dark energy field and the dark matter fluid arising from conformal or disformal  transformations of the metric, along with an exponential self-interacting potential. To this purpose we construct mock catalogues and perform a Markov Chain Monte Carlo analysis by considering ET and LISA standard sirens, and also their combination with Baryon Acoustic Oscillations (BAO) and Supernovae Ia (SNIa) data. We find that in all the four models considered, the accuracy on the $H_0$ parameter increases by one order of magnitude at 1$\sigma$ when compared to the SNIa+BAO data set, possibly shedding light in the future on the origin of the $H_0$-tension. The combination of standard sirens with SNIa+BAO allows to improve the accuracy on some coupling and exponential parameters, hinting at future prospects for constraining interactions in the dark sector. 
\end{abstract}

\maketitle

\tableofcontents



\section{Introduction}

Understanding the nature of the dark sector of the Universe is one of the greatest endeavours of Cosmology at the present. This comprises the weakly interacting dark matter (DM) - responsible for the formation and dynamics of structures in the Universe - and dark energy (DE) - the driver of the late time cosmic acceleration.  Together these components dominate about $95 \%$ of the energy budget of the Universe. In the $\Lambda$-cold-dark-matter ($\Lambda$CDM) scenario, i.e. the standard model of cosmology, DE is portrayed simply as a cosmological constant, $\Lambda$.  
This model comes with some theoretical issues \cite{Weinberg:1988cp,Martin:2012bt,Joyce:2014kja}, among which is the fact that fundamental theories do not properly account for the currently measured small value of the cosmological constant.
$\Lambda$CDM also requires a primordial inflationary period to explain the geometrical flatness, Cosmic Microwave Background (CMB) smoothness, and initial conditions for large-scale structures. More recently, observational tensions on the value of the cosmological parameters $H_0$ \cite{Planck:2015mrs,Planck:2018vyg,Riess:2011yx,Riess:2016jrr,Riess:2019cxk,BOSS:2014hwf,DiValentino:2020zio} and $\sigma_8$~\cite{deJong:2015wca,Hildebrandt:2016iqg,Kuijken:2015vca,FenechConti:2016oun,Joudaki:2019pmv,DiValentino:2020vvd}, measured by early- and late-Universe probes, increased the motivation to investigate alternative models of gravity \cite{CANTATA:2021ktz}.  

Consequently, alternative theories are explored by cosmologists in which $\Lambda$ is promoted to a dynamical DE scalar field, $\phi$, namely \textit{quintessence} \cite{Wetterich:1987fm,Peebles:1987ek} (see \cite{Tsujikawa:2013fta} for a review), which evolves in time according to its self-interaction potential.
While in the standard $\Lambda$CDM scenario the two dark components do not directly couple with each other, in a dynamical DE model one can instead consider that they experience some \textit{non-minimal} interaction. Such constructed models are referred to as \textit{coupled quintessence models} \cite{Wetterich:1994bg,Amendola:1999er}.
The dynamics of the field, along with the dark interaction, could provide a more natural explanation of the accelerated expansion, while also addressing the observational tensions \cite{Abdalla:2022yfr}. Nevertheless, the coupling can be formalised at the Lagrangian level through what is known as a conformal/disformal transformation of the metric tensor \cite{Brans:1961sx,Bekenstein:1992pj, Zumalacarregui:2012us, Zumalacarregui:2013pma, Bettoni:2013diz, vandebruck:2013yxa, vandebruck:2016cnh}. If this transformation depends directly on the quintessence field, then this is physically equivalent to considering that the DM particles propagate on the geodesics of the transformed metric, $\bar{g}_{\mu \nu}$. In the conformal case, this is achieved from a re-scaling of the metric and, consequently, of time- and space-like norms and intervals alike, while preserving the light cones:
 
 \begin{equation}
\bar{g}_{\mu \nu} =C (\phi) g_{\mu \nu} \mathcomma
\label{conf}
\end{equation}

\noindent where $C$ is the conformal function. These find important applications in modified gravity theories as they preserve the structure of Scalar-Tensor theories of the Brans-Dicke form \cite{Faraoni:1998qx}. Alternatively, one can consider that the metric transformation should depend on the first-order partial derivatives of the scalar field as well. This results in a disformal transformation:

\begin{equation}
 \bar{g}_{\mu \nu} =C ( \phi,X ) g_{\mu \nu} + D ( \phi) \partial_{\mu} \phi \partial_{\nu} \phi \mathcomma
\label{disf}
\end{equation}

\noindent where $C$ and $D$ are the conformal and disformal functions, respectively and $X = - g^{\mu \nu} \partial_\mu \phi \partial_\nu\phi/2 $. This gives rise to a more intricate scenario, with a distortion of the metric defined directionally according to the gradient of $\phi$.
First introduced by Bekenstein \cite{Bekenstein:1992pj}, the disformal transformations re-surged in the cosmological literature \cite{Zumalacarregui:2013pma} when, in analogy to the conformal case, it was shown that they preserve a more general class of Scalar-Tensor theories categorised in the Horndeski Lagrangian \cite{Bettoni:2013diz}.
Disformal transformations in cosmology arise naturally in brane-world models \cite{Koivisto:2013fta,vandebruck:2020fjo} and have been the focus of many theoretical proposals for the nature of the dark sector and their interactions \cite{Koivisto:2008ak,Zumalacarregui:2010wj,DeFelice:2011bh,Koivisto:2012za,Zumalacarregui:2012us,vandebruck:2015ida,Bettoni:2015wla,Sakstein:2014aca,Sakstein:2015jca,vandebruck:2016jgg}. The stability conditions for the functions $C$ and $D$ have been discussed in Refs.~\cite{Zumalacarregui:2012us,Zumalacarregui:2013pma,Bettoni:2013diz,Sakstein:2014aca} and the case in which $D \equiv D(\phi,X)$ has been discussed in Refs.~\cite{teixeira:2019hil,karwan:2016cnv}.

The coupling in the dark sector gives rise to an additional gravitational fifth force in the Universe between DM particles, mediated by the DE field. This new force leaves distinct features in the background equations, as well as signatures in the cosmological density perturbations that describe the formation of structures \cite{Gumjudpai:2005ry}. Although these deviations from the benchmark model are constrained to be small (especially at the background level), they are still expected to leave detectable, characteristic observational imprints that the data can probe.  These are essential to test the viability of such alternative models by identifying the range of validity of the parameter space and the robustness of its predictions.

In the past few years, we have witnessed the rise of gravitational wave (GW) astronomy as a new independent probe of gravitational effects \cite{Holz:2005df}. An accurate redshift-luminosity relation can be constructed when GW events are combined with an electromagnetic (EM) counterpart multi-messenger signal. These observations become \textit{standard sirens} \cite{Schutz:1986gp}, analogous to the standard candles used in local EM measurements.  So far, only one GW event, GW170817, with a corresponding EM counterpart, GRB170817A, has been detected by the Laser Interferometer Gravitational-Wave Observatory (LIGO)-Virgo and the International Gamma-ray Astrophysics Laboratory (INTEGRAL)-Fermi collaborations, respectively, and which originated from the merger of a binary pair of neutron stars \cite{LIGOScientific:2017vwq,LIGOScientific:2017zic}. This single combined detection had a strong impact on the allowed modifications to the gravitational interaction by ruling out many proposals \cite{Creminelli:2017sry,Baker:2017hug,Ezquiaga:2017ekz,Creminelli:2018xsv,Amendola:2017orw} with many other models further constrained \cite{LISACosmologyWorkingGroup:2019mwx,Belgacem:2018lbp,Allahyari:2021enz,Califano:2022syd,Ferreira:2022jcd}.  

Current GW detectors, (advanced) Virgo \cite{VIRGO:2014yos}, (advanced) LIGO \cite{LIGOScientific:2014pky}  and the Kamioka Gravitational Wave Detector (KAGRA) \cite{Somiya:2011np},
 are second-generation (2G) ground-based detectors, with another one under planning (2030), the Indian Initiative in Gravitational-wave Observations (IndIGO) \cite{IndIGO}.
The increasing number of detectors will boost the capabilities of GW astronomy both in the number of confirmed events (a larger volume of the Universe is covered) and sky localisation (a better triangulation
of the source), which will also aid in the search for a counterpart.  
However, 2G detectors are limited in their sensitivity and future third-generation (3G) ground-based detectors are designed to become more sensitive, precise and capable of probing a larger range of frequencies. Special emphasis should be given to the Einstein Telescope (ET), which is expected to improve the current sensitivity by a factor of 10 \cite{Punturo:2010zz}. ET will also extend the redshift range, \textit{e.g.} $z \sim 5$ for binary black-holes (BBHs) compared to $z\sim 0.5$ for 2G detectors \cite{Sathyaprakash:2012jk}. The number of detectable multi-messenger events is expected to reach tens of thousands of standard sirens \cite{Maggiore:2019uih}. While these ground-based detectors will cover a frequency band in the range $1 \lesssim f\lesssim 10^3$ Hz \cite{Cai:2016sby}, the upcoming space-based 3G detectors, such as the Laser Interferometer Space Antenna (LISA) \cite{LISA:2017pwj} will have a  peak sensitivity near $10^{-3}$ Hz and will be able to detect GW events beyond $z=20$, probing a wide range of targets. There are many proposals of 3G GW observatories, such as the DECi-hertz Interferometer Gravitational wave Observatory (DECIGO) \cite{Kawamura:2011zz}. However, we have opted to focus our analysis on ET and LISA covering ground- and space-based experiments. 

In this paper, we aim at forecasting the constraining power of future 3G detectors; in particular we will focus on ET and LISA. Given the potential of such missions we are interested in assessing their ability to constrain modifications to General Relativity (GR) as well as to  provide  complementary constraints on $H_0$ using standard sirens. We investigate four models characterised by coupling functions between the DM and DE fields: a conformally coupled quintessence field \cite{Amendola:1999er,Wetterich:1994bg}, characterised by a conformal coupling in the form of an exponential function of the scalar field; a kinetic model \cite{teixeira:2022sjr} with a conformal function given by a power law of the kinetic term of the scalar field; a purely disformally coupled quintessence field \cite{vandebruck:2015ida, vandebruck:2013yxa} with a constant disformal coupling and a mixed disformally coupled quintessence \cite{vandebruck:2016jgg,teixeira:2019hil} which combines the previous model with an exponential conformal coupling. All the scenarios considered are characterised by the same simple exponential potential which introduces one more free parameter. The models we consider differ considerably in the way the (effective) coupling between DM and DE evolves and, in particular, they differ in their background evolution. As such, they represent a well studied sample of models of interacting dark energy suitable for our analysis. We construct our pipeline following the methodology presented in \cite{zhao:2010sz,cai:2017aea, Li:2013lza,Nishizawa:2010xx} providing for the first time GWs forecasts on the free parameters of the four models in question.

This paper is organised as follows. We start by giving a brief introduction to  the physics of standard sirens in Sec.~\ref{sec:GWSS}. Sec.~\ref{sec:method} provides an overview of the methodology used and the details on the simulation of the standard siren events developed for this study, as well as a brief account of the data set combinations considered. We outline the criteria for particular catalogue choices, and discuss the sampling method employed for the forecasts. In Sec.~\ref{sec:res} we  introduce each of the four models under study and present the results of our analysis, emphasising their significant implications. Lastly, in Sec.~\ref{sec:sum}, we summarise our results and outline our concluding thoughts and future prospects.

\section{Gravitational Waves as Standard Sirens}\label{sec:GWSS}

Interferometers are sensitive to the strain, $h(t)$ from a GW event, which in the transverse-traceless gauge is described as, \cite{zhao:2010sz}  
\begin{align}
    h(t) = F_{\times}  (\theta_0, \phi_0, \psi) h_{\times}(t) + F_+(\theta_0, \phi_0, \psi) h_+(t)\mathcomma
\end{align}
where $\theta_0$ and $\phi_0$ define the initial location of the event relative to the detector in polar coordinates, $\psi$ is the polarisation of the GW event, and $t$ is cosmic time. We adopt a random sampling method in the range $[0-2\pi]$ for $\theta_0$ and $[0-\pi]$ for both $\phi_0$ and $\psi$. The factors $F_{\times,+}$ describe the antenna beam pattern function, 
\begin{align}
\begin{aligned}
    F_{\times}^{(1)} = \frac{\sqrt{3}}{2}&\left[\frac{1}{2}(1+\cos^2(\theta))\cos(2\phi)\cos(2\psi)\right.\label{eq:F} \\
    &\left.+ \cos(\theta)\sin(2\phi)\cos(2\psi)  \right]\mathcomma
    \\
    F_{+}^{(1)} = \frac{\sqrt{3}}{2}&\left[\frac{1}{2}(1+\cos^2(\theta))\cos(2\phi)\cos(2\psi)\right. \\
    &\left.- \cos(\theta)\sin(2\phi)\cos(2\psi)  \right]\mathperiod
\end{aligned}
\end{align}
The superscript number indicates which interferometer is being considered, \textit{e.g} LISA only has two separate interferometers and therefore $F^{(3)}=0$. 
Since the detectors are spatially distributed in an equilateral triangle formation, the other two antenna pattern functions relate to $F_{\times,+}^{(1)}$ as
\begin{equation} \label{ET F}
\begin{aligned}
    F_{\times,+}^{(1)}(\theta, \phi, \psi)&= F_{\times,+}^{(2)}(\theta, \phi+ \frac{2\pi}{3}, \psi)
    \\ &= F_{\times,+}^{(3)}(\theta, \phi+ \frac{4\pi}{3}, \psi) \mathperiod
\end{aligned}
\end{equation}
As LISA is sensitive to lower frequencies, and equivalently larger masses, it can detect GW events of inspiral mergers lasting over several months, during which the interferometer's position will change relative to the event. This change in position is accounted for following the method described in \cite{cai:2017aea}. The timescale of the event is described as
\begin{equation}
t=t_c-5(8\pi f)^{-8/3}M_c^{-5/3} \mathperiod
\end{equation}
Here $t_c$ is the time of the merger, $t$ indicates the time at which LISA detects the merger, $f$ is the frequency of the GW, and $M_c$ is the chirp mass. The location angles are updated accordingly:
\begin{align}
\theta&=\cos^{-1}\left[\frac{1}{2}\cos(\theta_0) \right.\label{eq:theta}\\
&\left.-\frac{\sqrt{3}}{2}\sin(\theta_0)\cos\left(\frac{2\pi t}{T}-\phi_0\right)           \right]\mathcomma
\nonumber \\
\phi&=\frac{2\pi t}{T} \label{eq:phi}\\
&-\tan\left[\frac{\sqrt{3}\cos(\theta_0)+\sin(\theta_0)\cos\left(\frac{2\pi t}{T}-\phi_0\right)}{2\sin(\theta_0)\cos\left(\frac{2\pi t}{T}-\phi_0\right)}	\right]\mathcomma
\nonumber
\end{align}
which, in turn, are used to update the beam pattern functions. Here we have specified the period, $T$, as the orbit around the Sun.

While the individual masses of the objects are not directly discernible, GW detectors are sensitive to the chirp mass, a collective mass quantity related to the frequency evolution of the signal emitted before the merger, during the inspiral phase of the binary \cite{Hilborn:2017liy}, defined as
\begin{equation}
M_c=(1+z)\left(\frac{(m_1\, m_2)^3}{m_1+m_2}\right)^{1/5}\mathcomma
\end{equation}
where $(1+z)$ is a conversion redshift factor from the physical to the observational chirp mass.

The Fourier transform of the strain using the stationary phase approximation \cite{Li:2013lza} reads
\begin{equation}
\mathcal{H}=\mathcal{A}f^{-7/6}e^{i\Psi(f)} \mathcomma
\label{eq:H}
\end{equation}
where $\Psi(f)$ is the phase of the waveform. Notice that when $\mathcal{H}$ is inserted into Eq.~\eqref{eq:SNR}, the exponential term disappears, meaning that the $\Psi(f)$ factor can be discarded for this analysis. $\mathcal{A}$ is the Fourier amplitude of the waveform,
\begin{align}
\mathcal{A}=&\, \frac{M_c^{5/6}}{d_L}\pi^{-2/3}\sqrt{\frac{5}{96}}\\
&\times\sqrt{[F_+(1+\cos^2(l))]^2 + (2F_\times\cos(l))^2}\mathcomma
\nonumber
\end{align}
where $d_L$ is the luminosity distance from the merger and $l$ is the inclination angle, which we sample randomly between $[0^\circ,20^\circ]$, as that is the maximum detection inclination range. 

LISA has been designed to effectively measure frequencies as low as $f_{min}=1\times10^{-4}\, \text{Hz}$, which is why it stands as a promising probe of extreme mass ratio inspiral (EMRI) and binary massive black hole (BMBH) mergers. For the purpose of the simulations, the upper bound frequency of LISA is determined by two quantities: the structure of LISA itself and the last stable orbit of the merging system. LISA can detect frequencies up to $f_{max}=c\,(2\pi L)^{-1}$, where $L$ is the length of LISA's interferometer arm, taken to be $2.5\, \text{Gm}$ and $c$ is the speed of light. Moreover, the total mass of an orbiting system is inversely proportional to its measured frequency, implying that even though massive mergers give rise to large detection amplitudes, the frequency will fall below $f_{min}$. Therefore if the last stable orbit frequency, $f_{LSO}=(6^{3/2}2\pi M_{obs})^{-1} $, with $M_{obs}$ being the observed total mass, is found to be lower than $f_{min}$, we disregard that simulated event. If otherwise it lies between $f_{min}$ and $f_{max}$ then $f_{LSO}$ becomes the new maximum frequency for that event. 

\section{Methodology and Data Sets}\label{sec:method}

Given the main objective of this study, we create simulated data that forecasts the potential future observations of standard siren events. Specifically, we focus on those that could be detected by ET and LISA. Below, we provide a concise overview of the samples we have generated along with the methodology and the data combinations used. 

\subsection{Simulated Cosmology}

To simulate GW catalogues from future probes of black hole mergers, the following cosmological quantities are required: the redshift of the merger, $z$, the value of the Hubble rate at merger, $H(z)$, its comoving and luminosity distance, $d_c(z)$ and $d_L(z)$ respectively, and the cosmic time between the merger and measurement, $t$.
For this purpose, we resort to the public Einstein-Boltzmann code \texttt{CLASS} code\footnote{\href{https://github.com/lesgourg/class_public}{https://github.com/lesgourg/class\_public}} \cite{lesgourgues2011cosmic,Blas_2011,lesgourgues2011cosmic2}, which we extend to accommodate general models of interacting dark energy. 
This new patch is then used to provide a \textit{mock Universe} adopting a flat $\Lambda$CDM as the fiducial model to simulate the GW data, according to the best-fit cosmological parameters  of the \textit{Planck} 2018 data release \cite{Planck:2018vyg}. These are: the Hubble parameter at present time, $H_0 = 67.32\, \text{km}\, \text{s}^{-1}\, \text{Mpc}^{-1}$, 
the density of baryons, $\Omega_bh^2 = 0.022383$ (with $h=H_0/100$) and the density of cold dark matter, $\Omega_ch^2 = 0.12011$. Furthermore, we are also interested in the derived quantity $\Omega_m^0 = \Omega_b + \Omega_c$, which for the fiducial \textit{Planck} case is $\Omega_m^0 = 0.3144$.

Provided with the background cosmology, we simulate the merger events to determine the redshift-luminosity relation. First, we generate a redshift distribution of events weighted by a probability distribution. The characteristics of these events, such as the chirp mass, are simulated using a uniform distribution. Although each instance of running the script will yield a different set of simulated data, the resulting conclusions will be unaltered as the fiducial parameters constrain the mock data.
Once the mergers have been simulated, we emulate the measurement process from the inspiral, yielding the errors associated with each event. As such, simulated data points are removed if they produce a signal-to-noise ratio below the threshold.

\subsection{Distribution of Simulated Merger Events}

ET is designed to probe a range of frequencies, $f$, similar to that of LIGO, thereby probing merger events of nearby compact objects such as binary neutron stars (BNS) in the mass range of $[1,2],[1,2]\, \text{M}\textsubscript{\(\odot\)}$, and black hole-neutron star binaries (BHNS) in the mass range $[3,10], [1,2]\, \text{M}\textsubscript{\(\odot\)}$, with the $[\cdot,\cdot]$ notation indicating the uniformly distributed mass ranges considered. Advanced LIGO claims a ratio of BHNS to BNS merger events of $\sim0.03$ \cite{LIGOScientific:2010weo}. The redshift probability distribution of these events is proportional to
\begin{equation}
    P\propto\frac{4\pi d_c(z) R(z)}{(1+z)H(z)} \mathcomma
\end{equation}
where the comoving distance and the Hubble parameter are taken at various redshifts determined by \texttt{CLASS}. $R(z)$ stands for the merger rate, which, at a linear approximation level, is \cite{Nishizawa:2010xx}
\begin{align}R=
    \begin{cases}
    1+2z \hspace{0.5cm} &\text{if    } z<1\mathcomma\\
    \frac{3}{4}(5-z) &\text{if    } 1\leq z<5\mathcomma\\
    0 &\text{otherwise}\mathperiod
    \end{cases}
\end{align}

On the other hand, LISA will target lower frequencies when compared with other proposed 3G detectors, implying sensitivity to events from larger mass binary systems since $f \propto M^{-1}$. Therefore we focus on simulating the detection of events from EMRIs and BMBHs in the ranges $[1-30],[10^4-10^8]\, \text{M}\textsubscript{\(\odot\)}$ \cite{Gair:2017ynp} and $[10^4-10^8],[10^4-10^8]\, \text{M}\textsubscript{\(\odot\)}$ \cite{Caprini:2016qxs}, respectively. The number of detected BMBH to EMRI events is estimated to follow a $2:1$ ratio according to the mission's proposal \cite{eLISA:2013xep,Amaro-Seoane:2012aqc}.

Although in principle LISA will also be able to probe mergers of binary intermediate-mass black holes (IMBHs) and binary compact objects, we opt to discard these from the simulations. This is due to the fact that there is no definitive observational proof of IMBH, and expected events from binary compact objects will only be observed at redshifts $z\approx 3$\cite{Mapelli:2010ht}. These events are insignificant since we are interested in the higher range of redshifts for our cosmology. 

Considering events involving BMBHs only, the redshift probability distributions are based on the histogram for the L6A2M5N2 mission specification \cite{Caprini:2016qxs} which considers three formation processes of BMBHs. We consider the light seed model (pop III) which attributes the formation of BMBHs to the remnants of population III stars around $z=15-20$. In \cite{Caprini:2016qxs}, two additional scenarios for massive black hole formation were investigated, namely delay and no delay scenarios. These cases involve the collapse of gas in a galactic centre at $z=15-20$, leading to the formation of a black hole through a heavy seed mechanism with and without a delay between galaxy merger and the merger of the central massive black hole. Further information on these scenarios can be found in Ref.~\cite{Klein:2015hvg}.

In our investigation we provide mock data and obtain forecasts for both the delay and no delay cases. However, the analysis reveals that the predicted constraining power from these models shows no actual improvement compared to the pop III case. Consequently, in this paper, we focus solely on the pop III model, as it proves sufficient to forecast the constraining power of LISA.

\subsection{Simulation of Measurements and Errors}

To simulate the errors associated with the standard siren catalogue, we follow the methodology of \cite{zhao:2010sz,cai:2017aea, Li:2013lza,Nishizawa:2010xx}. 
 An apparent detection of a GW event is assessed by evaluating the signal-to-noise ratio (SNR), $\rho$, and only confirmed if $\rho>8$. The SNR is defined as 
\begin{equation}
\rho_{1,2,3}^2 =4 \int_{f_{min}}^{f_{max}} df \frac{|\mathcal{H}|^2}{S_h}\mathcomma
\label{eq:SNR}
\end{equation}
where the number labels indicate the interferometer being considered. $\mathcal{H}$ has been  defined in Eq.~\eqref{eq:H} and $S_h$ is the noise power spectral density, an SNR weighting function that accounts for the particular properties of the instruments used.  For ET, in particular, $S_h$ is designed to follow 
\begin{align}
    \begin{aligned}
        S^{(\text{ET})}_h= S_0\left( x^{p_1} + a_1 x^{p_2} + 
         a_2\frac{1 + \sum^{6}_{n=1}b_n x^n}{1 + \sum^{4}_{m=1}c_m x^m}  \right) \mathcomma
    \end{aligned}
    \label{eq:S_ET}
\end{align}
where, $x = f/200$ Hz$^{-1}$, $S_0 = 1.449\times 10^{-52}$ \text{Hz}, 
        $p_1 = -4.05$,  $p_2 = -0.69$, $a_1 = 185.62$, $a_2 = 232.56$,
        $b_n = \left\{ 31.18, -64.72, 52.24, -42.16, 10.17, 11.53 \right\}$, and
        $c_m = \left\{13.58, -36.46, 18.56, 27.43 \right\}$, assuming a lower cutoff at $f = 1\, \text{Hz}$. On the other hand, for LISA, $S_h$ depends on the instrumental (or short) noise, $S_{inst}$, the noise from low-level acceleration, $S_{acc}$, and the confusion background noise, $S_{conf}$ \cite{Klein:2015hvg}:
\begin{align}
\begin{aligned}
  &S^{(\text{LISA})}_h =\frac{20}{3}\frac{4S_{acc} + S_{insta} + S_{conf}}{L^2} \left[1+ \left(\frac{fL}{0.81c} \right)  \right]  \mathcomma
\end{aligned}
\label{eq:S}
\end{align}
where 
$S_{acc}=9\times 10^{-30}/(2\pi f)^4( 1 +10^{-4}/f)$, $S_{inst}= 2.22\times10^{-23}$ and 
$S_{conf}=2.65\times10^{-23}$.
Therefore, the total SNR contribution for each detector is given by combining \eqref{eq:H} with either Eq.~\eqref{eq:S_ET} or Eq.~\eqref{eq:S} for the ET and LISA, respectively: 
\begin{equation}
\rho_{tot}=\sqrt{\rho_1^2+\rho_2^2 + \rho_3^2} \mathperiod
\end{equation}

The instrumental error in the luminosity distance is determined \textit{via} the Fisher matrix,
\begin{equation}
\sigma_{d_L}^{inst} \approx \left\langle \frac{\partial \mathcal{H}}{\partial d_L} , \frac{\partial \mathcal{H}}{\partial d_L} \right\rangle^{-\frac{1}{2}} \mathcomma
\end{equation}
following \cite{cai:2017aea}. Since $\mathcal{H}\propto d_L^{-1}$ this results simply in
\begin{equation}
\sigma_{d_L}^{inst}\approx \frac{2d_L}{\rho} \mathcomma
\label{eq:siginst}
\end{equation}
where the factor of $2$ accounts for the symmetry in the inclination angle, which actually ranges from $-20^\circ$ to $20^\circ$.
The error due to gravitational lensing is, 
\begin{equation}
\sigma_{d_L}^{len}= \frac{d_L}{2}\times 0.066\left[4(1- (1+z)^{1/4})  \right]^{1.8} \mathcomma
\label{eq:siglen}
\end{equation}
reduced by a half to account for both the merger and ringdown of the event.  

Being space-based, LISA is also subject to an error associated with the peculiar velocities of GW sources \cite{Tamanini:2016zlh}:
\begin{equation}
\sigma_{d_L}^{pec}=d_L\frac{\sqrt{\langle v^2\rangle}}{c}\left[1+\frac{c(1+z)}{Hd_L}\right] \mathcomma
\label{eq:sigpec}
\end{equation}
with an estimate of the peculiar velocity of the host galaxy with respect to the Hubble flow of $\sqrt{\langle v^2\rangle}=500\, \text{km}\, \text{s}^{-1}$.

Bringing all the contributions together, the total error in the luminosity distance is simply a combination of the errors in Eqs.~\eqref{eq:siginst}-\eqref{eq:sigpec}:
\begin{equation} \label{eq:tots}
\sigma_{d_L}=\sqrt{(\sigma_{d_L}^{inst })^2+(\sigma_{d_L}^{len })^2+(\sigma_{d_L}^{pec })^2} \mathperiod
\end{equation}

The simulation allows us to interpolate any number of events over a continuous redshift distribution in the range $0<z\lesssim 5$ for ET and $0<z\lesssim 10$ for LISA. However, 
the number of mergers detected by ET will depend on factors such as running costs and the complementary detection with other experiments \cite{zhao:2010sz}. ET is expected to report more than $10^4$ mergers yearly. However, due to the scarcity of EM counterpart signals, the predicted number of detectable mergers with an actual EM counterpart over the course of $10$ years is approximately $200$ \cite{Hou:2022rvk}.
According to \cite{Caprini:2016qxs}, LISA's number of detected mergers, for a $10$-year mission proposal, is $56$ events. 

\begin{figure}[t!]
\centering
\includegraphics[width=0.5\textwidth]{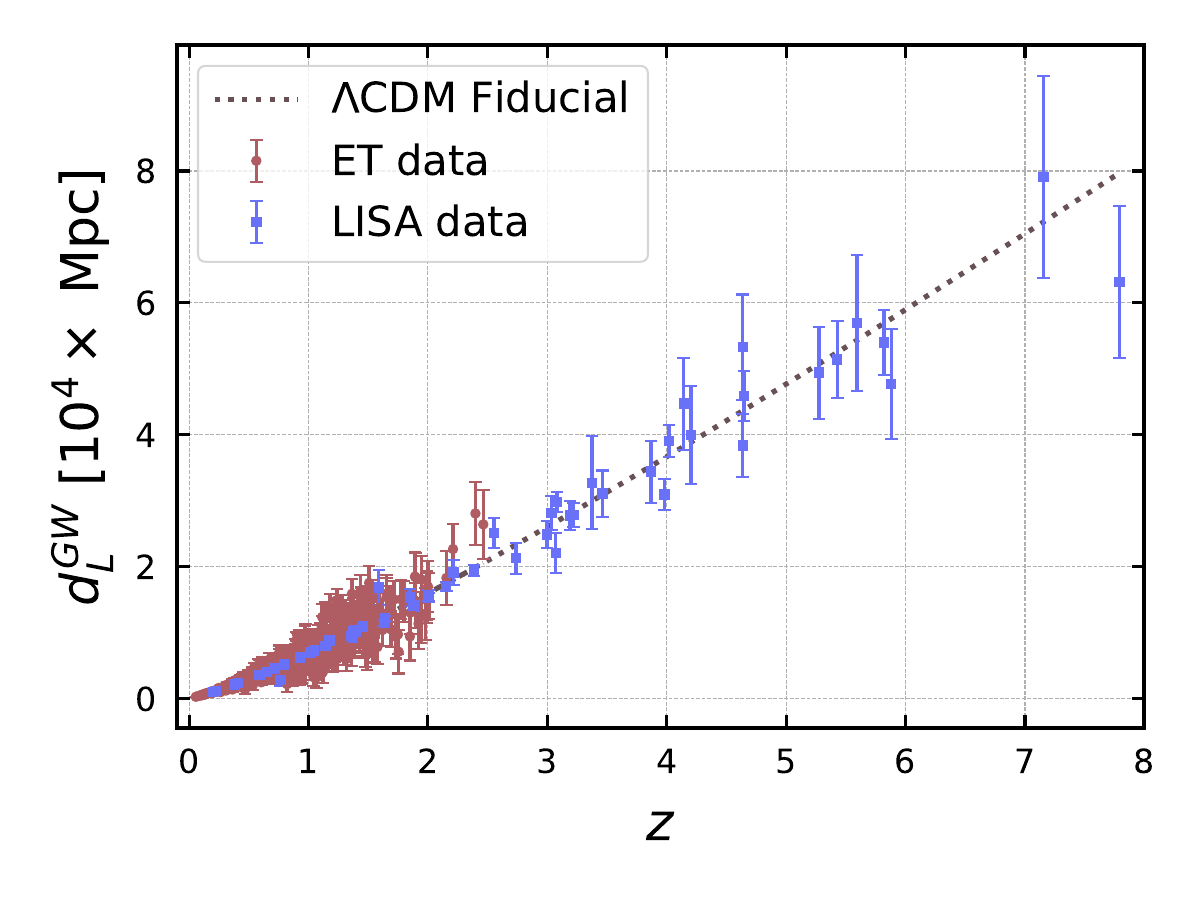}
 \caption{\label{fig:ETLISAmock} Mock data from ET (red circle markers) and LISA (blue square markers), for the fiducial model, $\Lambda$CDM, shown by the grey dotted line.}
\end{figure}

To incorporate uncertainty into the luminosity distance of each merger, we apply a Gaussian distribution centred around the background cosmology. The standard deviation for this distribution is set to the calculated errors, $\sigma_{d_L}$. This introduces artificial randomness around each merger, leading to a larger deviation from $\Lambda$CDM in LISA compared to ET. The reason for this difference is that LISA probes larger redshifts, which are associated with larger errors, resulting in a broader spread of the data, as depicted in Fig. \ref{fig:ETLISAmock}.

\subsection{Data Sets and Likelihoods}\label{sec:like}

To examine the fit of the simulated data to the coupled quintessence models considered in this study, we employ the Markov Chain Monte Carlo (MCMC) method, using samples generated from our modified version of \texttt{CLASS} interfaced with the \texttt{Monte Python} sampler \cite{Audren_2013, Brinckmann:2018cvx}. In particular, we resort to the Nested Sampling algorithm through the MultiNest\footnote{\href{https://github.com/farhanferoz/MultiNest}{https://github.com/farhanferoz/MultiNest}} \cite{Feroz:2007kg,Feroz:2008xx,Feroz:2013hea} and PyMultiNest\footnote{\href{https://github.com/JohannesBuchner/PyMultiNest}{https://github.com/JohannesBuchner/PyMultiNest}} \cite{Buchner:2014nha} packages to estimate observational constraints on the free parameters, instead of the traditional Metropolis-Hastings algorithm. The Metropolis-Hastings algorithm struggles to explore the full line of degeneracies between the parameters, resulting in false peaks in the posterior distribution, which the sampler cannot move away from. Nested sampling is able to explore the full extent of the degeneracies as it is much better suited for multi-model sampling (see \cref{sec:res}) and other more complicated distributions. Subsequently, we analyse the MCMC chains and present the results using the \texttt{GetDist}\footnote{\href{https://github.com/cmbant/getdist}{https://github.com/cmbant/getdist}} Python package \cite{Lewis:2019xzd}.

The likelihood function for the simulated data set of standard siren GW events is constructed according to the effective Gaussian distribution:
\begin{equation}
    \ln \mathcal{L}_{SS} = - \frac{1}{2} \sum_{i=1}^n \left[ \frac{d_{SS}^{\text{(obs)}} (z_i) - d_{SS} (z_i)}{\sigma_{d_{L,i}} } \right]^2 \mathcomma
\end{equation}
where $d_{SS}^{\text{(obs)}} (z)$ is the observed luminosity distance, which in this case corresponds to the samples generated according to the procedure outlined above; $d_{SS} (z)$ is the model-dependent theoretical prediction for the luminosity distance of the event, computed numerically with the modified \texttt{CLASS} code; $\sigma_{d_{L}}$ is the total error in the luminosity distance, as defined in Eq.~\eqref{eq:tots}; and $n$ is the number of observed events.

Since we want to forecast the constraining power of standard siren data probed by ET and LISA on coupled quintessence models, we assess the independent and combined constraints with \textit{current} background data. This allows for a direct comparison of whether GW catalogues will improve the constraints on $\{\Omega_m^0, H_0 \}$ and on the model-specific parameters, affecting the background evolution.
In particular, we include baryonic acoustic oscillations (BAO) data from the Sloan Digital Sky Survey (SDSS) DR7 Main Galaxy Sample \cite{Ross:2014qpa}, SDSS DR12 consensus release \cite{BOSS:2016hvq} and the 6dF Galaxy Survey \cite{Beutler:2011hx}, in combination with distance \textit{moduli} measurements of 1048 type Ia Supernova (SNIa) data from Pantheon \cite{Pan-STARRS1:2017jku}. This combined data set is referred to as ``SNIa+BAO". 

Our analysis involves a set of free sampling parameters, including the baseline $\Lambda$CDM cosmological parameters $(\Omega_m^0, H_0)$ and the parameters associated with each coupled quintessence model, for which we consider as fiducial value their $\Lambda$CDM limit.  The models discussed in Sec.~\ref{sec:res} reduce to $\Lambda$CDM in the following limits: $\lambda=0$ and $\beta=0$ for \ref{sec:casea}; $\lambda=0$ and $\alpha=0$ for \ref{sec:caseb}; $\lambda=0$ and $D_0=0$ for \ref{sec:casec}; $\lambda=0$, $\beta=0$ and $D_0=0$ for \ref{sec:cased}. We adopt flat priors for all parameters within the ranges specified in Table~\ref{tab:priors}.

\begin{table}[ht!]
\begin{center}
\begin{tabular}{|c|c|c|}
\hline
Model & Parameter                    & Prior \\
\hline
\multirow{4}{*}{All} & $\Omega_b h^2$                & $[0.018,0.03]$ \\
& $\Omega_c h^2$                & $[0.1,0.2]$ \\
& $h$                    & $[0.6,0.8]$ \\
& $\lambda$                           & $[0,2]$ \\
\hline
\ref{sec:casea} and \ref{sec:cased} & $\beta$                            & $[0,2]$ \\
\hline
\ref{sec:caseb} & $\alpha$                            & $[0,0.001]$ \\
\hline
\ref{sec:casec} and \ref{sec:cased} & $D_0/\text{meV}^{-1}$                            & $[0,2]$ \\
\hline %
\end{tabular}
\end{center}
\caption{Flat priors on the cosmological and model parameters sampled in Sec.~\ref{sec:res}.}
\label{tab:priors}
\end{table}

\section{Forecast Results}\label{sec:res}

In what follows, we employ the methodology and the data sets discussed in Sec.~\ref{sec:method} to investigate the power that LISA and ET standard sirens have in constraining the cosmological parameters $\{\Omega_m^0,H_0\}$, the model-dependent conformal and disformal coupling parameters and the steepness of the self-interacting potential. In particular, we consider four interacting DE models: a standard coupled quintessence model, a kinetically coupled model, a constant disformal model and a mixed conformal-disformal model. For each of the four scenarios, we provide a brief review of the theoretical framework before presenting the forecasts obtained considering the specifications and assumptions discussed in previous sections.

In each subsection that follows, we show the resulting 2D contours, and 1D marginalised posterior distributions for $\{H_0,\Omega_m^0\}$ plus the set of model-specific parameters (see Table~\ref{tab:priors}) for the cases of ET and LISA and their combination.
These plots also include a combination of SNIa+BAO, and for reference, the results of SNIa+BAO alone. The results are also summarised in a table with their corresponding $1\sigma$, identified in the text with the notation
$\{\sigma_{\text{p}}\}$ where $p$ is an index spanning over the model parameters. We also use
$\mathcal{F}_{\text{p}}^{(\text{i,j})} = \{\sigma_{\text{p}}^{(\text{j})}/\sigma_{\text{p}}^{(\text{i})}\} $ where here i and j stand for two different data sets, to denote the change in error for the specific parameter p.

\subsection{Conformal Coupling} \label{sec:casea}

\begin{table*}[ht!]
\begin{tabular}{|l|l|l|l|l|l|l|l|l|}
\hline
\multicolumn{9}{|c|}{Conformal Coupled Quintessence}  \\ \hline\hline
 Data sets &  $\Omega_m^0$ & $\sigma_{\Omega_m^0}$ & $H_0$ & $\sigma_{H_0}$ & $\beta$ & $\sigma_{\beta}$ & $\lambda$ & $\sigma_{\lambda}$\\ \hline \hline
 SNIa+BAO & $0.3019^{+0.0088}_{-0.0059}$ & $0.0074$ & $73.2^{+4.7}_{-3.5}$ & $4.1$ & $0.085^{+0.055}_{-0.043}$ & $0.049$ &  $0.42^{+0.20}_{-0.36}$ & $0.28$\\ 
 \hline\hline
ET & $0.307^{+0.011}_{-0.0050}$ & $0.0080$ & $67.49^{+0.39}_{-0.34}$ & $0.37$ & $0.115^{+0.060}_{-0.079}$ & $0.070$ & $0.50^{+0.26}_{-0.38}$ & $0.32$ \\
  ET+SNIa+BAO & $0.3046^{+0.0099}_{-0.0051}$ & $0.0075$ & $67.37\pm 0.36$ & $0.36$ & $0.063^{+0.033}_{-0.045}$ & $0.039$ & $0.49^{+0.26}_{-0.35}$ & $0.31$ \\
\hline
\hline 
   LISA  & $0.3039^{+0.0093}_{-0.0049}$ & $0.0071$ & $67.50^{+0.50}_{-0.44}$ & $0.47$ & $0.167^{+0.085}_{-0.11}$ & $0.098$ & $0.33^{+0.15}_{-0.32}$ & $0.24$ \\
   LISA+SNIa+BAO & $0.3028^{+0.0065}_{-0.0036}$ & $0.0051$ & $67.52\pm 0.37$ & $0.37$ & $0.048^{+0.025}_{-0.037}$ & $0.031$ & $0.33^{+0.15}_{-0.29}$ &  $0.22$ \\
  \hline
  \hline 
   ET+LISA  & $0.3079^{+0.0061}_{-0.0034}$ & $0.0048$ & $67.56\pm 0.26$ & $0.26$ & $0.178^{+0.099}_{-0.081}$ & $0.090$ & $0.30^{+0.13}_{-0.27}$ & $0.24$ \\
   ET+LISA+SNIa+BAO & $0.3044^{+0.0063}_{-0.0032}$ & $0.0048$ & $67.45\pm 0.28$ & $0.28$ & $0.052^{+0.028}_{-0.038}$ & $0.033$ & $0.35^{+0.17}_{-0.30}$ &  $0.24$ \\
  \hline
\end{tabular}
\caption{Marginalised constraints on cosmological and model parameters for the Conformal Coupled Quintessence model at 68\% C.L.}
\label{Tab:boundscq}
\end{table*}

The first model we consider is the conformal coupling model, for which
\begin{equation}
    C(\phi) = e^{2 \beta \phi/ M_{\rm Pl}}\quad \text{and}\quad V(\phi) = V_0 e^{- \lambda \phi/ M_{\rm Pl}} \mathcomma
\end{equation}
where $C(\phi)$ is defined according to Eq.~\eqref{conf} and $V(\phi)$ is the DE potential energy. The exponential parameters $\beta$ and $\lambda$ are constant dimensionless parameters and
$V_0$ is a constant with dimensions of (mass)$^4$ that sets the energy scale of the potential\footnote{For numerical purposes and to avoid degeneracies, $V_0$ is not taken to be a free parameter. Instead, it serves as a shooting parameter to set the fiducial value of $\Omega_\phi^0$ fulfilling the flatness condition.}. 
In such models, the mass of the DM particles becomes $\phi$--dependent and the DE field mediates a long-range force between DM particles so that the effective gravitational coupling is given by $G_{\rm eff} = G_N \left(1+2\beta^2 \right)$ \cite{Wetterich:1994bg,Farrar:2003uw,Amendola:2003wa}. The free parameters we are particularly interested in are the slope of the potential $\lambda$ and the coupling parameter $\beta$. Constraints on this model have been obtained in Ref.~\cite{vandeBruck:2016hpz} using background data only ($H(z)$, BAO and supernova Union2.1). Using these data, the authors found the following upper limits: $\beta<0.193$ and $\lambda<1.27$. In \cite{VanDeBruck:2017mua} stronger constraints have been obtained using \textit{Planck} data, BAO and SNIa data, also in line with Ref.~\cite{Gomez-Valent:2020mqn}, in which the authors found $\beta<0.0298$ and $\lambda<0.6$ for the $1\sigma$ upper limits. 

\begin{figure}[t!]
      \subfloat{\includegraphics[width=\linewidth]{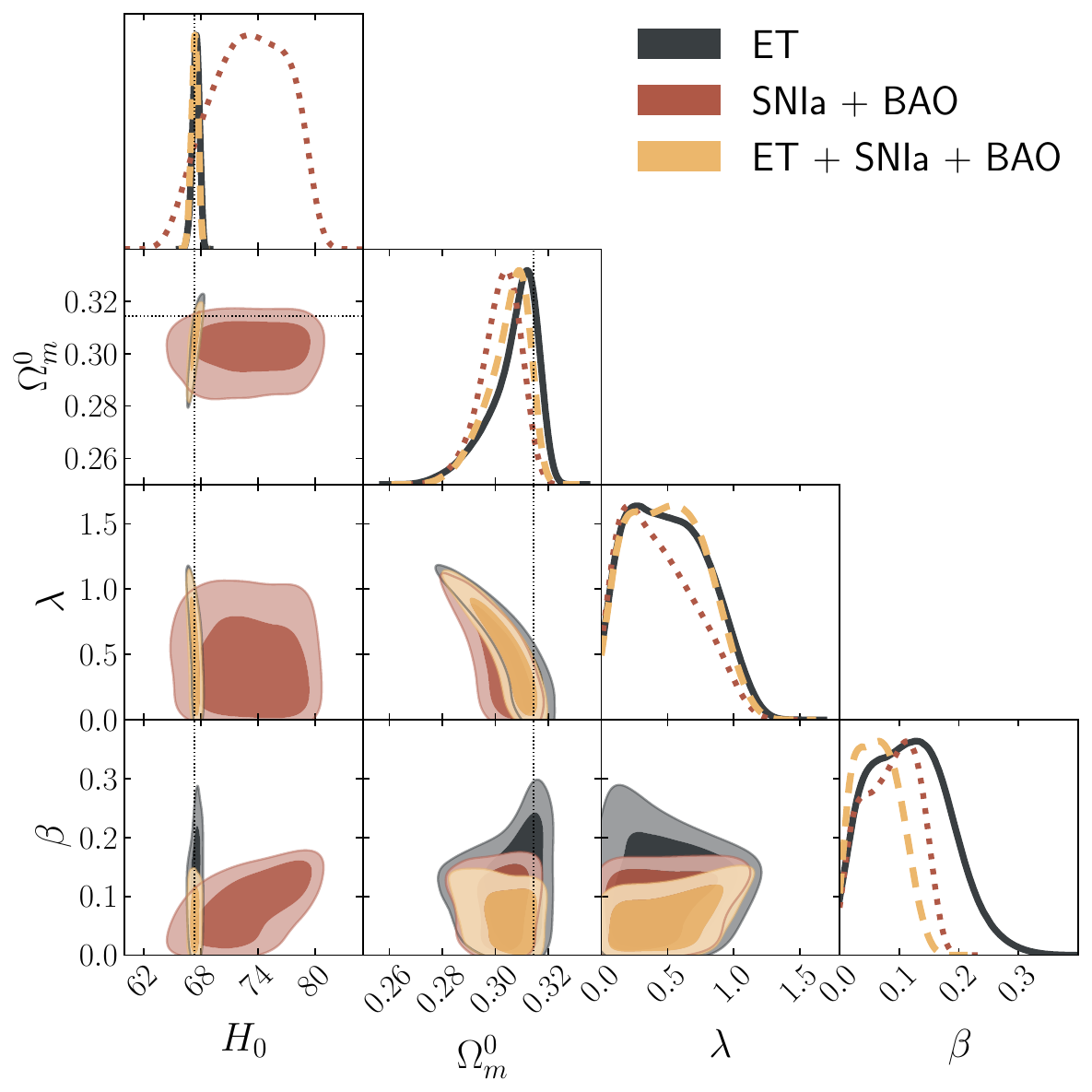}} 
  \caption{\label{fig:etCQ} 68\% and 95\% C.L. 2D contours and 1D marginalised posterior distributions for the parameters $\{H_0,\Omega_m^0, \lambda,\beta\}$ in the conformal coupled quintessence model with the ET mock data (charcoal filled line), SNIa+BAO data (red dotted line) and their combination (yellow dashed line). The dotted lines depict the fiducial values for the mock data $\{\Omega_m^0,H_0 \} = \{0.3144, 67.32\}$.}
\end{figure}

\begin{figure}[t!]
      \subfloat{\includegraphics[width=\linewidth]{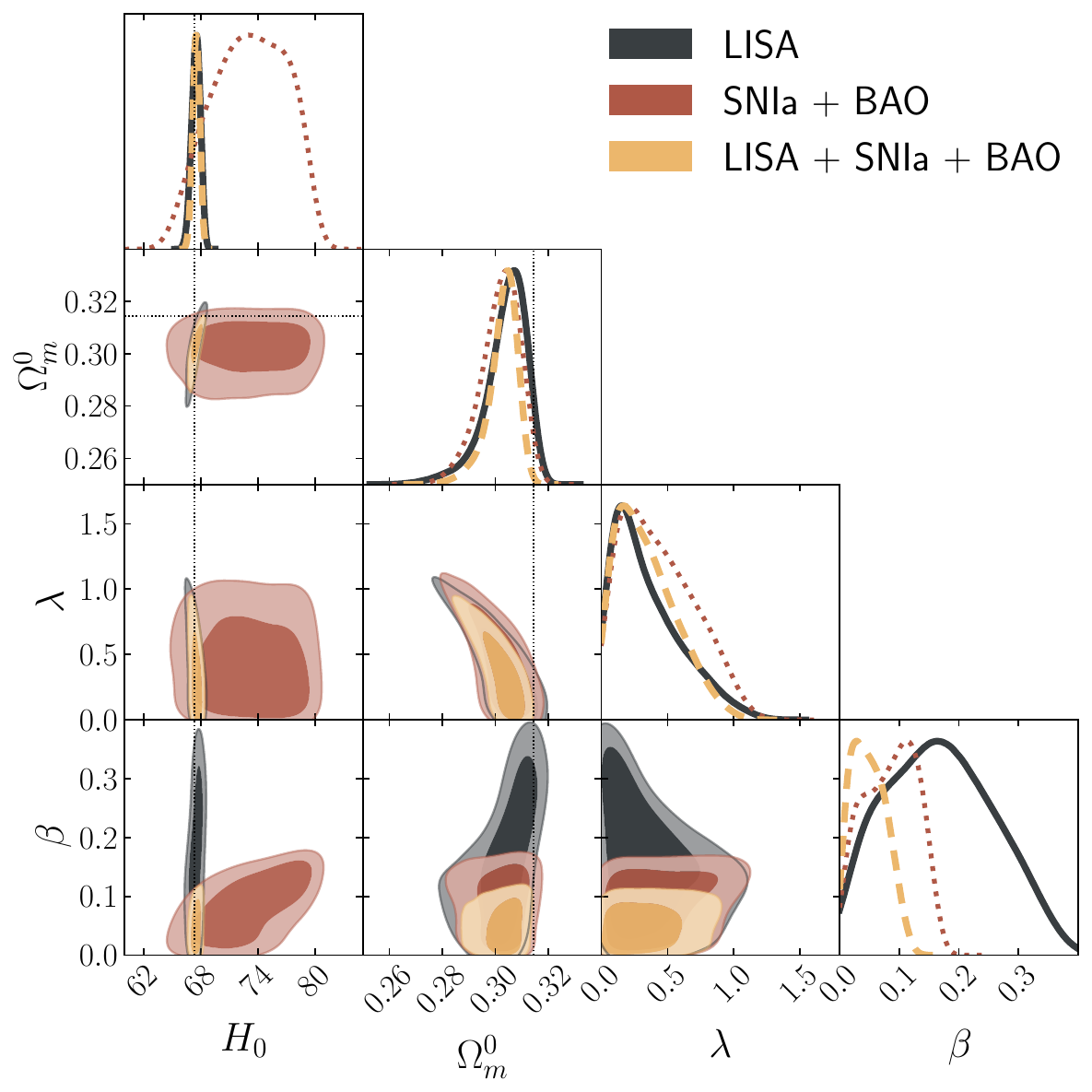}} 
  \caption{\label{fig:lpCQ} 68\% and 95\% C.L. 2D contours and 1D marginalised posterior distributions for the parameters $\{H_0,\Omega_m^0, \lambda,\beta\}$ in the conformal coupled quintessence model with LISA mock data (charcoal filled line), SNIa+BAO data (red dotted line) and their combination (yellow dashed line). The scale is the same as in Fig.~\ref{fig:etCQ} for comparison purposes, with the SNIa+BAO contours standing as the reference. The dotted lines depict the fiducial values for the mock data $\{\Omega_m^0,H_0 \} = \{0.3144, 67.32\}$.}
\end{figure}

\begin{figure}[t!]
      \subfloat{\includegraphics[width=\linewidth]{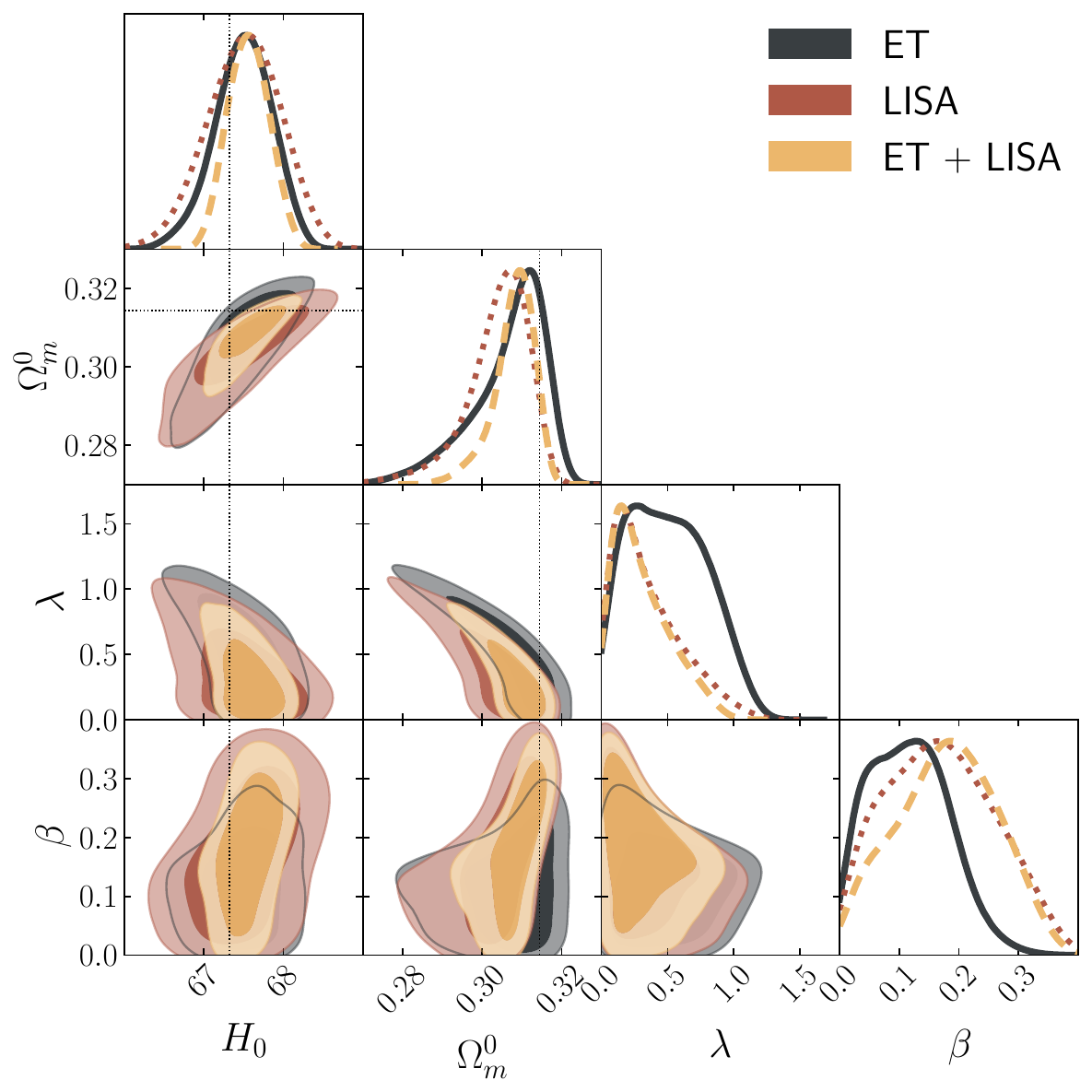}} 
  \caption{\label{fig:etlp2CQ} 68\% and 95\% C.L. 2D contours and 1D marginalised posterior distributions for the parameters $\{H_0,\Omega_m^0, \lambda,\beta\}$ in the conformal coupled quintessence model with ET mock data (charcoal filled line), LISA mock data (red dotted line) and their combination (yellow dashed line). The dotted lines depict the fiducial values for the mock data $\{\Omega_m^0,H_0 \} = \{0.3144, 67.32\}$.}
\end{figure}

According to the results in Figs.~\ref{fig:etCQ}, \ref{fig:lpCQ} and \ref{fig:etlp2CQ}, summarised in Table \ref{Tab:boundscq}, we comment on the resulting constraints for GW data sets compared with SNIa+BAO for the parameters $\{\Omega_m^0,H_0,\beta, \lambda\}$.
When ET standard sirens are considered, we find that the cosmological and model parameters can be constrained at $1\sigma$ with an accuracy $\{0.0080,0.37 ,0.0070,0.32\}$ for ET alone and $ \{0.0075, 0.36,0.039, 0.31\}$ for ET+SNIa+BAO, resulting in a change in error of $\mathcal{F}_{\Omega_m^0, H_0,\beta,\lambda}^{(\text{ET,ET+SNIa+BAO})} = \{0.94,0.97,0.56,0.97\}$. Thus, the forecasted constraints of ET+SNIa+BAO, compared to ET alone, have increased accuracy in all parameters shown by the reduction in $\sigma$. This trend is also present in the LISA data set with the cosmological and model parameters constrained with an accuracy of $ \{0.0071,  0.47, 0.098,0.24\}$ for LISA alone and $ \{0.0051, 0.37 ,0.031,0.22\}$ for LISA+SNIa+BAO, resulting in a reduction in $\sigma$ by a factor of $\mathcal{F}_{\Omega_m^0, H_0,\beta,\lambda}^{(\text{LISA,LISA+SNIa+BAO})} = \{0.72,0.79,0.32,0.92\}$. 
For the combination of just SNIa+BAO we obtain an accuracy of $ \{0.0074,4.1,0.049,0.28\}$. Compared to ET+SNIa+BAO and LISA+SNIa+BAO to SNIa+BAO, there is also a reduction in $\sigma$ for all parameters except one. The ET+SNIa+BAO data set results in a nominal increase in $\sigma_{\Omega_m^0}$ compared to SNIa+BAO. 
Comparing the errors of ET and LISA to SNIa+BAO we see only minor changes in the constraining power regarding $\Omega_m^0$, with ET performing slightly worse and LISA slightly better. A similar trend occurs for the parameter $\lambda$, with ET performing nominally worse and LISA better. However, particular attention should be given to the significant reduction in $\sigma_{H_0}$ when comparing ET and LISA to SNIa+BAO. There is a reduction in the error by a factor of $\mathcal{F}_{H_0}^{(\text{SNIa+BAO,ET})} = 0.090$ and $\mathcal{F}_{H_0}^{(\text{SNIa+BAO,LISA})} = 0.11$. Forecasting GWs will improve the constraints on $H_0$, suggesting that GWs will be critical in addressing the Hubble tension. On the other hand, we see the opposite effect with $\beta$, with an increase in the error by a factor of  $F_{\beta}^{(\text{SNIa+BAO,ET})} = 1.4$ and $\mathcal{F}_{\beta}^{(\text{SNIa+BAO,LISA})} = 2.0$. Nonetheless, when the background data is combined with ET and/or LISA, the constraints improve by $F_{\beta}^{(\text{ET,ET+SNIa+BAO})} = 0.56$ and $\mathcal{F}_{\beta}^{(\text{LISA,LISA+SNIa+BAO})} = 0.32$.
In comparing the constraining power of ET and LISA (see Fig. \ref{fig:etlp2CQ}), it is evident that they have comparable spreads for the cosmological parameters. An interesting feature we observe is that ET is more constraining in regard to $H_0$. We attribute this feature to the fact that the ET catalogue has more data points than LISA at low redshifts, as illustrated in Fig. \ref{fig:ETLISAmock}.

By combining GW data from LISA and ET, which implies an increase of data points over a wide range of redshifts, we predict an enhanced constraining power in the cosmological parameters, $\{H_0,~\Omega_m^0\}$, compared to the SNIa+BAO case, more precisely $\mathcal{F}_{\Omega_m^0, H_0}^{(\text{SNIa+BAO,ET+LISA})} = \{0.65,0.063\}$. However, for the model parameters $\beta$ and $\lambda$, we observe modifications to the constraining power with $\mathcal{F}_{\beta, \lambda}^{(\text{SNIa+BAO,ET+LISA})} = \{1.8,0.86\}$. The combination of ET+LISA with SNIa+BAO results in a negligible change of constraining power for $\Omega_m^0, H_0, \lambda$. Only the parameter $\beta$ is more constrained when the data sets are combined, with the $1\sigma$ reduced by almost a third. 

Compared to the current background constraints mentioned in the beginning of the section, we find in our analysis that the upper bounds at $1\sigma$ on the model parameters are improved in the following cases: $\beta< 0.14$ and $\lambda<0.62$ (SNIa+BAO); $\beta<0.175$ and $\lambda<0.76$ (ET); $\beta<0.096$ and $\lambda<0.75$ (ET+SNIa+BAO); $\lambda<0.48$ (LISA and LISA+SNIa+BAO) and $\beta<0.073$ (LISA+SNIa+BAO); $\lambda<0.43$ (ET+LISA); $\lambda<0.52$ and $\beta<0.08$ (ET+LISA+SNIa+BAO).

\subsection{Kinetic Conformal Coupling} \label{sec:caseb}

As an example of a coupled quintessence model in which the conformal function is less trivial, we focus on a pure dependence on derivatives of the scalar field through the kinetic term of $\phi$, $X= - \partial_{\mu} \phi \partial^{\mu} \phi /2$, to which we refer as the kinetic coupling. Such a setting has been proposed in \cite{Barros:2019rdv} (see references therein as well), and we focus on the particular example of a power law, as studied in \cite{teixeira:2022sjr}. Even though this model is proposed based on a Lagrangian framework ($\mathcal{L}_{\rm DM} \longrightarrow \left(X/M_{\rm Pl}^4 \right)^{\alpha} \mathcal{L}_{\rm DM}$), at the background level it is equivalent to the kinetic-dependent conformal transformation $\bar{g}_{\mu\nu} = C(X) g_{\mu\nu}$, with 

\begin{equation}
  C(X) = \left(M_{\rm Pl}^{-4} X \right)^{2\alpha} \quad \text{and}\quad V(\phi) = V_0 e^{- \lambda \phi / M_{\rm Pl}} \mathperiod
\end{equation}
where $\alpha$ is a dimensionless constant and a simple exponential potential has been assumed just like in the previous case, and the same considerations apply for $\lambda$ and $V_0$. 

In summary, an analysis based on \textit{Planck} and the SNIa+BAO background data in Ref. \cite{teixeira:2022sjr} reveals the power of BAO data in constraining $\Omega^0_m$, which is highly correlated with the steepness of the potential $\lambda$. The coupling parameter $\alpha$ is constrained to be of the order of $10^{-4}$. The constraints on the cosmological parameters are found to be compatible with the $\Lambda$CDM ones within the errors. 
Moreover, a positive correlation between $H_0$ and $\Omega^0_m$ is identified. While this trend is attributed to the evolution of the linear perturbations for non-vanishing $\alpha$, we find that it is still present for the background standard siren data sets.

\begin{table*}[ht!]
\centering
\begin{tabular}{|l|l|l|l|l|l|l|l|l|}
\hline
\multicolumn{9}{|c|}{Kinetic Coupled Quintessence}  \\ \hline\hline
 Data sets &  $\Omega_m^0$ & $\sigma_{\Omega_m^0}$ & $H_0$ & $\sigma_{H_0}$ & $10^4 \alpha$ & $\sigma_{10^4 \alpha}$ & $\lambda$ & $\sigma_{\lambda}$ \\ \hline \hline
 SNIa+BAO & $0.3016^{+0.0075}_{-0.0057}$ & $0.0066$ & $70.4\pm 3.1$ & $3.1$ & $5.1\pm 2.9$ & $2.9$ &  $0.34^{+0.16}_{-0.29}$ & $0.23$ \\ 
\hline
\hline
ET & $0.3067^{+0.0093}_{-0.0046}$ & $0.0070$ & $67.45\pm 0.36$ & $0.36$ & $4.8\pm 2.9$ & $2.9$ & $0.41^{+0.20}_{-0.31}$ & $0.26$ \\
  ET+SNIa+BAO & $0.3062^{+0.0074}_{-0.0043}$ & $0.0059$ & $67.36\pm 0.33$ & $0.33$ & $5.0\pm 2.9$ & $2.9$ & $0.37^{+0.19}_{-0.28}$ & $0.24$ \\
  \hline
\hline
   LISA  & $0.2997^{+0.0079}_{-0.0041}$ & $0.0060$ & $67.30\pm 0.39$ & $0.39$ & $4.9\pm 2.9$ & $2.9$ & $0.34^{+0.16}_{-0.30}$ & $0.23$ \\
   LISA+SNIa+BAO & $0.3024^{+0.0058}_{-0.0035}$ & $0.0047$ & $67.47\pm 0.36$ & $0.36$ & $5.0\pm 2.9$ & $2.9$ &  $0.29^{+0.13}_{-0.26}$ & $0.20$ \\
  \hline
  \hline
   ET+LISA & $0.3040^{+0.0058}_{-0.0031}$ & $0.0045$ & $67.42\pm 0.26$ & $0.26$ & $5.1\pm 2.9$ & $2.9$ & $0.31^{+0.15}_{-0.26}$ & $0.21$ \\
   ET+LISA+SNIa+BAO & $0.3040^{+0.0058}_{-0.0031}$ & $0.0045$ & $67.42\pm 0.27$ & $0.27$ & $4.9\pm 2.9$ & $2.9$ &  $0.29^{+0.14}_{-0.25}$ & $0.20$ \\
  \hline
\end{tabular}
\caption{Marginalised constraints on cosmological and model parameters for the Kinetic Model at 68\% C.L. }
\label{Tab:boundskcq}
\end{table*}

\begin{figure}[t!]
      \subfloat{\includegraphics[width=\linewidth]{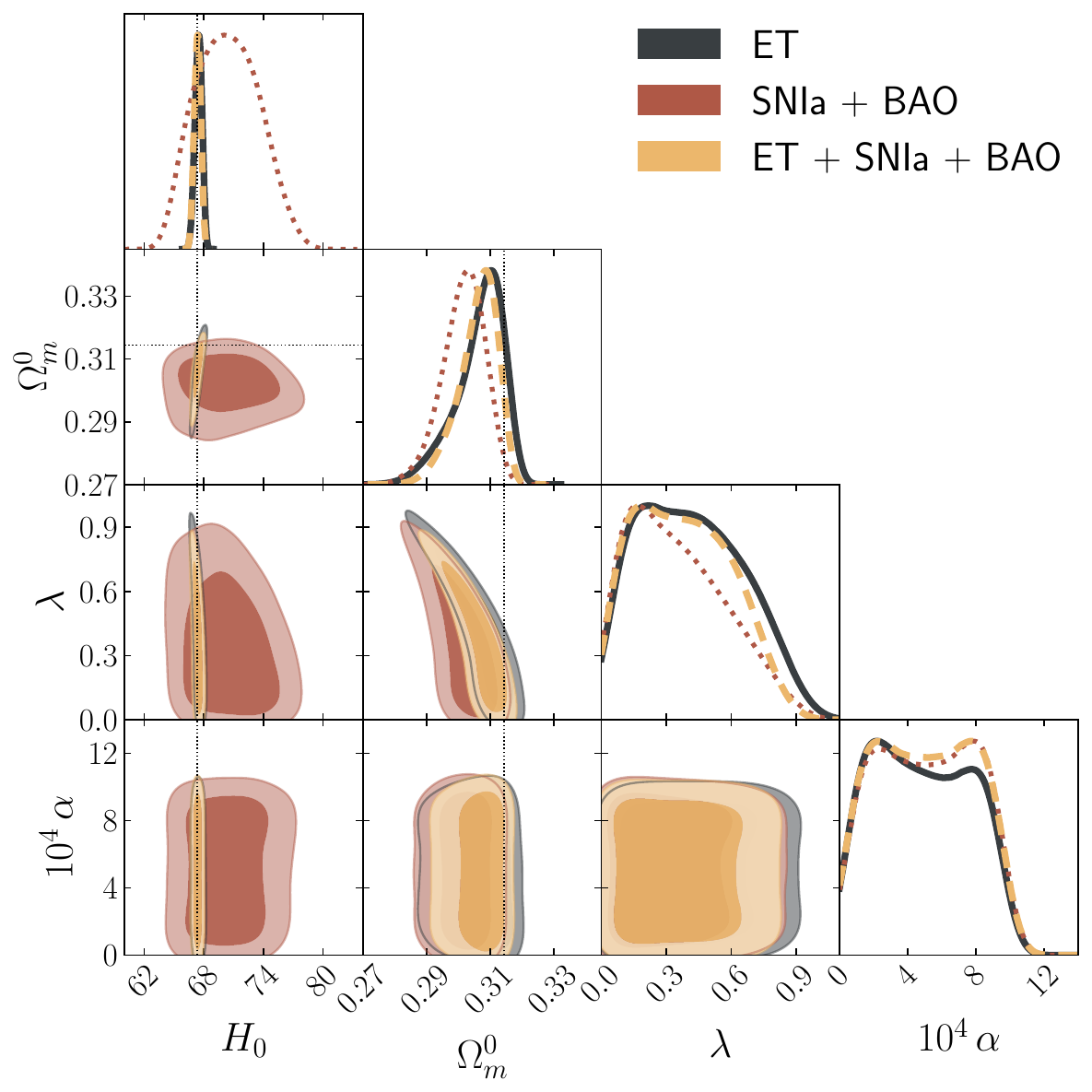}} 
  \caption{\label{fig:etk}68\% and 95\% C.L. 2D contours and 1D posterior distributions for the parameters $\{H_0,\Omega_m^0,\lambda,10^4\alpha\}$ in the kinetic conformal coupled quintessence model with ET (charcoal filled line), SNIa+BAO (red dotted line) data  and their combination (yellow dashed line). The dotted lines depict the fiducial values for the mock data $\{\Omega_m^0,H_0 \} = \{0.3144, 67.32\}$.}
\end{figure}

\begin{figure}[t!]
      \subfloat{\includegraphics[width=\linewidth]{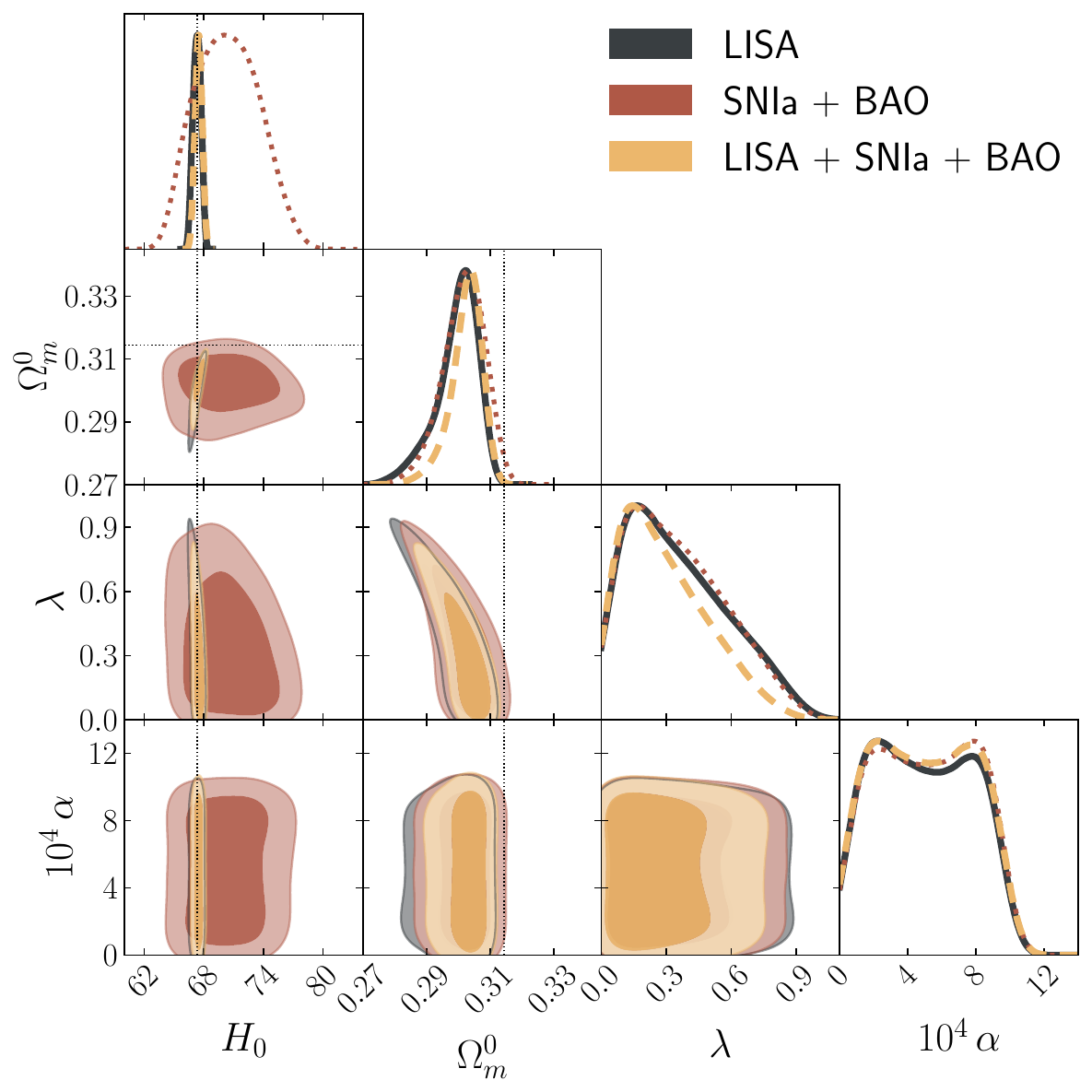}} 
  \caption{\label{fig:lpk} 68\% and 95\% C.L. 2D contours and 1D posterior distributions for the parameters $\{H_0,\Omega_m^0,\lambda,10^4\alpha\}$ in the kinetic conformal coupled quintessence model with LISA mock data (charcoal filled line), SNIa+BAO (red dotted line) data  and their combination (yellow dashed line). The scale is the same as in Fig.~\ref{fig:etk} for comparison purposes, with the SNIa+BAO contours standing as the reference. The dotted lines depict the fiducial values for the mock data $\{\Omega_m^0,H_0 \} = \{0.3144, 67.32\}$.}
\end{figure}

\begin{figure}[t!]
      \subfloat{\includegraphics[width=\linewidth]{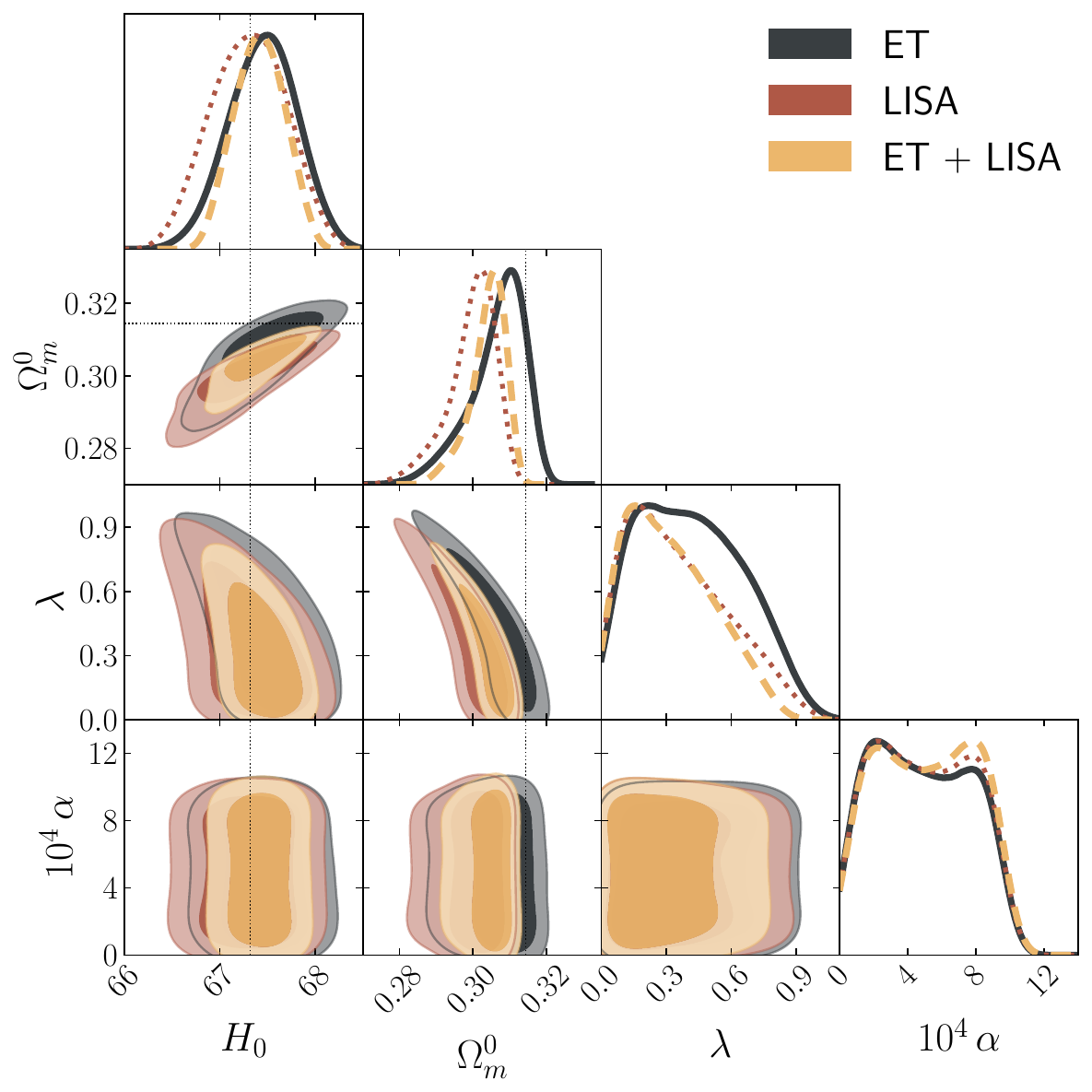}} 
  \caption{\label{fig:etlp2k} 68\% and 95\% C.L. 2D contours and 1D marginalised posterior distributions for the parameters $\{H_0,\Omega_m^0, \lambda,10^4\alpha\}$ in the kinetic conformal coupled quintessence model with ET mock data (charcoal filled line), LISA mock data (red dotted line) and their combination (yellow dashed line). The dotted lines depict the fiducial values for the mock data $\{\Omega_m^0,H_0 \} = \{0.3144, 67.32\}$.}
\end{figure}

From the results presented in Figures \ref{fig:etk}, \ref{fig:lpk} and \ref{fig:etlp2k}, and summarised in Table \ref{Tab:boundskcq}, we analyse the constraints on the parameters $\{\Omega_m^0,H_0,\lambda, 10^4\alpha\}$ for the same data sets as in the previous case.
When evaluating the errors from ET standard sirens and comparing them to SNIa+BAO data, we observe that for most parameters, ET's $1\sigma$ constraints are of the same order, apart from the $H_0$ parameter, which is improved by 1 order of magnitude. This reduction is quantified by the fractional change of $\mathcal{F}_{\Omega_m^0, H_0,\beta,\lambda}^{(\text{SNIa+BAO,ET})} = \{1.1,0.12,1.0,1.1\}$. When the data sets are combined (ET+SNIa+BAO), we find that the $1\sigma$ region is narrower for all parameters compared to ET alone, with $\mathcal{F}_{\Omega_m^0, H_0, \alpha, \lambda}^{(\text{ET, ET+SNIa+BAO})} = \{0.84, 0.92, 1.0, 0.92\}$. 
In the case of LISA standard sirens, we observe that all cosmological and model parameters are better or equally constrained by LISA alone compared to SNIa+BAO, with $\mathcal{F}_{\Omega_m^0, H_0, \alpha, \lambda}^{(\text{SNIa+BAO,LISA})} = \{0.91, 0.13, 1.0, 1.0\}$.
Combining LISA with SNIa+BAO, we find improved constraints with respect to the SNIa+BAO data set alone. Moreover, when comparing LISA+SNIa+BAO with LISA alone, the former shows an even better constraining power, with the most significant reduction in error observed for $\Omega^0_m$, with $\mathcal{F}_{\Omega_m^0, H_0, \alpha, \lambda}^{(\text{LISA,LISA+SNIa+BAO})} = \{0.78, 0.92, 1.0, 0.87\}$.
For both the ET and LISA data sets, the accuracy on $H_0$ can be improved by 1 order of magnitude ($0.36$ for ET and $0.39$ for LISA) compared to SNIa+BAO ($3.1$), as reported in \cref{sec:casea} as well. Interestingly, for all data sets and combinations, the accuracy of the model parameters remains largely unaffected, with the $1\sigma$ region for $\lambda$ showing only nominal changes and remaining unchanged for $\alpha$.
Comparing the constraining power of ET and LISA with their combination, ET+LISA, we see that the latter provides better constraining power for the cosmological parameters than any of the other data sets analysed. Regarding the model parameters, there seems to be a minimal change in accuracy compared to the single ET or LISA data sets. We do note that the GW combination provides better accuracy with respect to SNIa+BAO, with $\mathcal{F}_{\Omega_m^0, H_0, \alpha,\lambda}^{(\text{SNIa+BAO,ET+LISA})} = \{0.68,0.084, 1.0, 0.91\}$. The full combination of ET+LISA+SNIa+BAO, has a negligible change in the constraints when compared to ET+LISA for all parameters.

Comparing the accuracy of the constraints of the Kinetic model obtained in Ref. \cite{teixeira:2022sjr} with CMB TT, TE and EE  \textit{Planck} 2018, \textit{Planck} CMB lensing, BAO  and SNIa data, we note that the parameter $\alpha$ is better constrained by CMB data and its combination with BAO and SNIa by 1 order of magnitude when compared to our data combinations, given that the latter only depend on the background evolution. More precisely, we report $\sigma_{10^4\alpha}=2.9$ for all the data set combinations while in Ref. \cite{teixeira:2022sjr} this was reduced to $\sigma_{10^4\alpha}=0.95$ (Plk18), $0.84$ (Plk18+SNIa+BAO), and $0.7$ (Plk18+SNIa+BAO+Lensing).
Future ET and LISA catalogues will be able to constrain $\lambda$ at the same level as \textit{Planck} CMB data ($\sigma_\lambda=0.48$ with Plk18 and $\sigma_\lambda=0.2$ with both Plk18+SNIa+BAO and Plk18+SNIa+BAO+Lensing). On the other hand, the standard siren data will better constrain $H_0$ by 1 order of magnitude with respect to Plk18 ($\sigma_{H_0}=2.5$). CMB lensing data increase the constraint by 1 order of magnitude, namely with accuracy $\sigma_{H_0}=0.6$, which is of the same order of magnitude as the ET and LISA cases, even though the standard sirens perform better in terms of the relative error with $\sigma_{H_0} <0.4$ for all the combinations considered.

\subsection{Disformal Coupling} \label{sec:casec}

In the following we study the model with disformal coupling only, 

\begin{equation}
    C=1\mathcomma\ \ \ D= D_0^4\quad \text{and}\quad V(\phi) = V_0 e^{-\lambda \phi/ M_{\rm Pl}} \mathcomma
\end{equation}
in which case the conformal contribution vanishes and $D$ is simply a constant with dimensions of (mass)$^{-4}$ in Eq.~\eqref{disf} (and hence $D_0$ has units of (mass)$^{-1}$), and $V(\phi)$ follows the same considerations as in the previous cases. The constraints on this model have been obtained in \cite{vandeBruck:2016hpz} and \cite{VanDeBruck:2017mua}. It was found that using background data only ($H(z)$, BAO and supernova Union2.1 data) results in the following constraints: $D_0 > 0.07$ meV$^{-1}$ and $\lambda < 1.56$ at 95.4\% \cite{vandeBruck:2016hpz}. An upper bound can be obtained for $D_0$ with CMB data (including lensing) and BAO, SNIa, cosmic chronometers, cluster abundance, and $H_0$ priors  which is $D_0< 0.2500$ meV$^{-1}$ and a stringent upper limit for $\lambda$ is $< 0.6720$ at $1\sigma$ \cite{VanDeBruck:2017mua}.

\begin{table*}[ht!]
\centering
\begin{tabular}{|l|l|l|l|l|l|l|l|l|}
\hline
\multicolumn{9}{|c|}{Constant Disformal Coupled Quintessence}  \\ \hline\hline
 Data sets &  $\Omega_m^0$ & $\sigma_{\Omega_m^0}$ & $H_0$ &  $\sigma_{H_0}$ & $D_0/\text{meV}^{-1}$ & $\sigma_{D_0}$ & $\lambda$ & $\sigma_{\lambda}$ \\ \hline \hline
 SNIa+BAO & $0.315\pm 0.017$ & $0.017$ & $70.5\pm 3.1$ & $3.1$ & $1.20^{+0.65}_{-0.38}$ & $0.52$ & $0.87^{+0.59}_{-0.76}$ & $0.68$ \\ 
 \hline
 \hline
ET & $0.290^{+0.011}_{-0.013}$ & $0.012$ & $67.58^{+0.36}_{-0.27}$ & $0.32$ & $1.06\pm 0.51$ & $0.51$ & $1.06\pm 0.58$ & $0.58$\\
  ET+SNIa+BAO & $0.298^{+0.011}_{-0.014}$ & $0.013$ & $67.45\pm 0.31$ & $0.31$ & $1.15^{+0.66}_{-0.44}$ & $0.55$ & $0.92\pm 0.58$ & $0.58$\\
  \hline
  \hline
   LISA  & $0.320\pm 0.012$ & $0.012$ & $67.43\pm 0.33$ & $0.33$ & $1.22^{+0.64}_{-0.38}$ & $0.51$ & $0.87\pm 0.58$ & $0.58$ \\
   LISA+SNIa+BAO & $0.317\pm 0.012$ & $0.012$ & $67.52\pm 0.34$ & $0.34$ & $1.24^{+0.64}_{-0.36}$ & $0.50$ &  $0.86^{+0.65}_{-0.77}$ & $0.71$\\
  \hline
   \hline
   ET+LISA  & $0.3094^{+0.0087}_{-0.0099}$ & $0.0093$ & $67.49\pm 0.22$ & $0.22$ & $1.23^{+0.63}_{-0.36}$ & $0.50$ & $0.88\pm 0.58$ & $0.58$ \\
   ET+LISA+SNIa+BAO & $0.3100^{+0.0092}_{-0.0100}$ & $0.0096$ & $67.47\pm 0.25$ & $0.25$ & $1.24^{+0.63}_{-0.36}$ & $0.50$ &  $0.88^{+0.68}_{-0.77}$ & $0.73$\\
  \hline
\end{tabular}
\caption{Marginalised constraints on cosmological and model parameters for the Constant Disformal Coupled Quintessence Model at 68\% C.L. }
\label{Tab:boundscdcq}
\end{table*}

\begin{figure}[t!]
      \subfloat{\includegraphics[width=\linewidth]{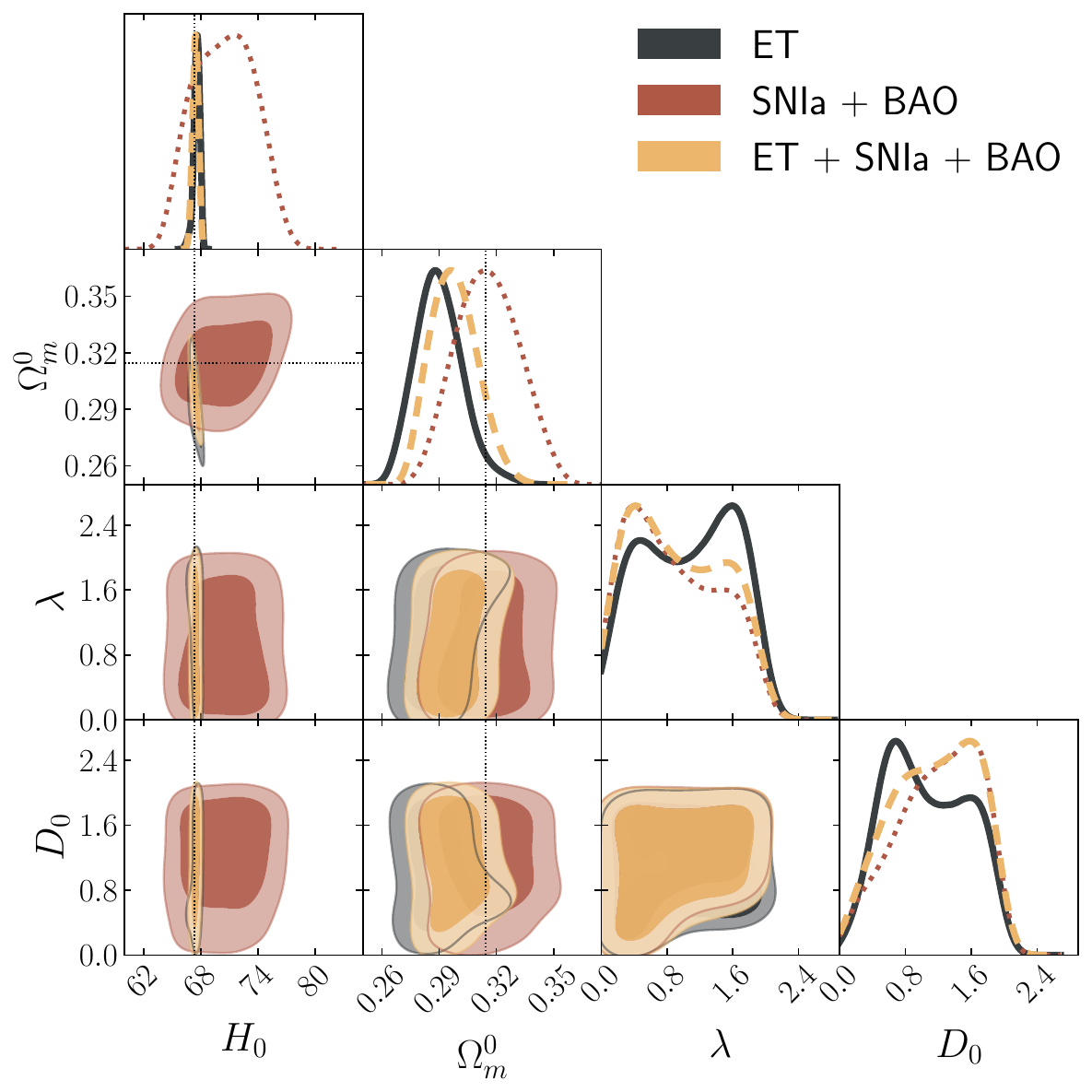}} 
  \caption{\label{fig:etdc} 
  68\% and 95\% C.L. 2D contours and 1D posterior distributions for the parameters $\{H_0,\Omega_m^0,\lambda, D_0\}$ in the constant disformal coupled quintessence model with ET (charcoal filled line), SNIa+BAO (red dotted line) data  and their combination (yellow dashed line). The dotted lines depict the fiducial values for the mock data $\{\Omega_m^0,H_0 \} = \{0.3144, 67.32\}$.}
\end{figure}

\begin{figure}[t!]
      \subfloat{\includegraphics[width=\linewidth]{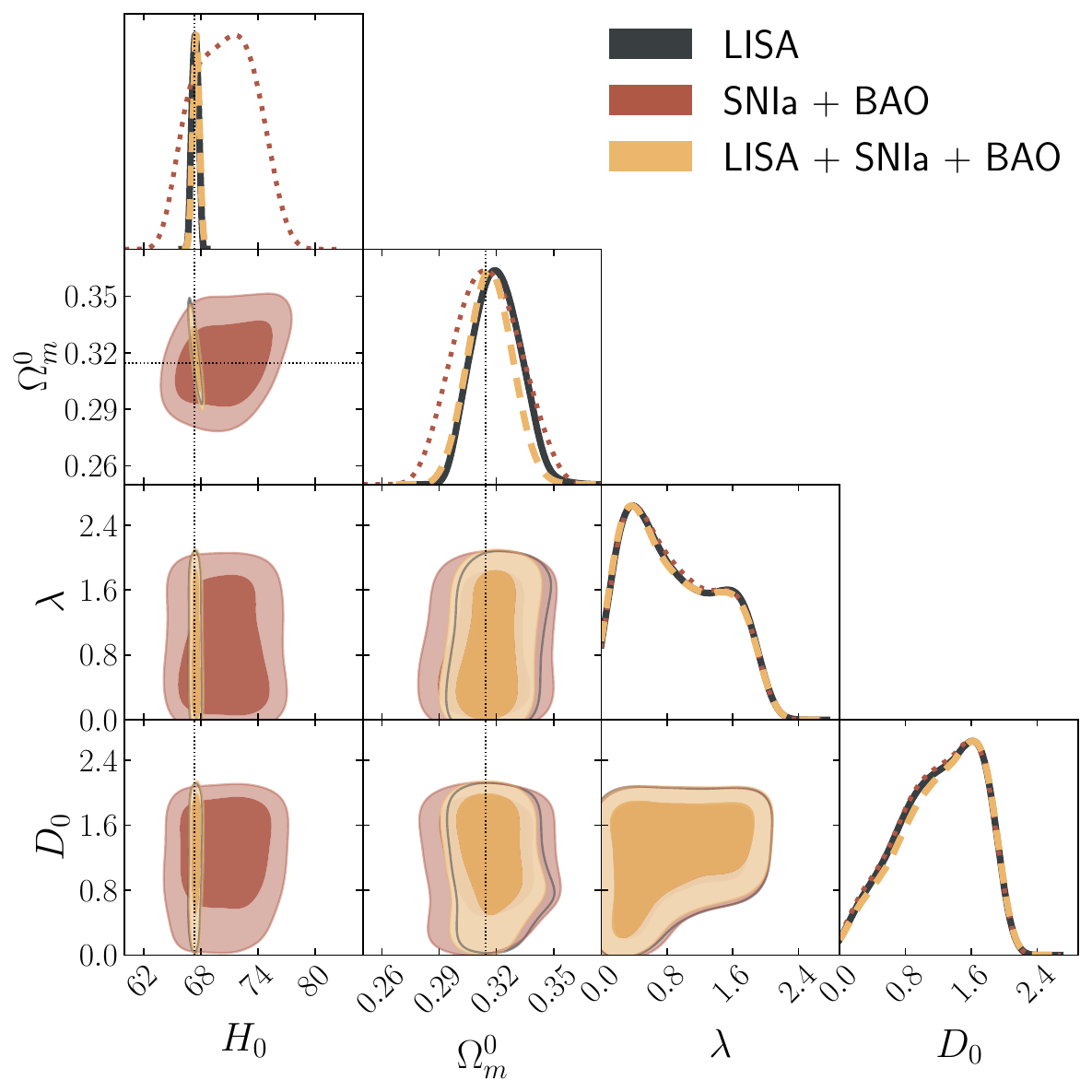}} 
  \caption{\label{fig:lpdc} 68\% and 95\% C.L. 2D contours and 1D posterior distributions for the parameters $\{H_0,\Omega_m^0,\lambda, D_0\}$ in the constant disformal coupled quintessence model with LISA (charcoal filled line), SNIa+BAO (red dotted line) data  and their combination (yellow dashed line). The scale is the same as in Fig.~\ref{fig:etdc} for comparison purposes, with the SNIa+BAO contours standing as the reference. The dotted lines depict the fiducial values for the mock data $\{\Omega_m^0,H_0 \} = \{0.3144, 67.32\}$.}
\end{figure}

\begin{figure}[t!]
      \subfloat{\includegraphics[width=\linewidth]{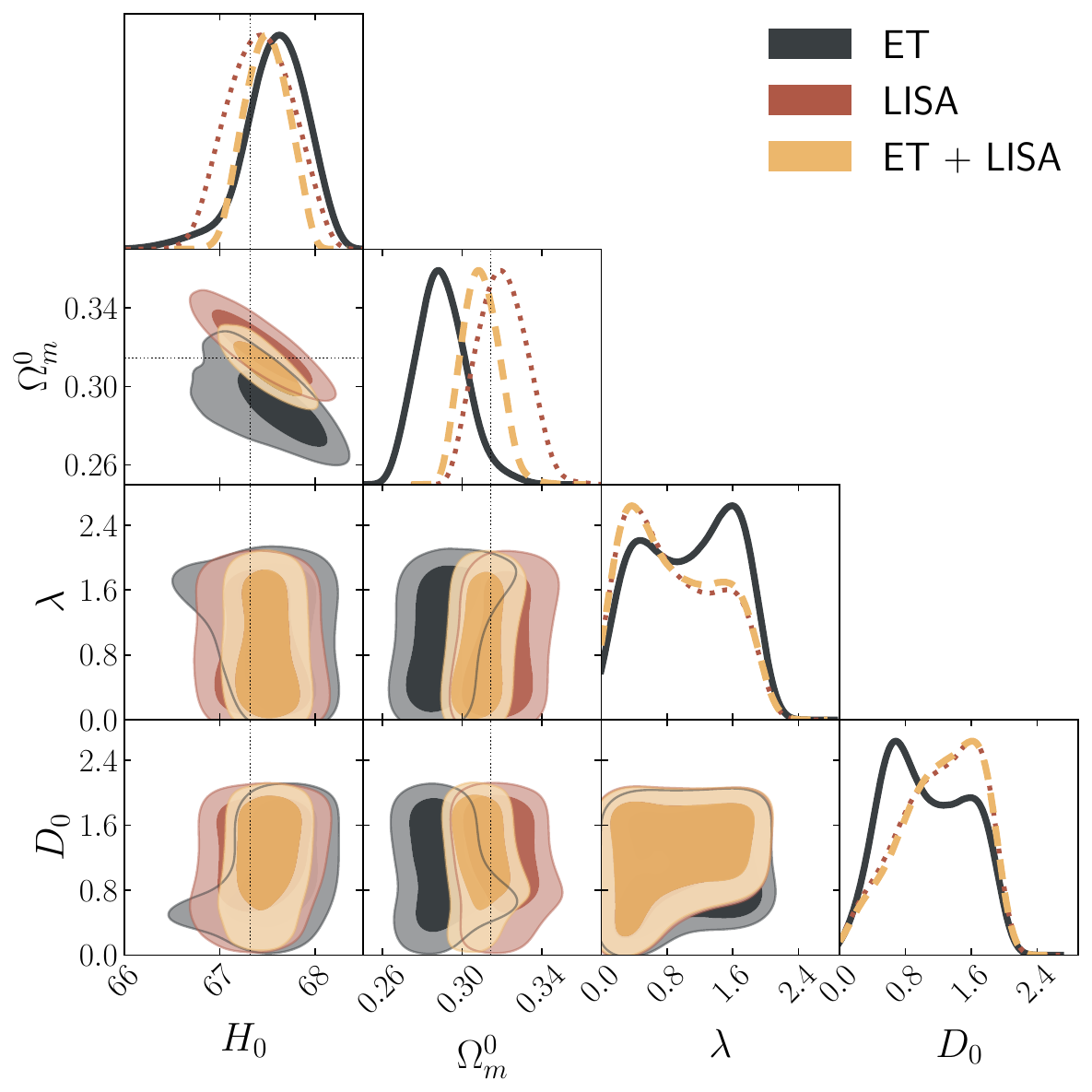}} 
  \caption{\label{fig:etlp2dc} 68\% and 95\% C.L. 2D contours and 1D marginalised posterior distributions for the parameters $\{H_0,\Omega_m^0,\lambda, D_0\}$ in the constant disformal coupled quintessence model with ET mock data (charcoal filled line), LISA mock data (red dotted line) and their combination (yellow dashed line). The dotted lines depict the fiducial values for the mock data $\{\Omega_m^0,H_0 \} = \{0.3144, 67.32\}$.}
\end{figure}

From Figs. \ref{fig:etdc}, \ref{fig:lpdc} and \ref{fig:etlp2dc}, summarised in Table \ref{Tab:boundscdcq}, we analyse the results for the parameters $\{\Omega_m^0,H_0,D_0,\lambda\}$  for the same data sets as in the previous cases.
In our analysis of the ET data set alone, we observe an improved accuracy for all cosmological and model parameters, $\{\Omega_m^0, H_0, D_0, \lambda\}$, compared to SNIa+BAO, with $\mathcal{F}_{\Omega_m^0, H_0, D_0, \lambda}^{(\text{SNIa+BAO,ET})} = \{0.71,0.10, 0.98,0.85\}$. As expected, the combination of data sets, ET+SNIa+BAO, also results in improved accuracy compared to SNIa+BAO. Compared to the ET data set alone, there are only minor changes in the parameters' accuracy, $\mathcal{F}_{\Omega_m^0, H_0, D_0, \lambda}^{(\text{ET,ET+SNIa+BAO})} = \{1.1,0.97, 1.1, 1.0\}$.
In the case of LISA standard sirens, we find that the cosmological and model parameters follow a similar accuracy trend with $\mathcal{F}_{\Omega_m^0, H_0, D_0, \lambda}^{(\text{SNIa+BAO,LISA})} = \{0.71,0.11, 0.98,0.85\}$. Moreover, the same is true for the combination LISA+SNIa+BAO, with $\mathcal{F}_{\Omega_m^0, H_0, D_0, \lambda}^{(\text{LISA,LISA+SNIa+BAO})} = \{1.0,1.0, 0.98, 1.22\}$. There is only a nominal change in the accuracy of parameters compared to that of LISA alone, apart from $\lambda$, which results in a larger $1\sigma$ region, with $\sigma_\lambda = 0.71$.
Regardless of the data combination, $\{\Omega_m^0,D_0,\lambda\}$ are constrained at the same level, with the parameter $\lambda$ having a slight improvement in accuracy for both ET and LISA (both have $\sigma_\lambda=0.58$ compared to $\sigma_\lambda=0.7$ for SNIa+BAO). The accuracy of the $H_0$ parameter is 1 order of magnitude better for both ET and LISA than SNIa+BAO.

There is no change in the model parameters for ET and LISA, and thus we see no noticeable change in the constraints for ET+LISA. However, there is an increase in accuracy for the cosmological parameters. As both ET and LISA improved the constraints compared to SNIa+BAO, we observe the expected result, that ET+LISA have further improved constraints with $\mathcal{F}_{\Omega_m^0, H_0, D_0,\lambda}^{(\text{SNIa+BAO,ET+LISA})} = \{0.55,0.071, 0.96, 0.85\}$. Following the trend with LISA and the combination with SNIa+BAO, we note that the combined data set ET+LISA+SNIa+BAO, has very little change in the accuracy compared to ET+LISA apart from the constraint for $\lambda$, which results in a worse accuracy than ET+LISA and SNIa+BAO with $\mathcal{F}_{\Omega_m^0, H_0, D_0,\lambda}^{(\text{ET+LISA,ET+LISA+SNIa+BAO})} = \{1.0,1.1,1.0,1.3\}$.

The first thing to be noted in comparison with the results reported in Ref.~\cite{vandeBruck:2016hpz} is that we were able to derive constraints at $68\%$ C.L. and not only at $95\%$ C.L. for all the model parameters, therefore providing better constraints in all the cases. Moreover, the constraining power on $H_0$ is largely improved from $\sigma_{H_0} \approx 2.2$ for the background data to $\sigma_{H_0} \approx 0.3$ in all the cases, including standard sirens. When compared with the results of CMB, CMB lensing and additional data in Ref.~\cite{VanDeBruck:2017mua}, which report only upper bounds for $\lambda$ and $D_0$, we see that both parameters are constrained at $68\%$ C.L. with standard sirens, with lower and upper bounds, in particular with more accommodating upper bounds, as this analysis includes only background data. Accordingly, the error in $H_0$ is brought to the same order of magnitude with $\sigma_{H_0} \approx 0.9$, which is still about 3 times larger than the one reported in this analysis.

\subsection{Mixed Conformal-Disformal Coupling}\label{sec:cased}

Finally, we discuss a model with a mixed coupling consisting of a conformal and a disformal part. Specifically, we consider 
\begin{align}
&C(\phi) = e^{2 \beta \phi/ M_{\rm Pl}}, \quad D(\phi) = D_0^4  \quad \mbox{and} \nonumber \\
&V(\phi) = V_0 e^{-\lambda \phi/ M_{\rm Pl}} \mathperiod
\end{align}
As for the disformal case, the constraints on such a model were discussed in \cite{vandeBruck:2016hpz} and \cite{VanDeBruck:2017mua}. For the same background data it has been reported that $D_0 >0.102$~meV$^{-1}$, $\beta <0.453$ and $\lambda<1.59$ at $95.4\%$ \cite{vandeBruck:2016hpz}. An example of the constraints including CMB data in \cite{VanDeBruck:2017mua} are $\beta \lesssim 0.17$ and $\lambda \lesssim 0.35$ at $1\sigma$, with the details depending on the data sets used, with the disformal coupling $D_0$ not always well constrained for this case, with lower bounds of $D_0 \gtrsim 0.35$~meV$^{-1}$ for some data combinations.

\begin{table*}[ht!]
\centering
\begin{tabular}{|l|l|l|l|l|l|l|l|l|l|l|}
\hline
\multicolumn{11}{|c|}{Mixed Conformal-Disformal Coupled Quintessence}  \\ \hline\hline
 Data sets &  $\Omega_m^0$ & $\sigma_{\Omega_m^0}$ & $H_0$ & $\sigma_{H_0}$ & $\beta$ & $\sigma_{\beta}$ & $D_0/\text{meV}^{-1}$ & $\sigma_{D_0}$ & $\lambda$ & $\sigma_{\lambda}$ \\ \hline \hline
 SNIa+BAO & $0.308^{+0.021}_{-0.015}$ & $0.018$ & $71.2\pm 3.3$ & $3.30$ & $1.01\pm 0.57$ & $0.57$ & $1.23^{+0.59}_{-0.43}$ & $0.51$ &  $0.98\pm 0.57$ & $0.57$\\ 
 \hline\hline
ET & $0.286^{+0.010}_{-0.012}$ & $0.011$ & $67.65\pm 0.29$ & $0.29$ & $0.85\pm 0.58$ & $0.58$ & $1.27^{+0.58}_{-0.35}$ & $0.47$ & $1.03\pm 0.58$ & $0.58$ \\
  ET+SNIa+BAO & $0.294^{+0.011}_{-0.013}$ & $0.012$ & $67.50\pm 0.30$ & $0.30$ & $0.92^{+0.66}_{-0.76}$ & $0.71$ & $1.32^{+0.53}_{-0.35}$ & $0.44$ & $0.97\pm 0.58$ & $0.58$\\
   \hline\hline
   LISA  & $0.310^{+0.017}_{-0.0087}$ & $0.013$ & $67.55^{+0.27}_{-0.31}$ & $0.29$ & $0.97\pm 0.56$ & $0.56$ & $1.15^{+0.63}_{-0.44}$ & $0.54$ & $1.01\pm 0.56$ & $0.56$\\
   LISA+SNIa+BAO & $0.310^{+0.016}_{-0.010}$ & $0.013$ & $67.59\pm 0.33$ & $0.33$ & $1.01\pm 0.58$ & $0.58$ & $1.25^{+0.53}_{-0.43}$ & $0.48$ &  $0.98\pm 0.57$ & $0.57$\\
  \hline
  \hline
   ET+LISA & $0.302^{+0.0120}_{-0.0058}$ & $0.0089$ & $67.54\pm 0.20$ & $0.20$ & $0.92\pm 0.55$ & $0.55$ & $1.09^{+0.76}_{-0.42}$ & $0.59$ & $1.05^{+0.71}_{-0.56}$ & $0.64$\\
   ET+LISA+SNIa+BAO & $0.304^{+0.0120}_{-0.0089}$ & $0.0105$ & $67.53\pm 0.24$ & $0.24$ & $0.98\pm 0.57$ & $0.57$ & $1.27^{+0.53}_{-0.41}$ & $0.47$ &  $0.97\pm 0.57$ & $0.57$\\
  \hline
\end{tabular}
\caption{Marginalised constraints on cosmological and model parameters for the Mixed Conformal-Disformal Coupled Quintessence Model at 68\% C.L. }
\label{Tab:boundsfdcq}
\end{table*}

\begin{figure}[t!]
      \subfloat{\includegraphics[width=\linewidth]{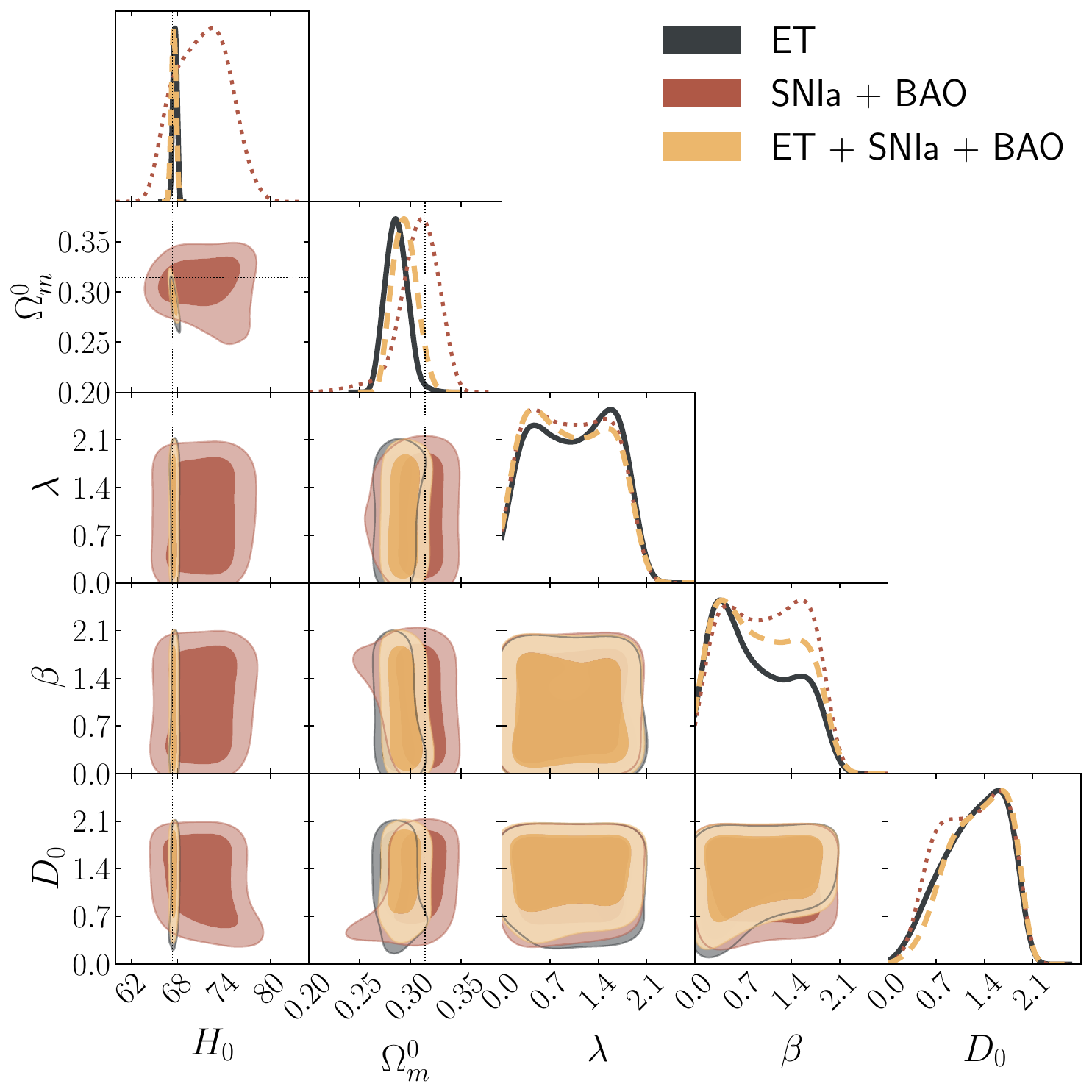}} 
  \caption{\label{fig:etdf} 68\% and 95\% C.L. 2D contours and 1D posterior distributions for the parameters $\{ H_0,\Omega_m^0,\lambda, \beta, D_0\}$ in the mixed conformal-disformal coupled quintessence model with ET (charcoal filled line), SNIa+BAO (red dotted line) data  and their combination (yellow dashed line). The dotted lines depict the fiducial values for the mock data $\{\Omega_m^0,H_0 \} = \{0.3144, 67.32\}$.}
\end{figure}

\begin{figure}[t!]
      \subfloat{\includegraphics[width=\linewidth]{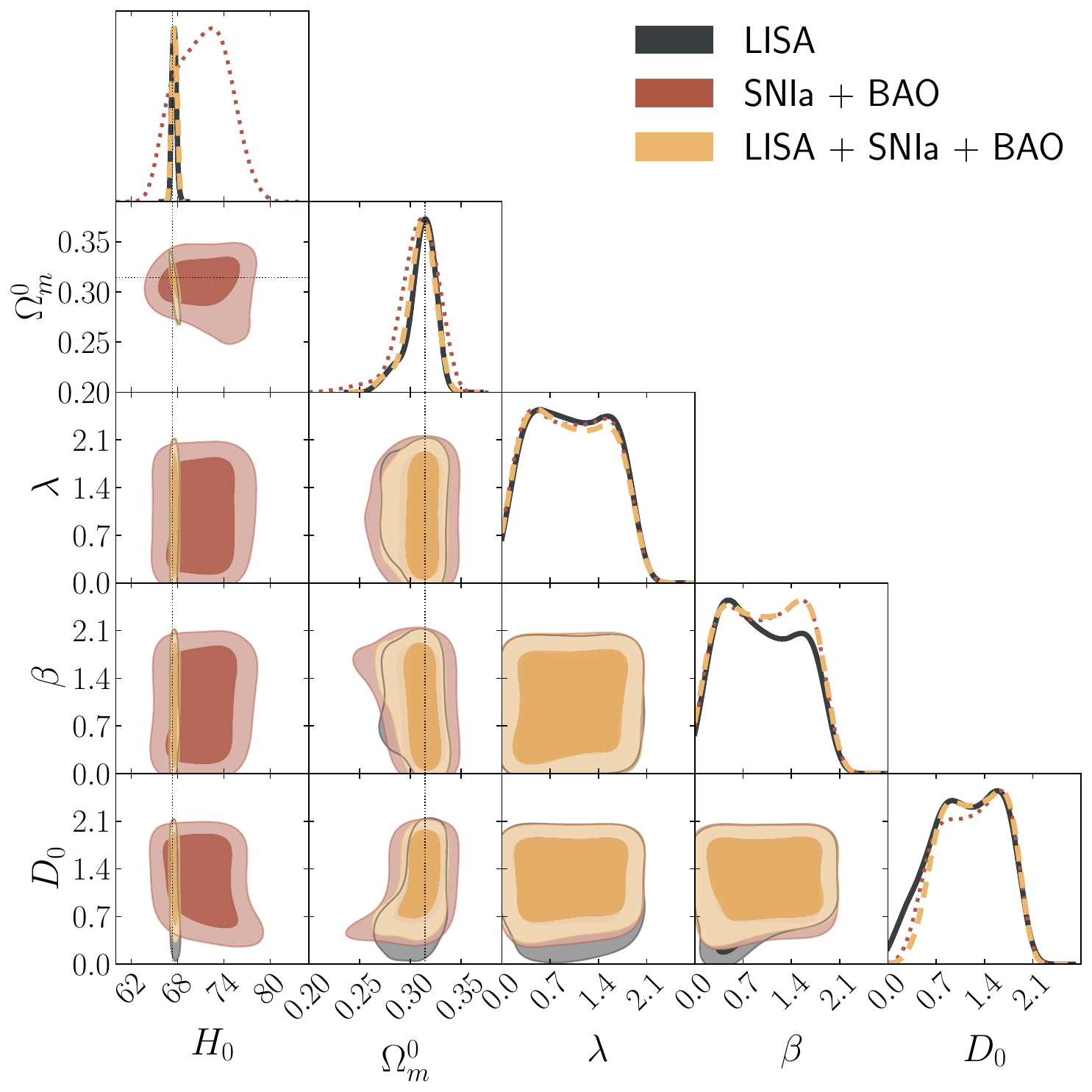}} 
  \caption{\label{fig:lpdf} 68\% and 95\% C.L. 2D contours and 1D posterior distributions for the parameters $\{ H_0,\Omega_m^0,\lambda, \beta, D_0\}$ in the mixed conformal-disformal coupled quintessence model with LISA (charcoal filled line), SNIa+BAO (red dotted line) data  and their combination (yellow dashed line). The scale is the same as in Fig.~\ref{fig:etdf} for comparison purposes, with the SNIa+BAO contours standing as the reference. The dotted lines depict the fiducial values for the mock data $\{\Omega_m^0,H_0 \} = \{0.3144, 67.32\}$.}
\end{figure}

\begin{figure}[t!]
      \subfloat{\includegraphics[width=\linewidth]{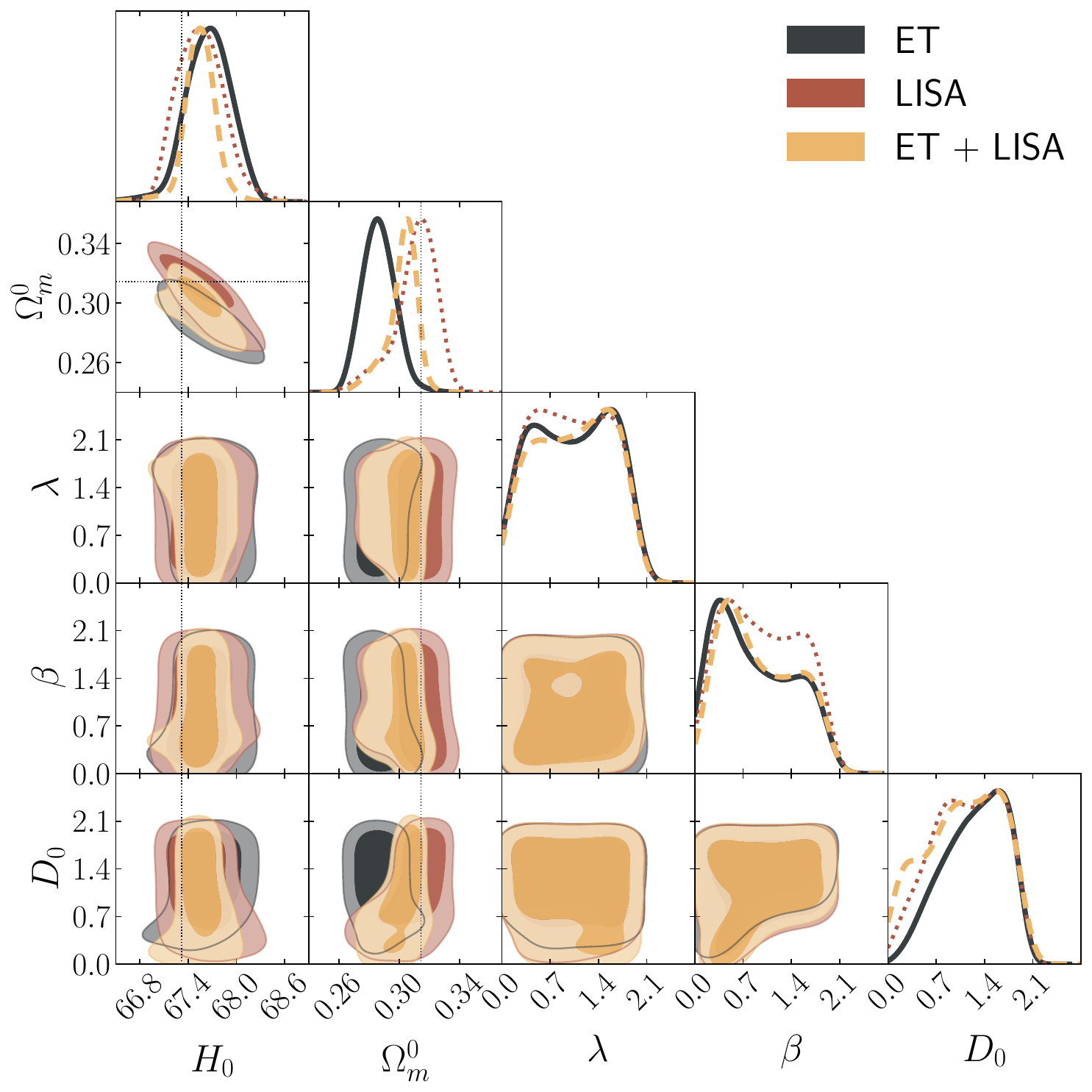}} 
  \caption{\label{fig:etlp2df} 68\% and 95\% C.L. 2D contours and 1D marginalised posterior distributions for the parameters $\{ H_0,\Omega_m^0,\lambda, \beta, D_0\}$ in the mixed conformal-disformal coupled quintessence model with ET mock data (charcoal filled line), LISA mock data (red dotted line) and their combination (yellow dashed line). The dotted lines depict the fiducial values for the mock data $\{\Omega_m^0,H_0 \} = \{0.3144, 67.32\}$.}
\end{figure}

In Figs. \ref{fig:etdf}, \ref{fig:lpdf} and \ref{fig:etlp2df} and Table \ref{Tab:boundsfdcq} we show the results for the parameters $\{\Omega_m^0, H_0, \beta, D_0, \lambda\}$ for the same data sets as before.  
In our analysis of the ET data set alone, we observe improved accuracy for the cosmological parameters, $\Omega_m^0$ and $H_0$, compared to the SNIa+BAO data set, with $\mathcal{F}_{\Omega_m^0, H_0}^{(\text{SNIa+BAO,ET})} = \{0.61,0.088\}$. The combined data sets, ET+SNIa+BAO, show comparable results, with a slight increase in accuracy compared to ET alone. For the model parameters, $\{\beta, D_0, \lambda$\}, ET compared with SNIa+BAO demonstrates close constraining power, with $\mathcal{F}_{\beta, D_0, \lambda}^{(\text{SNIa+BAO,ET})} = \{1.0,0.92,1.0\}$. However, the combined data set leads to an increase in the error in $\beta$ with $\sigma_\beta = 0.71$ for ET+SNIa+BAO. 
For the case of LISA standard sirens, we find that the cosmological parameters follow a similar trend as ET, with increased accuracy, $\mathcal{F}_{\Omega_m^0, H_0}^{(\text{SNIa+BAO,LISA})} = \{0.72,0.088\}$, with the combined data set showing a comparable trend. However, it is worth noting a noticeable reduction in accuracy for $H_0$ between LISA and the combined data set, $\mathcal{F}_{H_0}^{(\text{LISA, LISA+SNIa+BAO})} = \{1.1\}$.
Regarding the model parameters, $\{\beta, D_0, \lambda\}$, we find that unlike for ET, LISA alone exhibits increased accuracy compared to SNIa+BAO, except for $D_0$, $\mathcal{F}_{\beta, D_0, \lambda}^{(\text{SNIa+BAO,LISA})} = \{0.98,1.1,0.98\}$. The combination of the data sets results in comparable accuracy to LISA alone, with $\mathcal{F}_{\beta, D_0, \lambda}^{(\text{LISA,LISA+SNIa+BAO})} = \{1.0,0.89,1.0\}$. 
Similar to \cref{sec:casea,sec:caseb,sec:casec}, the combination of the GW data sets leads to a significant change in the accuracy of $\Omega_m^0$ and $H_0$ compared to the SNIa+BAO data sets, $\mathcal{F}_{\Omega_m^0,H_0}^{(\text{SNIa+BAO, ET+LISA})} = \{0.49,0.061\}$. The accuracy of the cosmological parameters is also slightly enhanced compared to ET or LISA alone. Regarding the model parameters, we find only a very small change compared to SNIa+BAO, $\mathcal{F}_{\beta,D_0,\lambda}^{(\text{SNIa+BAO, ET+LISA})}=\{0.96,1.2,1.1\}$, noting that both $D_0$ and $\lambda$ are slightly less constrained. Combining all of the data sets, we note that there is a similar trend as before, with the cosmological parameters exhibiting an enhanced constraint when compared to SNIa+BAO, while the model parameters remain mostly unchanged, $\mathcal{F}_{\Omega_m^0,H_0,\beta,D_0,\lambda}^{(\text{SNIa+BAO, ET+LISA+SNIa+BAO})}=\{0.58, 0.073,1.0,0.92,1.0\}$.

In summary, regardless of the combination of data sets considered, the constraints on ${\Omega_m^0, \beta, D_0, \lambda}$ are of the same order of magnitude as those obtained from SNIa+BAO. Additionally, we note that the accuracy on $H_0$ is improved by 1 order of magnitude for both ET and LISA compared to SNIa+BAO.

Similarly to the comparison in Sec.~\ref{sec:casec}, the main improvement in contrast with the results reported in Ref.~\cite{vandeBruck:2016hpz} is the fact that we can obtain constraints at $68\%$ C.L. for all the model parameters. The potential parameter $\lambda$ is constrained with upper bounds for the standard sirens at $1\sigma$ of the same order of the $2\sigma$ ones reported in the previous studies. Moreover, the constraining power on $H_0$ is largely improved from $\sigma_{H_0} \approx 2.1$ for the background data to $\sigma_{H_0} \approx 0.3$ in all the cases, including standard sirens. The comparison with results including CMB, CMB lensing and additional data in Ref.~\cite{VanDeBruck:2017mua}, which are either unable to constrain $D_0$ or find just a lower bound and report only upper bounds for $\lambda$ and $\beta$, shows that standard sirens successfully constrain the three model parameters at $68\%$ C.L. for all the combinations, which is a great improvement given that only background data has been considered. Including CMB data brings the error in $H_0$ to the same order of magnitude with $\sigma_{H_0} \approx 0.6$, which is still around 2 times larger than the ones reported in this analysis.

\section{Conclusions}\label{sec:sum}

In this paper, we have explored the potential of future GWs detectors, namely LISA and ET, to constrain conformal and disformal couplings between the dark energy and dark matter fluids. We have considered four models: conformal coupled quintessence, a kinetic model, constant disformal coupled quintessence and a mixed conformal-disformal model. All the cases considered have the same exponential potential of which we have constrained its slope, $\lambda$.

We have generated mock catalogues of standard siren events with ET and LISA specifications and with those we performed an MCMC analysis considering, separately, their combination with current SNIa+BAO data as well for reference. 
Under the assumption we have used to generate the mock data assuming a particular cosmology, we find the following: 

\begin{itemize}
\item The conformal coupled quintessence: the combinations of LISA+SNIa+BAO and ET+SNIa+BAO improve the constraints on both the slope parameter, $\lambda$, and the conformal coupling parameter $\beta$. The combination of ET+LISA with SNIa+BAO reduces the error in $\beta$ by one third.

\item The kinetic model: ET and LISA alone cannot improve the constraints on $\lambda$ or on the conformal exponential parameter $\alpha$. When LISA is combined with SNIa+BAO the accuracy slightly improves for the slope parameter and for the matter density $\Omega_m^0$.

\item The constant disformal coupled quintessence: all combinations can constrain the disformal parameter $D_0$ at $1\sigma$ with the same order of magnitude and a small improvement for LISA+SNIa+BAO. For the slope parameter instead, both ET, LISA and their combination perform better than SNIa+BAO. Moreover for the full catalogue combination of ET+LISA, the error in $\Omega_m^0$ can be reduced.

\item  The mixed conformal-disformal coupled quintessence: for all the parameters of the model there is not a significant improvement on the accuracy of their constraints when using ET or LISA data separately. A small reduction in the size of the $1\sigma$ region is reported only for the disformal parameter, $D_0$, in the full combinations. The error in $\Omega_m^0$ is slightly reduced for ET+LISA. 
\end{itemize}

Regardless of the model considered we found that the accuracy on the $H_0$ parameter increases by 1 order of magnitude at 1$\sigma$ when compared to the combination of BAO and SNIa data. This is promising in light of solving/understanding the $H_0$ tension. This improvement is also responsible for the increased accuracy on the constraints for the model parameters when we consider the full combinations that we just reviewed. 

Ultimately, our results show that future 3G detectors can improve our knowledge on DE-DM interaction and shed light on the $H_0$ tension. 

\acknowledgments

We thank M. Califano for useful discussions.
E.M.T. is supported by the grant SFRH/BD/143231/2019 from Funda\c{c}\~ao para a Ci\^encia e a Tecnologia (FCT) and acknowledges the support of the European Consortium for Astroparticle Theory (EuCAPT) in the form of an Exchange Travel Grant.
R.D. is supported by a STFC CDT studentship.
N.F. is supported by the Italian Ministry of University and Research (MUR) through the Rita Levi Montalcini project ``Tests of gravity on cosmic scales" with reference PGR19ILFGP.
C.v.d.B. is supported (in part) by the Lancaster–Manchester–Sheffield Consortium for Fundamental Physics under STFC grant: ST/T001038/1.
E.M.T, C.v.d.B. and N.F  also acknowledge the FCT project with ref. number PTDC/FIS-AST/0054/2021 and  the COST Action CosmoVerse, CA21136, supported by COST (European Cooperation in Science and Technology).

\bibliography{bibliography}

\begin{thebibliography}{104}%
\makeatletter
\providecommand \@ifxundefined [1]{%
 \@ifx{#1\undefined}
}%
\providecommand \@ifnum [1]{%
 \ifnum #1\expandafter \@firstoftwo
 \else \expandafter \@secondoftwo
 \fi
}%
\providecommand \@ifx [1]{%
 \ifx #1\expandafter \@firstoftwo
 \else \expandafter \@secondoftwo
 \fi
}%
\providecommand \natexlab [1]{#1}%
\providecommand \enquote  [1]{``#1''}%
\providecommand \bibnamefont  [1]{#1}%
\providecommand \bibfnamefont [1]{#1}%
\providecommand \citenamefont [1]{#1}%
\providecommand \href@noop [0]{\@secondoftwo}%
\providecommand \href [0]{\begingroup \@sanitize@url \@href}%
\providecommand \@href[1]{\@@startlink{#1}\@@href}%
\providecommand \@@href[1]{\endgroup#1\@@endlink}%
\providecommand \@sanitize@url [0]{\catcode `\\12\catcode `\$12\catcode
  `\&12\catcode `\#12\catcode `\^12\catcode `\_12\catcode `\%12\relax}%
\providecommand \@@startlink[1]{}%
\providecommand \@@endlink[0]{}%
\providecommand \url  [0]{\begingroup\@sanitize@url \@url }%
\providecommand \@url [1]{\endgroup\@href {#1}{\urlprefix }}%
\providecommand \urlprefix  [0]{URL }%
\providecommand \Eprint [0]{\href }%
\providecommand \doibase [0]{https://doi.org/}%
\providecommand \selectlanguage [0]{\@gobble}%
\providecommand \bibinfo  [0]{\@secondoftwo}%
\providecommand \bibfield  [0]{\@secondoftwo}%
\providecommand \translation [1]{[#1]}%
\providecommand \BibitemOpen [0]{}%
\providecommand \bibitemStop [0]{}%
\providecommand \bibitemNoStop [0]{.\EOS\space}%
\providecommand \EOS [0]{\spacefactor3000\relax}%
\providecommand \BibitemShut  [1]{\csname bibitem#1\endcsname}%
\let\auto@bib@innerbib\@empty
\bibitem [{\citenamefont {Weinberg}(1989)}]{Weinberg:1988cp}%
  \BibitemOpen
  \bibfield  {author} {\bibinfo {author} {\bibfnamefont {S.}~\bibnamefont
  {Weinberg}},\ }\bibfield  {title} {\bibinfo {title} {{The Cosmological
  Constant Problem}},\ }\href {https://doi.org/10.1103/RevModPhys.61.1}
  {\bibfield  {journal} {\bibinfo  {journal} {Rev. Mod. Phys.}\ }\textbf
  {\bibinfo {volume} {61}},\ \bibinfo {pages} {1} (\bibinfo {year}
  {1989})}\BibitemShut {NoStop}%
\bibitem [{\citenamefont {Martin}(2012)}]{Martin:2012bt}%
  \BibitemOpen
  \bibfield  {author} {\bibinfo {author} {\bibfnamefont {J.}~\bibnamefont
  {Martin}},\ }\bibfield  {title} {\bibinfo {title} {{Everything You Always
  Wanted To Know About The Cosmological Constant Problem (But Were Afraid To
  Ask)}},\ }\href {https://doi.org/10.1016/j.crhy.2012.04.008} {\bibfield
  {journal} {\bibinfo  {journal} {Comptes Rendus Physique}\ }\textbf {\bibinfo
  {volume} {13}},\ \bibinfo {pages} {566} (\bibinfo {year} {2012})},\ \Eprint
  {https://arxiv.org/abs/1205.3365} {arXiv:1205.3365 [astro-ph.CO]}
  \BibitemShut {NoStop}%
\bibitem [{\citenamefont {Joyce}\ \emph {et~al.}(2015)\citenamefont {Joyce},
  \citenamefont {Jain}, \citenamefont {Khoury},\ and\ \citenamefont
  {Trodden}}]{Joyce:2014kja}%
  \BibitemOpen
  \bibfield  {author} {\bibinfo {author} {\bibfnamefont {A.}~\bibnamefont
  {Joyce}}, \bibinfo {author} {\bibfnamefont {B.}~\bibnamefont {Jain}},
  \bibinfo {author} {\bibfnamefont {J.}~\bibnamefont {Khoury}},\ and\ \bibinfo
  {author} {\bibfnamefont {M.}~\bibnamefont {Trodden}},\ }\bibfield  {title}
  {\bibinfo {title} {{Beyond the Cosmological Standard Model}},\ }\href
  {https://doi.org/10.1016/j.physrep.2014.12.002} {\bibfield  {journal}
  {\bibinfo  {journal} {Phys. Rept.}\ }\textbf {\bibinfo {volume} {568}},\
  \bibinfo {pages} {1} (\bibinfo {year} {2015})},\ \Eprint
  {https://arxiv.org/abs/1407.0059} {arXiv:1407.0059 [astro-ph.CO]}
  \BibitemShut {NoStop}%
\bibitem [{\citenamefont {Adam}\ \emph {et~al.}(2016)\citenamefont {Adam} \emph
  {et~al.}}]{Planck:2015mrs}%
  \BibitemOpen
  \bibfield  {author} {\bibinfo {author} {\bibfnamefont {R.}~\bibnamefont
  {Adam}} \emph {et~al.} (\bibinfo {collaboration} {Planck}),\ }\bibfield
  {title} {\bibinfo {title} {{Planck 2015 results. I. Overview of products and
  scientific results}},\ }\href {https://doi.org/10.1051/0004-6361/201527101}
  {\bibfield  {journal} {\bibinfo  {journal} {Astron. Astrophys.}\ }\textbf
  {\bibinfo {volume} {594}},\ \bibinfo {pages} {A1} (\bibinfo {year} {2016})},\
  \Eprint {https://arxiv.org/abs/1502.01582} {arXiv:1502.01582 [astro-ph.CO]}
  \BibitemShut {NoStop}%
\bibitem [{\citenamefont {Aghanim}\ \emph {et~al.}(2020)\citenamefont {Aghanim}
  \emph {et~al.}}]{Planck:2018vyg}%
  \BibitemOpen
  \bibfield  {author} {\bibinfo {author} {\bibfnamefont {N.}~\bibnamefont
  {Aghanim}} \emph {et~al.} (\bibinfo {collaboration} {Planck}),\ }\bibfield
  {title} {\bibinfo {title} {{Planck 2018 results. VI. Cosmological
  parameters}},\ }\href {https://doi.org/10.1051/0004-6361/201833910}
  {\bibfield  {journal} {\bibinfo  {journal} {Astron. Astrophys.}\ }\textbf
  {\bibinfo {volume} {641}},\ \bibinfo {pages} {A6} (\bibinfo {year} {2020})},\
  \bibinfo {note} {[Erratum: Astron.Astrophys. 652, C4 (2021)]},\ \Eprint
  {https://arxiv.org/abs/1807.06209} {arXiv:1807.06209 [astro-ph.CO]}
  \BibitemShut {NoStop}%
\bibitem [{\citenamefont {Riess}\ \emph {et~al.}(2011)\citenamefont {Riess},
  \citenamefont {Macri}, \citenamefont {Casertano}, \citenamefont {Lampeitl},
  \citenamefont {Ferguson}, \citenamefont {Filippenko}, \citenamefont {Jha},
  \citenamefont {Li},\ and\ \citenamefont {Chornock}}]{Riess:2011yx}%
  \BibitemOpen
  \bibfield  {author} {\bibinfo {author} {\bibfnamefont {A.~G.}\ \bibnamefont
  {Riess}}, \bibinfo {author} {\bibfnamefont {L.}~\bibnamefont {Macri}},
  \bibinfo {author} {\bibfnamefont {S.}~\bibnamefont {Casertano}}, \bibinfo
  {author} {\bibfnamefont {H.}~\bibnamefont {Lampeitl}}, \bibinfo {author}
  {\bibfnamefont {H.~C.}\ \bibnamefont {Ferguson}}, \bibinfo {author}
  {\bibfnamefont {A.~V.}\ \bibnamefont {Filippenko}}, \bibinfo {author}
  {\bibfnamefont {S.~W.}\ \bibnamefont {Jha}}, \bibinfo {author} {\bibfnamefont
  {W.}~\bibnamefont {Li}},\ and\ \bibinfo {author} {\bibfnamefont
  {R.}~\bibnamefont {Chornock}},\ }\bibfield  {title} {\bibinfo {title} {{A 3\%
  Solution: Determination of the Hubble Constant with the Hubble Space
  Telescope and Wide Field Camera 3}},\ }\href
  {https://doi.org/10.1088/0004-637X/732/2/129} {\bibfield  {journal} {\bibinfo
   {journal} {Astrophys. J.}\ }\textbf {\bibinfo {volume} {730}},\ \bibinfo
  {pages} {119} (\bibinfo {year} {2011})},\ \bibinfo {note} {[Erratum:
  Astrophys.J. 732, 129 (2011)]},\ \Eprint {https://arxiv.org/abs/1103.2976}
  {arXiv:1103.2976 [astro-ph.CO]} \BibitemShut {NoStop}%
\bibitem [{\citenamefont {Riess}\ \emph {et~al.}(2016)\citenamefont {Riess}
  \emph {et~al.}}]{Riess:2016jrr}%
  \BibitemOpen
  \bibfield  {author} {\bibinfo {author} {\bibfnamefont {A.~G.}\ \bibnamefont
  {Riess}} \emph {et~al.},\ }\bibfield  {title} {\bibinfo {title} {{A 2.4\%
  Determination of the Local Value of the Hubble Constant}},\ }\href
  {https://doi.org/10.3847/0004-637X/826/1/56} {\bibfield  {journal} {\bibinfo
  {journal} {Astrophys. J.}\ }\textbf {\bibinfo {volume} {826}},\ \bibinfo
  {pages} {56} (\bibinfo {year} {2016})},\ \Eprint
  {https://arxiv.org/abs/1604.01424} {arXiv:1604.01424 [astro-ph.CO]}
  \BibitemShut {NoStop}%
\bibitem [{\citenamefont {Riess}\ \emph {et~al.}(2019)\citenamefont {Riess},
  \citenamefont {Casertano}, \citenamefont {Yuan}, \citenamefont {Macri},\ and\
  \citenamefont {Scolnic}}]{Riess:2019cxk}%
  \BibitemOpen
  \bibfield  {author} {\bibinfo {author} {\bibfnamefont {A.~G.}\ \bibnamefont
  {Riess}}, \bibinfo {author} {\bibfnamefont {S.}~\bibnamefont {Casertano}},
  \bibinfo {author} {\bibfnamefont {W.}~\bibnamefont {Yuan}}, \bibinfo {author}
  {\bibfnamefont {L.~M.}\ \bibnamefont {Macri}},\ and\ \bibinfo {author}
  {\bibfnamefont {D.}~\bibnamefont {Scolnic}},\ }\bibfield  {title} {\bibinfo
  {title} {{Large Magellanic Cloud Cepheid Standards Provide a 1\% Foundation
  for the Determination of the Hubble Constant and Stronger Evidence for
  Physics beyond $\Lambda$CDM}},\ }\href
  {https://doi.org/10.3847/1538-4357/ab1422} {\bibfield  {journal} {\bibinfo
  {journal} {Astrophys. J.}\ }\textbf {\bibinfo {volume} {876}},\ \bibinfo
  {pages} {85} (\bibinfo {year} {2019})},\ \Eprint
  {https://arxiv.org/abs/1903.07603} {arXiv:1903.07603 [astro-ph.CO]}
  \BibitemShut {NoStop}%
\bibitem [{\citenamefont {Delubac}\ \emph {et~al.}(2015)\citenamefont {Delubac}
  \emph {et~al.}}]{BOSS:2014hwf}%
  \BibitemOpen
  \bibfield  {author} {\bibinfo {author} {\bibfnamefont {T.}~\bibnamefont
  {Delubac}} \emph {et~al.} (\bibinfo {collaboration} {BOSS}),\ }\bibfield
  {title} {\bibinfo {title} {{Baryon acoustic oscillations in the
  Ly\ensuremath{\alpha} forest of BOSS DR11 quasars}},\ }\href
  {https://doi.org/10.1051/0004-6361/201423969} {\bibfield  {journal} {\bibinfo
   {journal} {Astron. Astrophys.}\ }\textbf {\bibinfo {volume} {574}},\
  \bibinfo {pages} {A59} (\bibinfo {year} {2015})},\ \Eprint
  {https://arxiv.org/abs/1404.1801} {arXiv:1404.1801 [astro-ph.CO]}
  \BibitemShut {NoStop}%
\bibitem [{\citenamefont {Di~Valentino}\ \emph
  {et~al.}(2021{\natexlab{a}})\citenamefont {Di~Valentino} \emph
  {et~al.}}]{DiValentino:2020zio}%
  \BibitemOpen
  \bibfield  {author} {\bibinfo {author} {\bibfnamefont {E.}~\bibnamefont
  {Di~Valentino}} \emph {et~al.},\ }\bibfield  {title} {\bibinfo {title}
  {{Snowmass2021 - Letter of interest cosmology intertwined II: The hubble
  constant tension}},\ }\href
  {https://doi.org/10.1016/j.astropartphys.2021.102605} {\bibfield  {journal}
  {\bibinfo  {journal} {Astropart. Phys.}\ }\textbf {\bibinfo {volume} {131}},\
  \bibinfo {pages} {102605} (\bibinfo {year} {2021}{\natexlab{a}})},\ \Eprint
  {https://arxiv.org/abs/2008.11284} {arXiv:2008.11284 [astro-ph.CO]}
  \BibitemShut {NoStop}%
\bibitem [{\citenamefont {de~Jong}\ \emph {et~al.}(2015)\citenamefont {de~Jong}
  \emph {et~al.}}]{deJong:2015wca}%
  \BibitemOpen
  \bibfield  {author} {\bibinfo {author} {\bibfnamefont {J.~T.~A.}\
  \bibnamefont {de~Jong}} \emph {et~al.},\ }\bibfield  {title} {\bibinfo
  {title} {{The first and second data releases of the Kilo-Degree Survey}},\
  }\href {https://doi.org/10.1051/0004-6361/201526601} {\bibfield  {journal}
  {\bibinfo  {journal} {Astron. Astrophys.}\ }\textbf {\bibinfo {volume}
  {582}},\ \bibinfo {pages} {A62} (\bibinfo {year} {2015})},\ \Eprint
  {https://arxiv.org/abs/1507.00742} {arXiv:1507.00742 [astro-ph.CO]}
  \BibitemShut {NoStop}%
\bibitem [{\citenamefont {Hildebrandt}\ \emph {et~al.}(2017)\citenamefont
  {Hildebrandt} \emph {et~al.}}]{Hildebrandt:2016iqg}%
  \BibitemOpen
  \bibfield  {author} {\bibinfo {author} {\bibfnamefont {H.}~\bibnamefont
  {Hildebrandt}} \emph {et~al.},\ }\bibfield  {title} {\bibinfo {title}
  {{KiDS-450: Cosmological parameter constraints from tomographic weak
  gravitational lensing}},\ }\href {https://doi.org/10.1093/mnras/stw2805}
  {\bibfield  {journal} {\bibinfo  {journal} {Mon. Not. Roy. Astron. Soc.}\
  }\textbf {\bibinfo {volume} {465}},\ \bibinfo {pages} {1454} (\bibinfo {year}
  {2017})},\ \Eprint {https://arxiv.org/abs/1606.05338} {arXiv:1606.05338
  [astro-ph.CO]} \BibitemShut {NoStop}%
\bibitem [{\citenamefont {Kuijken}\ \emph {et~al.}(2015)\citenamefont {Kuijken}
  \emph {et~al.}}]{Kuijken:2015vca}%
  \BibitemOpen
  \bibfield  {author} {\bibinfo {author} {\bibfnamefont {K.}~\bibnamefont
  {Kuijken}} \emph {et~al.},\ }\bibfield  {title} {\bibinfo {title}
  {{Gravitational Lensing Analysis of the Kilo Degree Survey}},\ }\href
  {https://doi.org/10.1093/mnras/stv2140} {\bibfield  {journal} {\bibinfo
  {journal} {Mon. Not. Roy. Astron. Soc.}\ }\textbf {\bibinfo {volume} {454}},\
  \bibinfo {pages} {3500} (\bibinfo {year} {2015})},\ \Eprint
  {https://arxiv.org/abs/1507.00738} {arXiv:1507.00738 [astro-ph.CO]}
  \BibitemShut {NoStop}%
\bibitem [{\citenamefont {Fenech~Conti}\ \emph {et~al.}(2017)\citenamefont
  {Fenech~Conti}, \citenamefont {Herbonnet}, \citenamefont {Hoekstra},
  \citenamefont {Merten}, \citenamefont {Miller},\ and\ \citenamefont
  {Viola}}]{FenechConti:2016oun}%
  \BibitemOpen
  \bibfield  {author} {\bibinfo {author} {\bibfnamefont {I.}~\bibnamefont
  {Fenech~Conti}}, \bibinfo {author} {\bibfnamefont {R.}~\bibnamefont
  {Herbonnet}}, \bibinfo {author} {\bibfnamefont {H.}~\bibnamefont {Hoekstra}},
  \bibinfo {author} {\bibfnamefont {J.}~\bibnamefont {Merten}}, \bibinfo
  {author} {\bibfnamefont {L.}~\bibnamefont {Miller}},\ and\ \bibinfo {author}
  {\bibfnamefont {M.}~\bibnamefont {Viola}},\ }\bibfield  {title} {\bibinfo
  {title} {{Calibration of weak-lensing shear in the Kilo-Degree Survey}},\
  }\href {https://doi.org/10.1093/mnras/stx200} {\bibfield  {journal} {\bibinfo
   {journal} {Mon. Not. Roy. Astron. Soc.}\ }\textbf {\bibinfo {volume}
  {467}},\ \bibinfo {pages} {1627} (\bibinfo {year} {2017})},\ \Eprint
  {https://arxiv.org/abs/1606.05337} {arXiv:1606.05337 [astro-ph.CO]}
  \BibitemShut {NoStop}%
\bibitem [{\citenamefont {Joudaki}\ \emph {et~al.}(2020)\citenamefont {Joudaki}
  \emph {et~al.}}]{Joudaki:2019pmv}%
  \BibitemOpen
  \bibfield  {author} {\bibinfo {author} {\bibfnamefont {S.}~\bibnamefont
  {Joudaki}} \emph {et~al.},\ }\bibfield  {title} {\bibinfo {title}
  {{KiDS+VIKING-450 and DES-Y1 combined: Cosmology with cosmic shear}},\ }\href
  {https://doi.org/10.1051/0004-6361/201936154} {\bibfield  {journal} {\bibinfo
   {journal} {Astron. Astrophys.}\ }\textbf {\bibinfo {volume} {638}},\
  \bibinfo {pages} {L1} (\bibinfo {year} {2020})},\ \Eprint
  {https://arxiv.org/abs/1906.09262} {arXiv:1906.09262 [astro-ph.CO]}
  \BibitemShut {NoStop}%
\bibitem [{\citenamefont {Di~Valentino}\ \emph
  {et~al.}(2021{\natexlab{b}})\citenamefont {Di~Valentino} \emph
  {et~al.}}]{DiValentino:2020vvd}%
  \BibitemOpen
  \bibfield  {author} {\bibinfo {author} {\bibfnamefont {E.}~\bibnamefont
  {Di~Valentino}} \emph {et~al.},\ }\bibfield  {title} {\bibinfo {title}
  {{Cosmology Intertwined III: $f \sigma_8$ and $S_8$}},\ }\href
  {https://doi.org/10.1016/j.astropartphys.2021.102604} {\bibfield  {journal}
  {\bibinfo  {journal} {Astropart. Phys.}\ }\textbf {\bibinfo {volume} {131}},\
  \bibinfo {pages} {102604} (\bibinfo {year} {2021}{\natexlab{b}})},\ \Eprint
  {https://arxiv.org/abs/2008.11285} {arXiv:2008.11285 [astro-ph.CO]}
  \BibitemShut {NoStop}%
\bibitem [{\citenamefont {Akrami}\ \emph {et~al.}(2021)\citenamefont {Akrami}
  \emph {et~al.}}]{CANTATA:2021ktz}%
  \BibitemOpen
  \bibfield  {author} {\bibinfo {author} {\bibfnamefont {Y.}~\bibnamefont
  {Akrami}} \emph {et~al.} (\bibinfo {collaboration} {CANTATA}),\ }\href
  {https://doi.org/10.1007/978-3-030-83715-0} {\emph {\bibinfo {title}
  {{Modified Gravity and Cosmology}: {An Update by the CANTATA Network}}}},\
  edited by\ \bibinfo {editor} {\bibfnamefont {E.~N.}\ \bibnamefont
  {Saridakis}}, \bibinfo {editor} {\bibfnamefont {R.}~\bibnamefont {Lazkoz}},
  \bibinfo {editor} {\bibfnamefont {V.}~\bibnamefont {Salzano}}, \bibinfo
  {editor} {\bibfnamefont {P.}~\bibnamefont {Vargas~Moniz}}, \bibinfo {editor}
  {\bibfnamefont {S.}~\bibnamefont {Capozziello}}, \bibinfo {editor}
  {\bibfnamefont {J.}~\bibnamefont {Beltr\'an~Jim\'enez}}, \bibinfo {editor}
  {\bibfnamefont {M.}~\bibnamefont {De~Laurentis}},\ and\ \bibinfo {editor}
  {\bibfnamefont {G.~J.}\ \bibnamefont {Olmo}}\ (\bibinfo  {publisher}
  {Springer},\ \bibinfo {year} {2021})\ \Eprint
  {https://arxiv.org/abs/2105.12582} {arXiv:2105.12582 [gr-qc]} \BibitemShut
  {NoStop}%
\bibitem [{\citenamefont {Wetterich}(1988)}]{Wetterich:1987fm}%
  \BibitemOpen
  \bibfield  {author} {\bibinfo {author} {\bibfnamefont {C.}~\bibnamefont
  {Wetterich}},\ }\bibfield  {title} {\bibinfo {title} {{Cosmology and the Fate
  of Dilatation Symmetry}},\ }\href
  {https://doi.org/10.1016/0550-3213(88)90193-9} {\bibfield  {journal}
  {\bibinfo  {journal} {Nucl. Phys. B}\ }\textbf {\bibinfo {volume} {302}},\
  \bibinfo {pages} {668} (\bibinfo {year} {1988})},\ \Eprint
  {https://arxiv.org/abs/1711.03844} {arXiv:1711.03844 [hep-th]} \BibitemShut
  {NoStop}%
\bibitem [{\citenamefont {Peebles}\ and\ \citenamefont
  {Ratra}(1988)}]{Peebles:1987ek}%
  \BibitemOpen
  \bibfield  {author} {\bibinfo {author} {\bibfnamefont {P.~J.~E.}\
  \bibnamefont {Peebles}}\ and\ \bibinfo {author} {\bibfnamefont
  {B.}~\bibnamefont {Ratra}},\ }\bibfield  {title} {\bibinfo {title}
  {{Cosmology with a Time Variable Cosmological Constant}},\ }\href
  {https://doi.org/10.1086/185100} {\bibfield  {journal} {\bibinfo  {journal}
  {Astrophys. J. Lett.}\ }\textbf {\bibinfo {volume} {325}},\ \bibinfo {pages}
  {L17} (\bibinfo {year} {1988})}\BibitemShut {NoStop}%
\bibitem [{\citenamefont {Tsujikawa}(2013)}]{Tsujikawa:2013fta}%
  \BibitemOpen
  \bibfield  {author} {\bibinfo {author} {\bibfnamefont {S.}~\bibnamefont
  {Tsujikawa}},\ }\bibfield  {title} {\bibinfo {title} {{Quintessence: A
  Review}},\ }\href {https://doi.org/10.1088/0264-9381/30/21/214003} {\bibfield
   {journal} {\bibinfo  {journal} {Class. Quant. Grav.}\ }\textbf {\bibinfo
  {volume} {30}},\ \bibinfo {pages} {214003} (\bibinfo {year} {2013})},\
  \Eprint {https://arxiv.org/abs/1304.1961} {arXiv:1304.1961 [gr-qc]}
  \BibitemShut {NoStop}%
\bibitem [{\citenamefont {Wetterich}(1995)}]{Wetterich:1994bg}%
  \BibitemOpen
  \bibfield  {author} {\bibinfo {author} {\bibfnamefont {C.}~\bibnamefont
  {Wetterich}},\ }\bibfield  {title} {\bibinfo {title} {{The Cosmon model for
  an asymptotically vanishing time dependent cosmological 'constant'}},\
  }\href@noop {} {\bibfield  {journal} {\bibinfo  {journal} {Astron.
  Astrophys.}\ }\textbf {\bibinfo {volume} {301}},\ \bibinfo {pages} {321}
  (\bibinfo {year} {1995})},\ \Eprint {https://arxiv.org/abs/hep-th/9408025}
  {arXiv:hep-th/9408025} \BibitemShut {NoStop}%
\bibitem [{\citenamefont {Amendola}(2000)}]{Amendola:1999er}%
  \BibitemOpen
  \bibfield  {author} {\bibinfo {author} {\bibfnamefont {L.}~\bibnamefont
  {Amendola}},\ }\bibfield  {title} {\bibinfo {title} {{Coupled
  quintessence}},\ }\href {https://doi.org/10.1103/PhysRevD.62.043511}
  {\bibfield  {journal} {\bibinfo  {journal} {Phys. Rev. D}\ }\textbf {\bibinfo
  {volume} {62}},\ \bibinfo {pages} {043511} (\bibinfo {year} {2000})},\
  \Eprint {https://arxiv.org/abs/astro-ph/9908023} {arXiv:astro-ph/9908023}
  \BibitemShut {NoStop}%
\bibitem [{\citenamefont {Abdalla}\ \emph {et~al.}(2022)\citenamefont {Abdalla}
  \emph {et~al.}}]{Abdalla:2022yfr}%
  \BibitemOpen
  \bibfield  {author} {\bibinfo {author} {\bibfnamefont {E.}~\bibnamefont
  {Abdalla}} \emph {et~al.},\ }\bibfield  {title} {\bibinfo {title} {{Cosmology
  intertwined: A review of the particle physics, astrophysics, and cosmology
  associated with the cosmological tensions and anomalies}},\ }\href
  {https://doi.org/10.1016/j.jheap.2022.04.002} {\bibfield  {journal} {\bibinfo
   {journal} {JHEAp}\ }\textbf {\bibinfo {volume} {34}},\ \bibinfo {pages} {49}
  (\bibinfo {year} {2022})},\ \Eprint {https://arxiv.org/abs/2203.06142}
  {arXiv:2203.06142 [astro-ph.CO]} \BibitemShut {NoStop}%
\bibitem [{\citenamefont {Brans}\ and\ \citenamefont
  {Dicke}(1961)}]{Brans:1961sx}%
  \BibitemOpen
  \bibfield  {author} {\bibinfo {author} {\bibfnamefont {C.}~\bibnamefont
  {Brans}}\ and\ \bibinfo {author} {\bibfnamefont {R.~H.}\ \bibnamefont
  {Dicke}},\ }\bibfield  {title} {\bibinfo {title} {{Mach's principle and a
  relativistic theory of gravitation}},\ }\href
  {https://doi.org/10.1103/PhysRev.124.925} {\bibfield  {journal} {\bibinfo
  {journal} {Phys. Rev.}\ }\textbf {\bibinfo {volume} {124}},\ \bibinfo {pages}
  {925} (\bibinfo {year} {1961})}\BibitemShut {NoStop}%
\bibitem [{\citenamefont {Bekenstein}(1993)}]{Bekenstein:1992pj}%
  \BibitemOpen
  \bibfield  {author} {\bibinfo {author} {\bibfnamefont {J.~D.}\ \bibnamefont
  {Bekenstein}},\ }\bibfield  {title} {\bibinfo {title} {{The Relation between
  physical and gravitational geometry}},\ }\href
  {https://doi.org/10.1103/PhysRevD.48.3641} {\bibfield  {journal} {\bibinfo
  {journal} {Phys. Rev.}\ }\textbf {\bibinfo {volume} {D48}},\ \bibinfo {pages}
  {3641} (\bibinfo {year} {1993})},\ \Eprint
  {https://arxiv.org/abs/gr-qc/9211017} {arXiv:gr-qc/9211017 [gr-qc]}
  \BibitemShut {NoStop}%
\bibitem [{\citenamefont {Zumalacarregui}\ \emph {et~al.}(2013)\citenamefont
  {Zumalacarregui}, \citenamefont {Koivisto},\ and\ \citenamefont
  {Mota}}]{Zumalacarregui:2012us}%
  \BibitemOpen
  \bibfield  {author} {\bibinfo {author} {\bibfnamefont {M.}~\bibnamefont
  {Zumalacarregui}}, \bibinfo {author} {\bibfnamefont {T.~S.}\ \bibnamefont
  {Koivisto}},\ and\ \bibinfo {author} {\bibfnamefont {D.~F.}\ \bibnamefont
  {Mota}},\ }\bibfield  {title} {\bibinfo {title} {{DBI Galileons in the
  Einstein Frame: Local Gravity and Cosmology}},\ }\href
  {https://doi.org/10.1103/PhysRevD.87.083010} {\bibfield  {journal} {\bibinfo
  {journal} {Phys. Rev.}\ }\textbf {\bibinfo {volume} {D87}},\ \bibinfo {pages}
  {083010} (\bibinfo {year} {2013})},\ \Eprint
  {https://arxiv.org/abs/1210.8016} {arXiv:1210.8016 [astro-ph.CO]}
  \BibitemShut {NoStop}%
\bibitem [{\citenamefont {Zumalacárregui}\ and\ \citenamefont
  {García-Bellido}(2014)}]{Zumalacarregui:2013pma}%
  \BibitemOpen
  \bibfield  {author} {\bibinfo {author} {\bibfnamefont {M.}~\bibnamefont
  {Zumalacárregui}}\ and\ \bibinfo {author} {\bibfnamefont {J.}~\bibnamefont
  {García-Bellido}},\ }\bibfield  {title} {\bibinfo {title} {{Transforming
  gravity: from derivative couplings to matter to second-order scalar-tensor
  theories beyond the Horndeski Lagrangian}},\ }\href
  {https://doi.org/10.1103/PhysRevD.89.064046} {\bibfield  {journal} {\bibinfo
  {journal} {Phys. Rev.}\ }\textbf {\bibinfo {volume} {D89}},\ \bibinfo {pages}
  {064046} (\bibinfo {year} {2014})},\ \Eprint
  {https://arxiv.org/abs/1308.4685} {arXiv:1308.4685 [gr-qc]} \BibitemShut
  {NoStop}%
\bibitem [{\citenamefont {Bettoni}\ and\ \citenamefont
  {Liberati}(2013)}]{Bettoni:2013diz}%
  \BibitemOpen
  \bibfield  {author} {\bibinfo {author} {\bibfnamefont {D.}~\bibnamefont
  {Bettoni}}\ and\ \bibinfo {author} {\bibfnamefont {S.}~\bibnamefont
  {Liberati}},\ }\bibfield  {title} {\bibinfo {title} {{Disformal invariance of
  second order scalar-tensor theories: Framing the Horndeski action}},\ }\href
  {https://doi.org/10.1103/PhysRevD.88.084020} {\bibfield  {journal} {\bibinfo
  {journal} {Phys. Rev.}\ }\textbf {\bibinfo {volume} {D88}},\ \bibinfo {pages}
  {084020} (\bibinfo {year} {2013})},\ \Eprint
  {https://arxiv.org/abs/1306.6724} {arXiv:1306.6724 [gr-qc]} \BibitemShut
  {NoStop}%
\bibitem [{\citenamefont {van~de Bruck}\ \emph {et~al.}(2013)\citenamefont
  {van~de Bruck}, \citenamefont {Morrice},\ and\ \citenamefont
  {Vu}}]{vandebruck:2013yxa}%
  \BibitemOpen
  \bibfield  {author} {\bibinfo {author} {\bibfnamefont {C.}~\bibnamefont
  {van~de Bruck}}, \bibinfo {author} {\bibfnamefont {J.}~\bibnamefont
  {Morrice}},\ and\ \bibinfo {author} {\bibfnamefont {S.}~\bibnamefont {Vu}},\
  }\bibfield  {title} {\bibinfo {title} {{Constraints on Nonconformal Couplings
  from the Properties of the Cosmic Microwave Background Radiation}},\ }\href
  {https://doi.org/10.1103/PhysRevLett.111.161302} {\bibfield  {journal}
  {\bibinfo  {journal} {Phys. Rev. Lett.}\ }\textbf {\bibinfo {volume} {111}},\
  \bibinfo {pages} {161302} (\bibinfo {year} {2013})},\ \Eprint
  {https://arxiv.org/abs/1303.1773} {arXiv:1303.1773 [astro-ph.CO]}
  \BibitemShut {NoStop}%
\bibitem [{\citenamefont {van~de Bruck}\ \emph
  {et~al.}(2016{\natexlab{a}})\citenamefont {van~de Bruck}, \citenamefont
  {Burrage},\ and\ \citenamefont {Morrice}}]{vandebruck:2016cnh}%
  \BibitemOpen
  \bibfield  {author} {\bibinfo {author} {\bibfnamefont {C.}~\bibnamefont
  {van~de Bruck}}, \bibinfo {author} {\bibfnamefont {C.}~\bibnamefont
  {Burrage}},\ and\ \bibinfo {author} {\bibfnamefont {J.}~\bibnamefont
  {Morrice}},\ }\bibfield  {title} {\bibinfo {title} {{Vacuum Cherenkov
  radiation and bremsstrahlung from disformal couplings}},\ }\href
  {https://doi.org/10.1088/1475-7516/2016/08/003} {\bibfield  {journal}
  {\bibinfo  {journal} {JCAP}\ }\textbf {\bibinfo {volume} {08}},\ \bibinfo
  {pages} {003}},\ \Eprint {https://arxiv.org/abs/1605.03567} {arXiv:1605.03567
  [gr-qc]} \BibitemShut {NoStop}%
\bibitem [{\citenamefont {Faraoni}\ \emph {et~al.}(1999)\citenamefont
  {Faraoni}, \citenamefont {Gunzig},\ and\ \citenamefont
  {Nardone}}]{Faraoni:1998qx}%
  \BibitemOpen
  \bibfield  {author} {\bibinfo {author} {\bibfnamefont {V.}~\bibnamefont
  {Faraoni}}, \bibinfo {author} {\bibfnamefont {E.}~\bibnamefont {Gunzig}},\
  and\ \bibinfo {author} {\bibfnamefont {P.}~\bibnamefont {Nardone}},\
  }\bibfield  {title} {\bibinfo {title} {{Conformal transformations in
  classical gravitational theories and in cosmology}},\ }\href@noop {}
  {\bibfield  {journal} {\bibinfo  {journal} {Fund. Cosmic Phys.}\ }\textbf
  {\bibinfo {volume} {20}},\ \bibinfo {pages} {121} (\bibinfo {year} {1999})},\
  \Eprint {https://arxiv.org/abs/gr-qc/9811047} {arXiv:gr-qc/9811047 [gr-qc]}
  \BibitemShut {NoStop}%
\bibitem [{\citenamefont {Koivisto}\ \emph {et~al.}(2014)\citenamefont
  {Koivisto}, \citenamefont {Wills},\ and\ \citenamefont
  {Zavala}}]{Koivisto:2013fta}%
  \BibitemOpen
  \bibfield  {author} {\bibinfo {author} {\bibfnamefont {T.}~\bibnamefont
  {Koivisto}}, \bibinfo {author} {\bibfnamefont {D.}~\bibnamefont {Wills}},\
  and\ \bibinfo {author} {\bibfnamefont {I.}~\bibnamefont {Zavala}},\
  }\bibfield  {title} {\bibinfo {title} {{Dark D-brane Cosmology}},\ }\href
  {https://doi.org/10.1088/1475-7516/2014/06/036} {\bibfield  {journal}
  {\bibinfo  {journal} {JCAP}\ }\textbf {\bibinfo {volume} {1406}},\ \bibinfo
  {pages} {036}},\ \Eprint {https://arxiv.org/abs/1312.2597} {arXiv:1312.2597
  [hep-th]} \BibitemShut {NoStop}%
\bibitem [{\citenamefont {van~de Bruck}\ and\ \citenamefont
  {Teixeira}(2020)}]{vandebruck:2020fjo}%
  \BibitemOpen
  \bibfield  {author} {\bibinfo {author} {\bibfnamefont {C.}~\bibnamefont
  {van~de Bruck}}\ and\ \bibinfo {author} {\bibfnamefont {E.~M.}\ \bibnamefont
  {Teixeira}},\ }\bibfield  {title} {\bibinfo {title} {{Dark D-Brane Cosmology:
  from background evolution to cosmological perturbations}},\ }\href
  {https://doi.org/10.1103/PhysRevD.102.103503} {\bibfield  {journal} {\bibinfo
   {journal} {Phys. Rev. D}\ }\textbf {\bibinfo {volume} {102}},\ \bibinfo
  {pages} {103503} (\bibinfo {year} {2020})},\ \Eprint
  {https://arxiv.org/abs/2007.15414} {arXiv:2007.15414 [gr-qc]} \BibitemShut
  {NoStop}%
\bibitem [{\citenamefont {Koivisto}(2008)}]{Koivisto:2008ak}%
  \BibitemOpen
  \bibfield  {author} {\bibinfo {author} {\bibfnamefont {T.~S.}\ \bibnamefont
  {Koivisto}},\ }\bibfield  {title} {\bibinfo {title} {{Disformal
  quintessence}},\ }\Eprint {https://arxiv.org/abs/0811.1957} {arXiv:0811.1957
  [astro-ph]}  (\bibinfo {year} {2008})\BibitemShut {NoStop}%
\bibitem [{\citenamefont {Zumalacarregui}\ \emph {et~al.}(2010)\citenamefont
  {Zumalacarregui}, \citenamefont {Koivisto}, \citenamefont {Mota},\ and\
  \citenamefont {Ruiz-Lapuente}}]{Zumalacarregui:2010wj}%
  \BibitemOpen
  \bibfield  {author} {\bibinfo {author} {\bibfnamefont {M.}~\bibnamefont
  {Zumalacarregui}}, \bibinfo {author} {\bibfnamefont {T.~S.}\ \bibnamefont
  {Koivisto}}, \bibinfo {author} {\bibfnamefont {D.~F.}\ \bibnamefont {Mota}},\
  and\ \bibinfo {author} {\bibfnamefont {P.}~\bibnamefont {Ruiz-Lapuente}},\
  }\bibfield  {title} {\bibinfo {title} {{Disformal Scalar Fields and the Dark
  Sector of the Universe}},\ }\href
  {https://doi.org/10.1088/1475-7516/2010/05/038} {\bibfield  {journal}
  {\bibinfo  {journal} {JCAP}\ }\textbf {\bibinfo {volume} {1005}},\ \bibinfo
  {pages} {038}},\ \Eprint {https://arxiv.org/abs/1004.2684} {arXiv:1004.2684
  [astro-ph.CO]} \BibitemShut {NoStop}%
\bibitem [{\citenamefont {De~Felice}\ and\ \citenamefont
  {Tsujikawa}(2012)}]{DeFelice:2011bh}%
  \BibitemOpen
  \bibfield  {author} {\bibinfo {author} {\bibfnamefont {A.}~\bibnamefont
  {De~Felice}}\ and\ \bibinfo {author} {\bibfnamefont {S.}~\bibnamefont
  {Tsujikawa}},\ }\bibfield  {title} {\bibinfo {title} {{Conditions for the
  cosmological viability of the most general scalar-tensor theories and their
  applications to extended Galileon dark energy models}},\ }\href
  {https://doi.org/10.1088/1475-7516/2012/02/007} {\bibfield  {journal}
  {\bibinfo  {journal} {JCAP}\ }\textbf {\bibinfo {volume} {1202}},\ \bibinfo
  {pages} {007}},\ \Eprint {https://arxiv.org/abs/1110.3878} {arXiv:1110.3878
  [gr-qc]} \BibitemShut {NoStop}%
\bibitem [{\citenamefont {Koivisto}\ \emph {et~al.}(2012)\citenamefont
  {Koivisto}, \citenamefont {Mota},\ and\ \citenamefont
  {Zumalacarregui}}]{Koivisto:2012za}%
  \BibitemOpen
  \bibfield  {author} {\bibinfo {author} {\bibfnamefont {T.~S.}\ \bibnamefont
  {Koivisto}}, \bibinfo {author} {\bibfnamefont {D.~F.}\ \bibnamefont {Mota}},\
  and\ \bibinfo {author} {\bibfnamefont {M.}~\bibnamefont {Zumalacarregui}},\
  }\bibfield  {title} {\bibinfo {title} {{Screening Modifications of Gravity
  through Disformally Coupled Fields}},\ }\href
  {https://doi.org/10.1103/PhysRevLett.109.241102} {\bibfield  {journal}
  {\bibinfo  {journal} {Phys. Rev. Lett.}\ }\textbf {\bibinfo {volume} {109}},\
  \bibinfo {pages} {241102} (\bibinfo {year} {2012})},\ \Eprint
  {https://arxiv.org/abs/1205.3167} {arXiv:1205.3167 [astro-ph.CO]}
  \BibitemShut {NoStop}%
\bibitem [{\citenamefont {van~de Bruck}\ and\ \citenamefont
  {Morrice}(2015)}]{vandebruck:2015ida}%
  \BibitemOpen
  \bibfield  {author} {\bibinfo {author} {\bibfnamefont {C.}~\bibnamefont
  {van~de Bruck}}\ and\ \bibinfo {author} {\bibfnamefont {J.}~\bibnamefont
  {Morrice}},\ }\bibfield  {title} {\bibinfo {title} {{Disformal couplings and
  the dark sector of the universe}},\ }\href
  {https://doi.org/10.1088/1475-7516/2015/04/036} {\bibfield  {journal}
  {\bibinfo  {journal} {JCAP}\ }\textbf {\bibinfo {volume} {04}},\ \bibinfo
  {pages} {036}},\ \Eprint {https://arxiv.org/abs/1501.03073} {arXiv:1501.03073
  [gr-qc]} \BibitemShut {NoStop}%
\bibitem [{\citenamefont {Bettoni}\ and\ \citenamefont
  {Liberati}(2015)}]{Bettoni:2015wla}%
  \BibitemOpen
  \bibfield  {author} {\bibinfo {author} {\bibfnamefont {D.}~\bibnamefont
  {Bettoni}}\ and\ \bibinfo {author} {\bibfnamefont {S.}~\bibnamefont
  {Liberati}},\ }\bibfield  {title} {\bibinfo {title} {{Dynamics of
  non-minimally coupled perfect fluids}},\ }\href
  {https://doi.org/10.1088/1475-7516/2015/08/023} {\bibfield  {journal}
  {\bibinfo  {journal} {JCAP}\ }\textbf {\bibinfo {volume} {1508}}\bibfield
  {number} {\bibinfo  {number} { (08)},\ \bibinfo {pages} {023}},\ }\Eprint
  {https://arxiv.org/abs/1502.06613} {arXiv:1502.06613 [gr-qc]} \BibitemShut
  {NoStop}%
\bibitem [{\citenamefont {Sakstein}(2015)}]{Sakstein:2014aca}%
  \BibitemOpen
  \bibfield  {author} {\bibinfo {author} {\bibfnamefont {J.}~\bibnamefont
  {Sakstein}},\ }\bibfield  {title} {\bibinfo {title} {{Towards Viable
  Cosmological Models of Disformal Theories of Gravity}},\ }\href
  {https://doi.org/10.1103/PhysRevD.91.024036} {\bibfield  {journal} {\bibinfo
  {journal} {Phys. Rev.}\ }\textbf {\bibinfo {volume} {D91}},\ \bibinfo {pages}
  {024036} (\bibinfo {year} {2015})},\ \Eprint
  {https://arxiv.org/abs/1409.7296} {arXiv:1409.7296 [astro-ph.CO]}
  \BibitemShut {NoStop}%
\bibitem [{\citenamefont {Sakstein}\ and\ \citenamefont
  {Verner}(2015)}]{Sakstein:2015jca}%
  \BibitemOpen
  \bibfield  {author} {\bibinfo {author} {\bibfnamefont {J.}~\bibnamefont
  {Sakstein}}\ and\ \bibinfo {author} {\bibfnamefont {S.}~\bibnamefont
  {Verner}},\ }\bibfield  {title} {\bibinfo {title} {{Disformal Gravity
  Theories: A Jordan Frame Analysis}},\ }\href
  {https://doi.org/10.1103/PhysRevD.92.123005} {\bibfield  {journal} {\bibinfo
  {journal} {Phys. Rev.}\ }\textbf {\bibinfo {volume} {D92}},\ \bibinfo {pages}
  {123005} (\bibinfo {year} {2015})},\ \Eprint
  {https://arxiv.org/abs/1509.05679} {arXiv:1509.05679 [gr-qc]} \BibitemShut
  {NoStop}%
\bibitem [{\citenamefont {van~de Bruck}\ \emph
  {et~al.}(2016{\natexlab{b}})\citenamefont {van~de Bruck}, \citenamefont
  {Mifsud}, \citenamefont {Mimoso},\ and\ \citenamefont
  {Nunes}}]{vandebruck:2016jgg}%
  \BibitemOpen
  \bibfield  {author} {\bibinfo {author} {\bibfnamefont {C.}~\bibnamefont
  {van~de Bruck}}, \bibinfo {author} {\bibfnamefont {J.}~\bibnamefont
  {Mifsud}}, \bibinfo {author} {\bibfnamefont {J.~P.}\ \bibnamefont {Mimoso}},\
  and\ \bibinfo {author} {\bibfnamefont {N.~J.}\ \bibnamefont {Nunes}},\
  }\bibfield  {title} {\bibinfo {title} {{Generalized dark energy interactions
  with multiple fluids}},\ }\href
  {https://doi.org/10.1088/1475-7516/2016/11/031} {\bibfield  {journal}
  {\bibinfo  {journal} {JCAP}\ }\textbf {\bibinfo {volume} {1611}}\bibfield
  {number} {\bibinfo  {number} { (11)},\ \bibinfo {pages} {031}},\ }\Eprint
  {https://arxiv.org/abs/1605.03834} {arXiv:1605.03834 [gr-qc]} \BibitemShut
  {NoStop}%
\bibitem [{\citenamefont {Teixeira}\ \emph {et~al.}(2020)\citenamefont
  {Teixeira}, \citenamefont {Nunes},\ and\ \citenamefont
  {Nunes}}]{teixeira:2019hil}%
  \BibitemOpen
  \bibfield  {author} {\bibinfo {author} {\bibfnamefont {E.~M.}\ \bibnamefont
  {Teixeira}}, \bibinfo {author} {\bibfnamefont {A.}~\bibnamefont {Nunes}},\
  and\ \bibinfo {author} {\bibfnamefont {N.~J.}\ \bibnamefont {Nunes}},\
  }\bibfield  {title} {\bibinfo {title} {{Disformally Coupled Quintessence}},\
  }\href {https://doi.org/10.1103/PhysRevD.101.083506} {\bibfield  {journal}
  {\bibinfo  {journal} {Phys. Rev. D}\ }\textbf {\bibinfo {volume} {101}},\
  \bibinfo {pages} {083506} (\bibinfo {year} {2020})},\ \Eprint
  {https://arxiv.org/abs/1912.13348} {arXiv:1912.13348 [gr-qc]} \BibitemShut
  {NoStop}%
\bibitem [{\citenamefont {Karwan}\ and\ \citenamefont
  {Sapa}(2017)}]{karwan:2016cnv}%
  \BibitemOpen
  \bibfield  {author} {\bibinfo {author} {\bibfnamefont {K.}~\bibnamefont
  {Karwan}}\ and\ \bibinfo {author} {\bibfnamefont {S.}~\bibnamefont {Sapa}},\
  }\bibfield  {title} {\bibinfo {title} {{Dynamics of the universe with
  disformal coupling between the dark sectors}},\ }\href
  {https://doi.org/10.1140/epjc/s10052-017-4924-4} {\bibfield  {journal}
  {\bibinfo  {journal} {Eur. Phys. J.}\ }\textbf {\bibinfo {volume} {C77}},\
  \bibinfo {pages} {352} (\bibinfo {year} {2017})},\ \Eprint
  {https://arxiv.org/abs/1611.05324} {arXiv:1611.05324 [gr-qc]} \BibitemShut
  {NoStop}%
\bibitem [{\citenamefont {Gumjudpai}\ \emph {et~al.}(2005)\citenamefont
  {Gumjudpai}, \citenamefont {Naskar}, \citenamefont {Sami},\ and\
  \citenamefont {Tsujikawa}}]{Gumjudpai:2005ry}%
  \BibitemOpen
  \bibfield  {author} {\bibinfo {author} {\bibfnamefont {B.}~\bibnamefont
  {Gumjudpai}}, \bibinfo {author} {\bibfnamefont {T.}~\bibnamefont {Naskar}},
  \bibinfo {author} {\bibfnamefont {M.}~\bibnamefont {Sami}},\ and\ \bibinfo
  {author} {\bibfnamefont {S.}~\bibnamefont {Tsujikawa}},\ }\bibfield  {title}
  {\bibinfo {title} {{Coupled dark energy: Towards a general description of the
  dynamics}},\ }\href {https://doi.org/10.1088/1475-7516/2005/06/007}
  {\bibfield  {journal} {\bibinfo  {journal} {JCAP}\ }\textbf {\bibinfo
  {volume} {0506}},\ \bibinfo {pages} {007}},\ \Eprint
  {https://arxiv.org/abs/hep-th/0502191} {arXiv:hep-th/0502191 [hep-th]}
  \BibitemShut {NoStop}%
\bibitem [{\citenamefont {Holz}\ and\ \citenamefont
  {Hughes}(2005)}]{Holz:2005df}%
  \BibitemOpen
  \bibfield  {author} {\bibinfo {author} {\bibfnamefont {D.~E.}\ \bibnamefont
  {Holz}}\ and\ \bibinfo {author} {\bibfnamefont {S.~A.}\ \bibnamefont
  {Hughes}},\ }\bibfield  {title} {\bibinfo {title} {{Using gravitational-wave
  standard sirens}},\ }\href {https://doi.org/10.1086/431341} {\bibfield
  {journal} {\bibinfo  {journal} {Astrophys. J.}\ }\textbf {\bibinfo {volume}
  {629}},\ \bibinfo {pages} {15} (\bibinfo {year} {2005})},\ \Eprint
  {https://arxiv.org/abs/astro-ph/0504616} {arXiv:astro-ph/0504616}
  \BibitemShut {NoStop}%
\bibitem [{\citenamefont {Schutz}(1986)}]{Schutz:1986gp}%
  \BibitemOpen
  \bibfield  {author} {\bibinfo {author} {\bibfnamefont {B.~F.}\ \bibnamefont
  {Schutz}},\ }\bibfield  {title} {\bibinfo {title} {{Determining the Hubble
  Constant from Gravitational Wave Observations}},\ }\href
  {https://doi.org/10.1038/323310a0} {\bibfield  {journal} {\bibinfo  {journal}
  {Nature}\ }\textbf {\bibinfo {volume} {323}},\ \bibinfo {pages} {310}
  (\bibinfo {year} {1986})}\BibitemShut {NoStop}%
\bibitem [{\citenamefont {Abbott}\ \emph
  {et~al.}(2017{\natexlab{a}})\citenamefont {Abbott} \emph
  {et~al.}}]{LIGOScientific:2017vwq}%
  \BibitemOpen
  \bibfield  {author} {\bibinfo {author} {\bibfnamefont {B.~P.}\ \bibnamefont
  {Abbott}} \emph {et~al.} (\bibinfo {collaboration} {LIGO Scientific,
  Virgo}),\ }\bibfield  {title} {\bibinfo {title} {{GW170817: Observation of
  Gravitational Waves from a Binary Neutron Star Inspiral}},\ }\href
  {https://doi.org/10.1103/PhysRevLett.119.161101} {\bibfield  {journal}
  {\bibinfo  {journal} {Phys. Rev. Lett.}\ }\textbf {\bibinfo {volume} {119}},\
  \bibinfo {pages} {161101} (\bibinfo {year} {2017}{\natexlab{a}})},\ \Eprint
  {https://arxiv.org/abs/1710.05832} {arXiv:1710.05832 [gr-qc]} \BibitemShut
  {NoStop}%
\bibitem [{\citenamefont {Abbott}\ \emph
  {et~al.}(2017{\natexlab{b}})\citenamefont {Abbott} \emph
  {et~al.}}]{LIGOScientific:2017zic}%
  \BibitemOpen
  \bibfield  {author} {\bibinfo {author} {\bibfnamefont {B.~P.}\ \bibnamefont
  {Abbott}} \emph {et~al.} (\bibinfo {collaboration} {LIGO Scientific, Virgo,
  Fermi-GBM, INTEGRAL}),\ }\bibfield  {title} {\bibinfo {title} {{Gravitational
  Waves and Gamma-rays from a Binary Neutron Star Merger: GW170817 and GRB
  170817A}},\ }\href {https://doi.org/10.3847/2041-8213/aa920c} {\bibfield
  {journal} {\bibinfo  {journal} {Astrophys. J. Lett.}\ }\textbf {\bibinfo
  {volume} {848}},\ \bibinfo {pages} {L13} (\bibinfo {year}
  {2017}{\natexlab{b}})},\ \Eprint {https://arxiv.org/abs/1710.05834}
  {arXiv:1710.05834 [astro-ph.HE]} \BibitemShut {NoStop}%
\bibitem [{\citenamefont {Creminelli}\ and\ \citenamefont
  {Vernizzi}(2017)}]{Creminelli:2017sry}%
  \BibitemOpen
  \bibfield  {author} {\bibinfo {author} {\bibfnamefont {P.}~\bibnamefont
  {Creminelli}}\ and\ \bibinfo {author} {\bibfnamefont {F.}~\bibnamefont
  {Vernizzi}},\ }\bibfield  {title} {\bibinfo {title} {{Dark Energy after
  GW170817 and GRB170817A}},\ }\href
  {https://doi.org/10.1103/PhysRevLett.119.251302} {\bibfield  {journal}
  {\bibinfo  {journal} {Phys. Rev. Lett.}\ }\textbf {\bibinfo {volume} {119}},\
  \bibinfo {pages} {251302} (\bibinfo {year} {2017})},\ \Eprint
  {https://arxiv.org/abs/1710.05877} {arXiv:1710.05877 [astro-ph.CO]}
  \BibitemShut {NoStop}%
\bibitem [{\citenamefont {Baker}\ \emph {et~al.}(2017)\citenamefont {Baker},
  \citenamefont {Bellini}, \citenamefont {Ferreira}, \citenamefont {Lagos},
  \citenamefont {Noller},\ and\ \citenamefont {Sawicki}}]{Baker:2017hug}%
  \BibitemOpen
  \bibfield  {author} {\bibinfo {author} {\bibfnamefont {T.}~\bibnamefont
  {Baker}}, \bibinfo {author} {\bibfnamefont {E.}~\bibnamefont {Bellini}},
  \bibinfo {author} {\bibfnamefont {P.~G.}\ \bibnamefont {Ferreira}}, \bibinfo
  {author} {\bibfnamefont {M.}~\bibnamefont {Lagos}}, \bibinfo {author}
  {\bibfnamefont {J.}~\bibnamefont {Noller}},\ and\ \bibinfo {author}
  {\bibfnamefont {I.}~\bibnamefont {Sawicki}},\ }\bibfield  {title} {\bibinfo
  {title} {{Strong constraints on cosmological gravity from GW170817 and GRB
  170817A}},\ }\href {https://doi.org/10.1103/PhysRevLett.119.251301}
  {\bibfield  {journal} {\bibinfo  {journal} {Phys. Rev. Lett.}\ }\textbf
  {\bibinfo {volume} {119}},\ \bibinfo {pages} {251301} (\bibinfo {year}
  {2017})},\ \Eprint {https://arxiv.org/abs/1710.06394} {arXiv:1710.06394
  [astro-ph.CO]} \BibitemShut {NoStop}%
\bibitem [{\citenamefont {Ezquiaga}\ and\ \citenamefont
  {Zumalac\'arregui}(2017)}]{Ezquiaga:2017ekz}%
  \BibitemOpen
  \bibfield  {author} {\bibinfo {author} {\bibfnamefont {J.~M.}\ \bibnamefont
  {Ezquiaga}}\ and\ \bibinfo {author} {\bibfnamefont {M.}~\bibnamefont
  {Zumalac\'arregui}},\ }\bibfield  {title} {\bibinfo {title} {{Dark Energy
  After GW170817: Dead Ends and the Road Ahead}},\ }\href
  {https://doi.org/10.1103/PhysRevLett.119.251304} {\bibfield  {journal}
  {\bibinfo  {journal} {Phys. Rev. Lett.}\ }\textbf {\bibinfo {volume} {119}},\
  \bibinfo {pages} {251304} (\bibinfo {year} {2017})},\ \Eprint
  {https://arxiv.org/abs/1710.05901} {arXiv:1710.05901 [astro-ph.CO]}
  \BibitemShut {NoStop}%
\bibitem [{\citenamefont {Creminelli}\ \emph {et~al.}(2018)\citenamefont
  {Creminelli}, \citenamefont {Lewandowski}, \citenamefont {Tambalo},\ and\
  \citenamefont {Vernizzi}}]{Creminelli:2018xsv}%
  \BibitemOpen
  \bibfield  {author} {\bibinfo {author} {\bibfnamefont {P.}~\bibnamefont
  {Creminelli}}, \bibinfo {author} {\bibfnamefont {M.}~\bibnamefont
  {Lewandowski}}, \bibinfo {author} {\bibfnamefont {G.}~\bibnamefont
  {Tambalo}},\ and\ \bibinfo {author} {\bibfnamefont {F.}~\bibnamefont
  {Vernizzi}},\ }\bibfield  {title} {\bibinfo {title} {{Gravitational Wave
  Decay into Dark Energy}},\ }\href
  {https://doi.org/10.1088/1475-7516/2018/12/025} {\bibfield  {journal}
  {\bibinfo  {journal} {JCAP}\ }\textbf {\bibinfo {volume} {12}},\ \bibinfo
  {pages} {025}},\ \Eprint {https://arxiv.org/abs/1809.03484} {arXiv:1809.03484
  [astro-ph.CO]} \BibitemShut {NoStop}%
\bibitem [{\citenamefont {Amendola}\ \emph {et~al.}(2018)\citenamefont
  {Amendola}, \citenamefont {Kunz}, \citenamefont {Saltas},\ and\ \citenamefont
  {Sawicki}}]{Amendola:2017orw}%
  \BibitemOpen
  \bibfield  {author} {\bibinfo {author} {\bibfnamefont {L.}~\bibnamefont
  {Amendola}}, \bibinfo {author} {\bibfnamefont {M.}~\bibnamefont {Kunz}},
  \bibinfo {author} {\bibfnamefont {I.~D.}\ \bibnamefont {Saltas}},\ and\
  \bibinfo {author} {\bibfnamefont {I.}~\bibnamefont {Sawicki}},\ }\bibfield
  {title} {\bibinfo {title} {{Fate of Large-Scale Structure in Modified Gravity
  After GW170817 and GRB170817A}},\ }\href
  {https://doi.org/10.1103/PhysRevLett.120.131101} {\bibfield  {journal}
  {\bibinfo  {journal} {Phys. Rev. Lett.}\ }\textbf {\bibinfo {volume} {120}},\
  \bibinfo {pages} {131101} (\bibinfo {year} {2018})},\ \Eprint
  {https://arxiv.org/abs/1711.04825} {arXiv:1711.04825 [astro-ph.CO]}
  \BibitemShut {NoStop}%
\bibitem [{\citenamefont {Belgacem}\ \emph {et~al.}(2019)\citenamefont
  {Belgacem} \emph {et~al.}}]{LISACosmologyWorkingGroup:2019mwx}%
  \BibitemOpen
  \bibfield  {author} {\bibinfo {author} {\bibfnamefont {E.}~\bibnamefont
  {Belgacem}} \emph {et~al.} (\bibinfo {collaboration} {LISA Cosmology Working
  Group}),\ }\bibfield  {title} {\bibinfo {title} {{Testing modified gravity at
  cosmological distances with LISA standard sirens}},\ }\href
  {https://doi.org/10.1088/1475-7516/2019/07/024} {\bibfield  {journal}
  {\bibinfo  {journal} {JCAP}\ }\textbf {\bibinfo {volume} {07}},\ \bibinfo
  {pages} {024}},\ \Eprint {https://arxiv.org/abs/1906.01593} {arXiv:1906.01593
  [astro-ph.CO]} \BibitemShut {NoStop}%
\bibitem [{\citenamefont {Belgacem}\ \emph {et~al.}(2018)\citenamefont
  {Belgacem}, \citenamefont {Dirian}, \citenamefont {Foffa},\ and\
  \citenamefont {Maggiore}}]{Belgacem:2018lbp}%
  \BibitemOpen
  \bibfield  {author} {\bibinfo {author} {\bibfnamefont {E.}~\bibnamefont
  {Belgacem}}, \bibinfo {author} {\bibfnamefont {Y.}~\bibnamefont {Dirian}},
  \bibinfo {author} {\bibfnamefont {S.}~\bibnamefont {Foffa}},\ and\ \bibinfo
  {author} {\bibfnamefont {M.}~\bibnamefont {Maggiore}},\ }\bibfield  {title}
  {\bibinfo {title} {{Modified gravitational-wave propagation and standard
  sirens}},\ }\href {https://doi.org/10.1103/PhysRevD.98.023510} {\bibfield
  {journal} {\bibinfo  {journal} {Phys. Rev. D}\ }\textbf {\bibinfo {volume}
  {98}},\ \bibinfo {pages} {023510} (\bibinfo {year} {2018})},\ \Eprint
  {https://arxiv.org/abs/1805.08731} {arXiv:1805.08731 [gr-qc]} \BibitemShut
  {NoStop}%
\bibitem [{\citenamefont {Allahyari}\ \emph {et~al.}(2022)\citenamefont
  {Allahyari}, \citenamefont {Nunes},\ and\ \citenamefont
  {Mota}}]{Allahyari:2021enz}%
  \BibitemOpen
  \bibfield  {author} {\bibinfo {author} {\bibfnamefont {A.}~\bibnamefont
  {Allahyari}}, \bibinfo {author} {\bibfnamefont {R.~C.}\ \bibnamefont
  {Nunes}},\ and\ \bibinfo {author} {\bibfnamefont {D.~F.}\ \bibnamefont
  {Mota}},\ }\bibfield  {title} {\bibinfo {title} {{No slip gravity in light of
  LISA standard sirens}},\ }\href {https://doi.org/10.1093/mnras/stac1445}
  {\bibfield  {journal} {\bibinfo  {journal} {Mon. Not. Roy. Astron. Soc.}\
  }\textbf {\bibinfo {volume} {514}},\ \bibinfo {pages} {1274} (\bibinfo {year}
  {2022})},\ \Eprint {https://arxiv.org/abs/2110.07634} {arXiv:2110.07634
  [astro-ph.CO]} \BibitemShut {NoStop}%
\bibitem [{\citenamefont {Califano}\ \emph {et~al.}(2023)\citenamefont
  {Califano}, \citenamefont {de~Martino}, \citenamefont {Vernieri},\ and\
  \citenamefont {Capozziello}}]{Califano:2022syd}%
  \BibitemOpen
  \bibfield  {author} {\bibinfo {author} {\bibfnamefont {M.}~\bibnamefont
  {Califano}}, \bibinfo {author} {\bibfnamefont {I.}~\bibnamefont
  {de~Martino}}, \bibinfo {author} {\bibfnamefont {D.}~\bibnamefont
  {Vernieri}},\ and\ \bibinfo {author} {\bibfnamefont {S.}~\bibnamefont
  {Capozziello}},\ }\bibfield  {title} {\bibinfo {title} {{Exploiting the
  Einstein Telescope to solve the Hubble tension}},\ }\href
  {https://doi.org/10.1103/PhysRevD.107.123519} {\bibfield  {journal} {\bibinfo
   {journal} {Phys. Rev. D}\ }\textbf {\bibinfo {volume} {107}},\ \bibinfo
  {pages} {123519} (\bibinfo {year} {2023})},\ \Eprint
  {https://arxiv.org/abs/2208.13999} {arXiv:2208.13999 [astro-ph.CO]}
  \BibitemShut {NoStop}%
\bibitem [{\citenamefont {Ferreira}\ \emph {et~al.}(2022)\citenamefont
  {Ferreira}, \citenamefont {Barreiro}, \citenamefont {Mimoso},\ and\
  \citenamefont {Nunes}}]{Ferreira:2022jcd}%
  \BibitemOpen
  \bibfield  {author} {\bibinfo {author} {\bibfnamefont {J.}~\bibnamefont
  {Ferreira}}, \bibinfo {author} {\bibfnamefont {T.}~\bibnamefont {Barreiro}},
  \bibinfo {author} {\bibfnamefont {J.}~\bibnamefont {Mimoso}},\ and\ \bibinfo
  {author} {\bibfnamefont {N.~J.}\ \bibnamefont {Nunes}},\ }\bibfield  {title}
  {\bibinfo {title} {{Forecasting F(Q) cosmology with \ensuremath{\Lambda}CDM
  background using standard sirens}},\ }\href
  {https://doi.org/10.1103/PhysRevD.105.123531} {\bibfield  {journal} {\bibinfo
   {journal} {Phys. Rev. D}\ }\textbf {\bibinfo {volume} {105}},\ \bibinfo
  {pages} {123531} (\bibinfo {year} {2022})},\ \Eprint
  {https://arxiv.org/abs/2203.13788} {arXiv:2203.13788 [astro-ph.CO]}
  \BibitemShut {NoStop}%
\bibitem [{\citenamefont {Acernese}\ \emph {et~al.}(2015)\citenamefont
  {Acernese} \emph {et~al.}}]{VIRGO:2014yos}%
  \BibitemOpen
  \bibfield  {author} {\bibinfo {author} {\bibfnamefont {F.}~\bibnamefont
  {Acernese}} \emph {et~al.} (\bibinfo {collaboration} {VIRGO}),\ }\bibfield
  {title} {\bibinfo {title} {{Advanced Virgo: a second-generation
  interferometric gravitational wave detector}},\ }\href
  {https://doi.org/10.1088/0264-9381/32/2/024001} {\bibfield  {journal}
  {\bibinfo  {journal} {Class. Quant. Grav.}\ }\textbf {\bibinfo {volume}
  {32}},\ \bibinfo {pages} {024001} (\bibinfo {year} {2015})},\ \Eprint
  {https://arxiv.org/abs/1408.3978} {arXiv:1408.3978 [gr-qc]} \BibitemShut
  {NoStop}%
\bibitem [{\citenamefont {Aasi}\ \emph {et~al.}(2015)\citenamefont {Aasi} \emph
  {et~al.}}]{LIGOScientific:2014pky}%
  \BibitemOpen
  \bibfield  {author} {\bibinfo {author} {\bibfnamefont {J.}~\bibnamefont
  {Aasi}} \emph {et~al.} (\bibinfo {collaboration} {LIGO Scientific}),\
  }\bibfield  {title} {\bibinfo {title} {{Advanced LIGO}},\ }\href
  {https://doi.org/10.1088/0264-9381/32/7/074001} {\bibfield  {journal}
  {\bibinfo  {journal} {Class. Quant. Grav.}\ }\textbf {\bibinfo {volume}
  {32}},\ \bibinfo {pages} {074001} (\bibinfo {year} {2015})},\ \Eprint
  {https://arxiv.org/abs/1411.4547} {arXiv:1411.4547 [gr-qc]} \BibitemShut
  {NoStop}%
\bibitem [{\citenamefont {Somiya}(2012)}]{Somiya:2011np}%
  \BibitemOpen
  \bibfield  {author} {\bibinfo {author} {\bibfnamefont {K.}~\bibnamefont
  {Somiya}} (\bibinfo {collaboration} {KAGRA}),\ }\bibfield  {title} {\bibinfo
  {title} {{Detector configuration of KAGRA: The Japanese cryogenic
  gravitational-wave detector}},\ }\href
  {https://doi.org/10.1088/0264-9381/29/12/124007} {\bibfield  {journal}
  {\bibinfo  {journal} {Class. Quant. Grav.}\ }\textbf {\bibinfo {volume}
  {29}},\ \bibinfo {pages} {124007} (\bibinfo {year} {2012})},\ \Eprint
  {https://arxiv.org/abs/1111.7185} {arXiv:1111.7185 [gr-qc]} \BibitemShut
  {NoStop}%
\bibitem [{\citenamefont {Iyer}(2015)}]{IndIGO}%
  \BibitemOpen
  \bibfield  {author} {\bibinfo {author} {\bibfnamefont {B.}~\bibnamefont
  {Iyer}} (\bibinfo {collaboration} {IndIGO}),\ }\href@noop {} {\bibinfo
  {title} {{Technical Report}}} (\bibinfo {year} {2015}),\ \Eprint
  {https://arxiv.org/abs/{https://dcc.ligo.org/LIGO-M1100296/public}}
  {{https://dcc.ligo.org/LIGO-M1100296/public}} \BibitemShut {NoStop}%
\bibitem [{\citenamefont {Punturo}\ \emph {et~al.}(2010)\citenamefont {Punturo}
  \emph {et~al.}}]{Punturo:2010zz}%
  \BibitemOpen
  \bibfield  {author} {\bibinfo {author} {\bibfnamefont {M.}~\bibnamefont
  {Punturo}} \emph {et~al.},\ }\bibfield  {title} {\bibinfo {title} {{The
  Einstein Telescope: A third-generation gravitational wave observatory}},\
  }\href {https://doi.org/10.1088/0264-9381/27/19/194002} {\bibfield  {journal}
  {\bibinfo  {journal} {Class. Quant. Grav.}\ }\textbf {\bibinfo {volume}
  {27}},\ \bibinfo {pages} {194002} (\bibinfo {year} {2010})}\BibitemShut
  {NoStop}%
\bibitem [{\citenamefont {Sathyaprakash}\ \emph {et~al.}(2012)\citenamefont
  {Sathyaprakash} \emph {et~al.}}]{Sathyaprakash:2012jk}%
  \BibitemOpen
  \bibfield  {author} {\bibinfo {author} {\bibfnamefont {B.}~\bibnamefont
  {Sathyaprakash}} \emph {et~al.},\ }\bibfield  {title} {\bibinfo {title}
  {{Scientific Objectives of Einstein Telescope}},\ }\href
  {https://doi.org/10.1088/0264-9381/29/12/124013} {\bibfield  {journal}
  {\bibinfo  {journal} {Class. Quant. Grav.}\ }\textbf {\bibinfo {volume}
  {29}},\ \bibinfo {pages} {124013} (\bibinfo {year} {2012})},\ \bibinfo {note}
  {[Erratum: Class.Quant.Grav. 30, 079501 (2013)]},\ \Eprint
  {https://arxiv.org/abs/1206.0331} {arXiv:1206.0331 [gr-qc]} \BibitemShut
  {NoStop}%
\bibitem [{\citenamefont {Maggiore}\ \emph {et~al.}(2020)\citenamefont
  {Maggiore} \emph {et~al.}}]{Maggiore:2019uih}%
  \BibitemOpen
  \bibfield  {author} {\bibinfo {author} {\bibfnamefont {M.}~\bibnamefont
  {Maggiore}} \emph {et~al.},\ }\bibfield  {title} {\bibinfo {title} {{Science
  Case for the Einstein Telescope}},\ }\href
  {https://doi.org/10.1088/1475-7516/2020/03/050} {\bibfield  {journal}
  {\bibinfo  {journal} {JCAP}\ }\textbf {\bibinfo {volume} {03}},\ \bibinfo
  {pages} {050}},\ \Eprint {https://arxiv.org/abs/1912.02622} {arXiv:1912.02622
  [astro-ph.CO]} \BibitemShut {NoStop}%
\bibitem [{\citenamefont {Cai}\ and\ \citenamefont {Yang}(2017)}]{Cai:2016sby}%
  \BibitemOpen
  \bibfield  {author} {\bibinfo {author} {\bibfnamefont {R.-G.}\ \bibnamefont
  {Cai}}\ and\ \bibinfo {author} {\bibfnamefont {T.}~\bibnamefont {Yang}},\
  }\bibfield  {title} {\bibinfo {title} {{Estimating cosmological parameters by
  the simulated data of gravitational waves from the Einstein Telescope}},\
  }\href {https://doi.org/10.1103/PhysRevD.95.044024} {\bibfield  {journal}
  {\bibinfo  {journal} {Phys. Rev. D}\ }\textbf {\bibinfo {volume} {95}},\
  \bibinfo {pages} {044024} (\bibinfo {year} {2017})},\ \Eprint
  {https://arxiv.org/abs/1608.08008} {arXiv:1608.08008 [astro-ph.CO]}
  \BibitemShut {NoStop}%
\bibitem [{\citenamefont {Amaro-Seoane}\ \emph {et~al.}(2017)\citenamefont
  {Amaro-Seoane} \emph {et~al.}}]{LISA:2017pwj}%
  \BibitemOpen
  \bibfield  {author} {\bibinfo {author} {\bibfnamefont {P.}~\bibnamefont
  {Amaro-Seoane}} \emph {et~al.} (\bibinfo {collaboration} {LISA}),\ }\bibfield
   {title} {\bibinfo {title} {{Laser Interferometer Space Antenna}},\
  }\href@noop {} {\  (\bibinfo {year} {2017})},\ \Eprint
  {https://arxiv.org/abs/1702.00786} {arXiv:1702.00786 [astro-ph.IM]}
  \BibitemShut {NoStop}%
\bibitem [{\citenamefont {Kawamura}\ \emph {et~al.}(2011)\citenamefont
  {Kawamura} \emph {et~al.}}]{Kawamura:2011zz}%
  \BibitemOpen
  \bibfield  {author} {\bibinfo {author} {\bibfnamefont {S.}~\bibnamefont
  {Kawamura}} \emph {et~al.},\ }\bibfield  {title} {\bibinfo {title} {{The
  Japanese space gravitational wave antenna: DECIGO}},\ }\href
  {https://doi.org/10.1088/0264-9381/28/9/094011} {\bibfield  {journal}
  {\bibinfo  {journal} {Class. Quant. Grav.}\ }\textbf {\bibinfo {volume}
  {28}},\ \bibinfo {pages} {094011} (\bibinfo {year} {2011})}\BibitemShut
  {NoStop}%
\bibitem [{\citenamefont {Teixeira}\ \emph {et~al.}(2022)\citenamefont
  {Teixeira}, \citenamefont {Barros}, \citenamefont {Ferreira},\ and\
  \citenamefont {Frusciante}}]{teixeira:2022sjr}%
  \BibitemOpen
  \bibfield  {author} {\bibinfo {author} {\bibfnamefont {E.~M.}\ \bibnamefont
  {Teixeira}}, \bibinfo {author} {\bibfnamefont {B.~J.}\ \bibnamefont
  {Barros}}, \bibinfo {author} {\bibfnamefont {V.~M.~C.}\ \bibnamefont
  {Ferreira}},\ and\ \bibinfo {author} {\bibfnamefont {N.}~\bibnamefont
  {Frusciante}},\ }\bibfield  {title} {\bibinfo {title} {{Dissecting
  kinetically coupled quintessence: phenomenology and observational tests}},\
  }\href {https://doi.org/10.1088/1475-7516/2022/11/059} {\bibfield  {journal}
  {\bibinfo  {journal} {JCAP}\ }\textbf {\bibinfo {volume} {11}},\ \bibinfo
  {pages} {059}},\ \Eprint {https://arxiv.org/abs/2207.13682} {arXiv:2207.13682
  [astro-ph.CO]} \BibitemShut {NoStop}%
\bibitem [{\citenamefont {Zhao}\ \emph {et~al.}(2011)\citenamefont {Zhao},
  \citenamefont {Van Den~Broeck}, \citenamefont {Baskaran},\ and\ \citenamefont
  {Li}}]{zhao:2010sz}%
  \BibitemOpen
  \bibfield  {author} {\bibinfo {author} {\bibfnamefont {W.}~\bibnamefont
  {Zhao}}, \bibinfo {author} {\bibfnamefont {C.}~\bibnamefont {Van
  Den~Broeck}}, \bibinfo {author} {\bibfnamefont {D.}~\bibnamefont
  {Baskaran}},\ and\ \bibinfo {author} {\bibfnamefont {T.~G.~F.}\ \bibnamefont
  {Li}},\ }\bibfield  {title} {\bibinfo {title} {{Determination of Dark Energy
  by the Einstein Telescope: Comparing with CMB, BAO and SNIa Observations}},\
  }\href {https://doi.org/10.1103/PhysRevD.83.023005} {\bibfield  {journal}
  {\bibinfo  {journal} {Phys. Rev.}\ }\textbf {\bibinfo {volume} {D83}},\
  \bibinfo {pages} {023005} (\bibinfo {year} {2011})},\ \Eprint
  {https://arxiv.org/abs/1009.0206} {arXiv:1009.0206 [astro-ph.CO]}
  \BibitemShut {NoStop}%
\bibitem [{\citenamefont {Cai}\ \emph {et~al.}(2018)\citenamefont {Cai},
  \citenamefont {Liu}, \citenamefont {Liu}, \citenamefont {Wang},\ and\
  \citenamefont {Yang}}]{cai:2017aea}%
  \BibitemOpen
  \bibfield  {author} {\bibinfo {author} {\bibfnamefont {R.-G.}\ \bibnamefont
  {Cai}}, \bibinfo {author} {\bibfnamefont {T.-B.}\ \bibnamefont {Liu}},
  \bibinfo {author} {\bibfnamefont {X.-W.}\ \bibnamefont {Liu}}, \bibinfo
  {author} {\bibfnamefont {S.-J.}\ \bibnamefont {Wang}},\ and\ \bibinfo
  {author} {\bibfnamefont {T.}~\bibnamefont {Yang}},\ }\bibfield  {title}
  {\bibinfo {title} {{Probing cosmic anisotropy with gravitational waves as
  standard sirens}},\ }\href {https://doi.org/10.1103/PhysRevD.97.103005}
  {\bibfield  {journal} {\bibinfo  {journal} {Phys. Rev.}\ }\textbf {\bibinfo
  {volume} {D97}},\ \bibinfo {pages} {103005} (\bibinfo {year} {2018})},\
  \Eprint {https://arxiv.org/abs/1712.00952} {arXiv:1712.00952 [astro-ph.CO]}
  \BibitemShut {NoStop}%
\bibitem [{\citenamefont {Li}(2013)}]{Li:2013lza}%
  \BibitemOpen
  \bibfield  {author} {\bibinfo {author} {\bibfnamefont {T.~G.~F.}\
  \bibnamefont {Li}},\ }\emph {\bibinfo {title} {{Extracting Physics from
  Gravitational Waves: Testing the Strong-field Dynamics of General Relativity
  and Inferring the Large-scale Structure of the Universe}}},\ \href
  {https://gwic.ligo.org/thesisprize/2013/tgfli_thesis.pdf} {Ph.D. thesis},\
  \bibinfo  {school} {Vrije U., Amsterdam} (\bibinfo {year} {2013})\BibitemShut
  {NoStop}%
\bibitem [{\citenamefont {Nishizawa}\ \emph {et~al.}(2011)\citenamefont
  {Nishizawa}, \citenamefont {Taruya},\ and\ \citenamefont
  {Saito}}]{Nishizawa:2010xx}%
  \BibitemOpen
  \bibfield  {author} {\bibinfo {author} {\bibfnamefont {A.}~\bibnamefont
  {Nishizawa}}, \bibinfo {author} {\bibfnamefont {A.}~\bibnamefont {Taruya}},\
  and\ \bibinfo {author} {\bibfnamefont {S.}~\bibnamefont {Saito}},\ }\bibfield
   {title} {\bibinfo {title} {{Tracing the redshift evolution of Hubble
  parameter with gravitational-wave standard sirens}},\ }\href
  {https://doi.org/10.1103/PhysRevD.83.084045} {\bibfield  {journal} {\bibinfo
  {journal} {Phys. Rev. D}\ }\textbf {\bibinfo {volume} {83}},\ \bibinfo
  {pages} {084045} (\bibinfo {year} {2011})},\ \Eprint
  {https://arxiv.org/abs/1011.5000} {arXiv:1011.5000 [astro-ph.CO]}
  \BibitemShut {NoStop}%
\bibitem [{\citenamefont {Hilborn}(2018)}]{Hilborn:2017liy}%
  \BibitemOpen
  \bibfield  {author} {\bibinfo {author} {\bibfnamefont {R.~C.}\ \bibnamefont
  {Hilborn}},\ }\bibfield  {title} {\bibinfo {title} {{Gravitational waves from
  orbiting binaries without general relativity}},\ }\href
  {https://doi.org/10.1119/1.5020984} {\bibfield  {journal} {\bibinfo
  {journal} {Am. J. Phys.}\ }\textbf {\bibinfo {volume} {86}},\ \bibinfo
  {pages} {186} (\bibinfo {year} {2018})},\ \Eprint
  {https://arxiv.org/abs/1710.04635} {arXiv:1710.04635 [physics.ed-ph]}
  \BibitemShut {NoStop}%
\bibitem [{\citenamefont
  {Lesgourgues}(2011{\natexlab{a}})}]{lesgourgues2011cosmic}%
  \BibitemOpen
  \bibfield  {author} {\bibinfo {author} {\bibfnamefont {J.}~\bibnamefont
  {Lesgourgues}},\ }\href@noop {} {\bibinfo {title} {The cosmic linear
  anisotropy solving system (class) i: Overview}} (\bibinfo {year}
  {2011}{\natexlab{a}}),\ \Eprint {https://arxiv.org/abs/1104.2932}
  {arXiv:1104.2932 [astro-ph.IM]} \BibitemShut {NoStop}%
\bibitem [{\citenamefont {Blas}\ \emph {et~al.}(2011)\citenamefont {Blas},
  \citenamefont {Lesgourgues},\ and\ \citenamefont {Tram}}]{Blas_2011}%
  \BibitemOpen
  \bibfield  {author} {\bibinfo {author} {\bibfnamefont {D.}~\bibnamefont
  {Blas}}, \bibinfo {author} {\bibfnamefont {J.}~\bibnamefont {Lesgourgues}},\
  and\ \bibinfo {author} {\bibfnamefont {T.}~\bibnamefont {Tram}},\ }\bibfield
  {title} {\bibinfo {title} {The cosmic linear anisotropy solving system
  (class). part ii: Approximation schemes},\ }\href
  {https://doi.org/10.1088/1475-7516/2011/07/034} {\bibfield  {journal}
  {\bibinfo  {journal} {Journal of Cosmology and Astroparticle Physics}\
  }\textbf {\bibinfo {volume} {2011}}\bibinfo  {number} { (07)},\ \bibinfo
  {pages} {034–034}}\BibitemShut {NoStop}%
\bibitem [{\citenamefont
  {Lesgourgues}(2011{\natexlab{b}})}]{lesgourgues2011cosmic2}%
  \BibitemOpen
\bibfield  {number} {  }\bibfield  {author} {\bibinfo {author} {\bibfnamefont
  {J.}~\bibnamefont {Lesgourgues}},\ }\href@noop {} {\bibinfo {title} {The
  cosmic linear anisotropy solving system (class) iii: Comparision with camb
  for lambdacdm}} (\bibinfo {year} {2011}{\natexlab{b}}),\ \Eprint
  {https://arxiv.org/abs/1104.2934} {arXiv:1104.2934 [astro-ph.CO]}
  \BibitemShut {NoStop}%
\bibitem [{\citenamefont {Abadie}\ \emph {et~al.}(2010)\citenamefont {Abadie}
  \emph {et~al.}}]{LIGOScientific:2010weo}%
  \BibitemOpen
  \bibfield  {author} {\bibinfo {author} {\bibfnamefont {J.}~\bibnamefont
  {Abadie}} \emph {et~al.} (\bibinfo {collaboration} {LIGO Scientific}),\
  }\bibfield  {title} {\bibinfo {title} {{Calibration of the LIGO Gravitational
  Wave Detectors in the Fifth Science Run}},\ }\href
  {https://doi.org/10.1016/j.nima.2010.07.089} {\bibfield  {journal} {\bibinfo
  {journal} {Nucl. Instrum. Meth. A}\ }\textbf {\bibinfo {volume} {624}},\
  \bibinfo {pages} {223} (\bibinfo {year} {2010})},\ \Eprint
  {https://arxiv.org/abs/1007.3973} {arXiv:1007.3973 [gr-qc]} \BibitemShut
  {NoStop}%
\bibitem [{\citenamefont {Gair}\ \emph {et~al.}(2017)\citenamefont {Gair},
  \citenamefont {Babak}, \citenamefont {Sesana}, \citenamefont {Amaro-Seoane},
  \citenamefont {Barausse}, \citenamefont {Berry}, \citenamefont {Berti},\ and\
  \citenamefont {Sopuerta}}]{Gair:2017ynp}%
  \BibitemOpen
  \bibfield  {author} {\bibinfo {author} {\bibfnamefont {J.~R.}\ \bibnamefont
  {Gair}}, \bibinfo {author} {\bibfnamefont {S.}~\bibnamefont {Babak}},
  \bibinfo {author} {\bibfnamefont {A.}~\bibnamefont {Sesana}}, \bibinfo
  {author} {\bibfnamefont {P.}~\bibnamefont {Amaro-Seoane}}, \bibinfo {author}
  {\bibfnamefont {E.}~\bibnamefont {Barausse}}, \bibinfo {author}
  {\bibfnamefont {C.~P.~L.}\ \bibnamefont {Berry}}, \bibinfo {author}
  {\bibfnamefont {E.}~\bibnamefont {Berti}},\ and\ \bibinfo {author}
  {\bibfnamefont {C.}~\bibnamefont {Sopuerta}},\ }\bibfield  {title} {\bibinfo
  {title} {{Prospects for observing extreme-mass-ratio inspirals with LISA}},\
  }\bibfield  {booktitle} {\emph {\bibinfo {booktitle} {{Proceedings, 11th
  International LISA Symposium: Zurich, Switzerland, September 5-9, 2016}}},\
  }\href {https://doi.org/10.1088/1742-6596/840/1/012021} {\bibfield  {journal}
  {\bibinfo  {journal} {J. Phys. Conf. Ser.}\ }\textbf {\bibinfo {volume}
  {840}},\ \bibinfo {pages} {012021} (\bibinfo {year} {2017})},\ \Eprint
  {https://arxiv.org/abs/1704.00009} {arXiv:1704.00009 [astro-ph.GA]}
  \BibitemShut {NoStop}%
\bibitem [{\citenamefont {Caprini}\ and\ \citenamefont
  {Tamanini}(2016)}]{Caprini:2016qxs}%
  \BibitemOpen
  \bibfield  {author} {\bibinfo {author} {\bibfnamefont {C.}~\bibnamefont
  {Caprini}}\ and\ \bibinfo {author} {\bibfnamefont {N.}~\bibnamefont
  {Tamanini}},\ }\bibfield  {title} {\bibinfo {title} {{Constraining early and
  interacting dark energy with gravitational wave standard sirens: the
  potential of the eLISA mission}},\ }\href
  {https://doi.org/10.1088/1475-7516/2016/10/006} {\bibfield  {journal}
  {\bibinfo  {journal} {JCAP}\ }\textbf {\bibinfo {volume} {1610}}\bibfield
  {number} {\bibinfo  {number} { (10)},\ \bibinfo {pages} {006}},\ }\Eprint
  {https://arxiv.org/abs/1607.08755} {arXiv:1607.08755 [astro-ph.CO]}
  \BibitemShut {NoStop}%
\bibitem [{\citenamefont {Seoane}\ \emph {et~al.}(2013)\citenamefont {Seoane}
  \emph {et~al.}}]{eLISA:2013xep}%
  \BibitemOpen
  \bibfield  {author} {\bibinfo {author} {\bibfnamefont {P.~A.}\ \bibnamefont
  {Seoane}} \emph {et~al.} (\bibinfo {collaboration} {eLISA}),\ }\bibfield
  {title} {\bibinfo {title} {{The Gravitational Universe}},\ }\Eprint
  {https://arxiv.org/abs/1305.5720} {arXiv:1305.5720 [astro-ph.CO]}  (\bibinfo
  {year} {2013})\BibitemShut {NoStop}%
\bibitem [{\citenamefont {Amaro-Seoane}\ \emph {et~al.}(2013)\citenamefont
  {Amaro-Seoane} \emph {et~al.}}]{Amaro-Seoane:2012aqc}%
  \BibitemOpen
  \bibfield  {author} {\bibinfo {author} {\bibfnamefont {P.}~\bibnamefont
  {Amaro-Seoane}} \emph {et~al.},\ }\bibfield  {title} {\bibinfo {title}
  {{eLISA/NGO: Astrophysics and cosmology in the gravitational-wave millihertz
  regime}},\ }\href@noop {} {\bibfield  {journal} {\bibinfo  {journal} {GW
  Notes}\ }\textbf {\bibinfo {volume} {6}},\ \bibinfo {pages} {4} (\bibinfo
  {year} {2013})},\ \Eprint {https://arxiv.org/abs/1201.3621} {arXiv:1201.3621
  [astro-ph.CO]} \BibitemShut {NoStop}%
\bibitem [{\citenamefont {Mapelli}\ \emph {et~al.}(2010)\citenamefont
  {Mapelli}, \citenamefont {Huwyler}, \citenamefont {Mayer}, \citenamefont
  {Jetzer},\ and\ \citenamefont {Vecchio}}]{Mapelli:2010ht}%
  \BibitemOpen
  \bibfield  {author} {\bibinfo {author} {\bibfnamefont {M.}~\bibnamefont
  {Mapelli}}, \bibinfo {author} {\bibfnamefont {C.}~\bibnamefont {Huwyler}},
  \bibinfo {author} {\bibfnamefont {L.}~\bibnamefont {Mayer}}, \bibinfo
  {author} {\bibfnamefont {P.}~\bibnamefont {Jetzer}},\ and\ \bibinfo {author}
  {\bibfnamefont {A.}~\bibnamefont {Vecchio}},\ }\bibfield  {title} {\bibinfo
  {title} {{Gravitational waves from intermediate-mass black holes in young
  clusters}},\ }\href {https://doi.org/10.1088/0004-637X/719/2/987} {\bibfield
  {journal} {\bibinfo  {journal} {Astrophys. J.}\ }\textbf {\bibinfo {volume}
  {719}},\ \bibinfo {pages} {987} (\bibinfo {year} {2010})},\ \Eprint
  {https://arxiv.org/abs/1006.1664} {arXiv:1006.1664 [astro-ph.CO]}
  \BibitemShut {NoStop}%
\bibitem [{\citenamefont {Klein}\ \emph {et~al.}(2016)\citenamefont {Klein}
  \emph {et~al.}}]{Klein:2015hvg}%
  \BibitemOpen
  \bibfield  {author} {\bibinfo {author} {\bibfnamefont {A.}~\bibnamefont
  {Klein}} \emph {et~al.},\ }\bibfield  {title} {\bibinfo {title} {{Science
  with the space-based interferometer eLISA: Supermassive black hole
  binaries}},\ }\href {https://doi.org/10.1103/PhysRevD.93.024003} {\bibfield
  {journal} {\bibinfo  {journal} {Phys. Rev.}\ }\textbf {\bibinfo {volume}
  {D93}},\ \bibinfo {pages} {024003} (\bibinfo {year} {2016})},\ \Eprint
  {https://arxiv.org/abs/1511.05581} {arXiv:1511.05581 [gr-qc]} \BibitemShut
  {NoStop}%
\bibitem [{\citenamefont {Tamanini}\ \emph {et~al.}(2016)\citenamefont
  {Tamanini}, \citenamefont {Caprini}, \citenamefont {Barausse}, \citenamefont
  {Sesana}, \citenamefont {Klein},\ and\ \citenamefont
  {Petiteau}}]{Tamanini:2016zlh}%
  \BibitemOpen
  \bibfield  {author} {\bibinfo {author} {\bibfnamefont {N.}~\bibnamefont
  {Tamanini}}, \bibinfo {author} {\bibfnamefont {C.}~\bibnamefont {Caprini}},
  \bibinfo {author} {\bibfnamefont {E.}~\bibnamefont {Barausse}}, \bibinfo
  {author} {\bibfnamefont {A.}~\bibnamefont {Sesana}}, \bibinfo {author}
  {\bibfnamefont {A.}~\bibnamefont {Klein}},\ and\ \bibinfo {author}
  {\bibfnamefont {A.}~\bibnamefont {Petiteau}},\ }\bibfield  {title} {\bibinfo
  {title} {{Science with the space-based interferometer eLISA. III: Probing the
  expansion of the Universe using gravitational wave standard sirens}},\ }\href
  {https://doi.org/10.1088/1475-7516/2016/04/002} {\bibfield  {journal}
  {\bibinfo  {journal} {JCAP}\ }\textbf {\bibinfo {volume} {1604}}\bibfield
  {number} {\bibinfo  {number} { (04)},\ \bibinfo {pages} {002}},\ }\Eprint
  {https://arxiv.org/abs/1601.07112} {arXiv:1601.07112 [astro-ph.CO]}
  \BibitemShut {NoStop}%
\bibitem [{\citenamefont {Hou}\ \emph {et~al.}(2023)\citenamefont {Hou},
  \citenamefont {Qi}, \citenamefont {Han}, \citenamefont {Zhang}, \citenamefont
  {Cao},\ and\ \citenamefont {Zhang}}]{Hou:2022rvk}%
  \BibitemOpen
  \bibfield  {author} {\bibinfo {author} {\bibfnamefont {W.-T.}\ \bibnamefont
  {Hou}}, \bibinfo {author} {\bibfnamefont {J.-Z.}\ \bibnamefont {Qi}},
  \bibinfo {author} {\bibfnamefont {T.}~\bibnamefont {Han}}, \bibinfo {author}
  {\bibfnamefont {J.-F.}\ \bibnamefont {Zhang}}, \bibinfo {author}
  {\bibfnamefont {S.}~\bibnamefont {Cao}},\ and\ \bibinfo {author}
  {\bibfnamefont {X.}~\bibnamefont {Zhang}},\ }\bibfield  {title} {\bibinfo
  {title} {{Prospects for constraining interacting dark energy models from
  gravitational wave and gamma ray burst joint observation}},\ }\href
  {https://doi.org/10.1088/1475-7516/2023/05/017} {\bibfield  {journal}
  {\bibinfo  {journal} {JCAP}\ }\textbf {\bibinfo {volume} {05}},\ \bibinfo
  {pages} {017}},\ \Eprint {https://arxiv.org/abs/2211.10087} {arXiv:2211.10087
  [astro-ph.CO]} \BibitemShut {NoStop}%
\bibitem [{\citenamefont {Audren}\ \emph {et~al.}(2013)\citenamefont {Audren},
  \citenamefont {Lesgourgues}, \citenamefont {Benabed},\ and\ \citenamefont
  {Prunet}}]{Audren_2013}%
  \BibitemOpen
  \bibfield  {author} {\bibinfo {author} {\bibfnamefont {B.}~\bibnamefont
  {Audren}}, \bibinfo {author} {\bibfnamefont {J.}~\bibnamefont {Lesgourgues}},
  \bibinfo {author} {\bibfnamefont {K.}~\bibnamefont {Benabed}},\ and\ \bibinfo
  {author} {\bibfnamefont {S.}~\bibnamefont {Prunet}},\ }\bibfield  {title}
  {\bibinfo {title} {Conservative constraints on early cosmology with
  montepython},\ }\href {https://doi.org/10.1088/1475-7516/2013/02/001}
  {\bibfield  {journal} {\bibinfo  {journal} {Journal of Cosmology and
  Astroparticle Physics}\ }\textbf {\bibinfo {volume} {2013}}\bibinfo  {number}
  { (02)},\ \bibinfo {pages} {001}}\BibitemShut {NoStop}%
\bibitem [{\citenamefont {Brinckmann}\ and\ \citenamefont
  {Lesgourgues}(2018)}]{Brinckmann:2018cvx}%
  \BibitemOpen
\bibfield  {number} {  }\bibfield  {author} {\bibinfo {author} {\bibfnamefont
  {T.}~\bibnamefont {Brinckmann}}\ and\ \bibinfo {author} {\bibfnamefont
  {J.}~\bibnamefont {Lesgourgues}},\ }\href@noop {} {\bibinfo {title}
  {{MontePython 3: boosted MCMC sampler and other features}}} (\bibinfo {year}
  {2018}),\ \Eprint {https://arxiv.org/abs/1804.07261} {arXiv:1804.07261
  [astro-ph.CO]} \BibitemShut {NoStop}%
\bibitem [{\citenamefont {Feroz}\ and\ \citenamefont
  {Hobson}(2008)}]{Feroz:2007kg}%
  \BibitemOpen
  \bibfield  {author} {\bibinfo {author} {\bibfnamefont {F.}~\bibnamefont
  {Feroz}}\ and\ \bibinfo {author} {\bibfnamefont {M.~P.}\ \bibnamefont
  {Hobson}},\ }\bibfield  {title} {\bibinfo {title} {{Multimodal nested
  sampling: an efficient and robust alternative to MCMC methods for
  astronomical data analysis}},\ }\href
  {https://doi.org/10.1111/j.1365-2966.2007.12353.x} {\bibfield  {journal}
  {\bibinfo  {journal} {Mon. Not. Roy. Astron. Soc.}\ }\textbf {\bibinfo
  {volume} {384}},\ \bibinfo {pages} {449} (\bibinfo {year} {2008})},\ \Eprint
  {https://arxiv.org/abs/0704.3704} {arXiv:0704.3704 [astro-ph]} \BibitemShut
  {NoStop}%
\bibitem [{\citenamefont {Feroz}\ \emph {et~al.}(2009)\citenamefont {Feroz},
  \citenamefont {Hobson},\ and\ \citenamefont {Bridges}}]{Feroz:2008xx}%
  \BibitemOpen
  \bibfield  {author} {\bibinfo {author} {\bibfnamefont {F.}~\bibnamefont
  {Feroz}}, \bibinfo {author} {\bibfnamefont {M.~P.}\ \bibnamefont {Hobson}},\
  and\ \bibinfo {author} {\bibfnamefont {M.}~\bibnamefont {Bridges}},\
  }\bibfield  {title} {\bibinfo {title} {{MultiNest: an efficient and robust
  Bayesian inference tool for cosmology and particle physics}},\ }\href
  {https://doi.org/10.1111/j.1365-2966.2009.14548.x} {\bibfield  {journal}
  {\bibinfo  {journal} {Mon. Not. Roy. Astron. Soc.}\ }\textbf {\bibinfo
  {volume} {398}},\ \bibinfo {pages} {1601} (\bibinfo {year} {2009})},\ \Eprint
  {https://arxiv.org/abs/0809.3437} {arXiv:0809.3437 [astro-ph]} \BibitemShut
  {NoStop}%
\bibitem [{\citenamefont {Feroz}\ \emph {et~al.}(2019)\citenamefont {Feroz},
  \citenamefont {Hobson}, \citenamefont {Cameron},\ and\ \citenamefont
  {Pettitt}}]{Feroz:2013hea}%
  \BibitemOpen
  \bibfield  {author} {\bibinfo {author} {\bibfnamefont {F.}~\bibnamefont
  {Feroz}}, \bibinfo {author} {\bibfnamefont {M.~P.}\ \bibnamefont {Hobson}},
  \bibinfo {author} {\bibfnamefont {E.}~\bibnamefont {Cameron}},\ and\ \bibinfo
  {author} {\bibfnamefont {A.~N.}\ \bibnamefont {Pettitt}},\ }\bibfield
  {title} {\bibinfo {title} {{Importance Nested Sampling and the MultiNest
  Algorithm}},\ }\href {https://doi.org/10.21105/astro.1306.2144} {\bibfield
  {journal} {\bibinfo  {journal} {Open J. Astrophys.}\ }\textbf {\bibinfo
  {volume} {2}},\ \bibinfo {pages} {10} (\bibinfo {year} {2019})},\ \Eprint
  {https://arxiv.org/abs/1306.2144} {arXiv:1306.2144 [astro-ph.IM]}
  \BibitemShut {NoStop}%
\bibitem [{\citenamefont {Buchner}\ \emph {et~al.}(2014)\citenamefont
  {Buchner}, \citenamefont {Georgakakis}, \citenamefont {Nandra}, \citenamefont
  {Hsu}, \citenamefont {Rangel}, \citenamefont {Brightman}, \citenamefont
  {Merloni}, \citenamefont {Salvato}, \citenamefont {Donley},\ and\
  \citenamefont {Kocevski}}]{Buchner:2014nha}%
  \BibitemOpen
  \bibfield  {author} {\bibinfo {author} {\bibfnamefont {J.}~\bibnamefont
  {Buchner}}, \bibinfo {author} {\bibfnamefont {A.}~\bibnamefont
  {Georgakakis}}, \bibinfo {author} {\bibfnamefont {K.}~\bibnamefont {Nandra}},
  \bibinfo {author} {\bibfnamefont {L.}~\bibnamefont {Hsu}}, \bibinfo {author}
  {\bibfnamefont {C.}~\bibnamefont {Rangel}}, \bibinfo {author} {\bibfnamefont
  {M.}~\bibnamefont {Brightman}}, \bibinfo {author} {\bibfnamefont
  {A.}~\bibnamefont {Merloni}}, \bibinfo {author} {\bibfnamefont
  {M.}~\bibnamefont {Salvato}}, \bibinfo {author} {\bibfnamefont
  {J.}~\bibnamefont {Donley}},\ and\ \bibinfo {author} {\bibfnamefont
  {D.}~\bibnamefont {Kocevski}},\ }\bibfield  {title} {\bibinfo {title} {{X-ray
  spectral modelling of the AGN obscuring region in the CDFS: Bayesian model
  selection and catalogue}},\ }\href
  {https://doi.org/10.1051/0004-6361/201322971} {\bibfield  {journal} {\bibinfo
   {journal} {Astron. Astrophys.}\ }\textbf {\bibinfo {volume} {564}},\
  \bibinfo {pages} {A125} (\bibinfo {year} {2014})},\ \Eprint
  {https://arxiv.org/abs/1402.0004} {arXiv:1402.0004 [astro-ph.HE]}
  \BibitemShut {NoStop}%
\bibitem [{\citenamefont {Lewis}(2019)}]{Lewis:2019xzd}%
  \BibitemOpen
  \bibfield  {author} {\bibinfo {author} {\bibfnamefont {A.}~\bibnamefont
  {Lewis}},\ }\href {https://getdist.readthedocs.io} {\bibinfo {title}
  {{GetDist: a Python package for analysing Monte Carlo samples}}} (\bibinfo
  {year} {2019}),\ \Eprint {https://arxiv.org/abs/1910.13970} {arXiv:1910.13970
  [astro-ph.IM]} \BibitemShut {NoStop}%
\bibitem [{\citenamefont {Ross}\ \emph {et~al.}(2015)\citenamefont {Ross},
  \citenamefont {Samushia}, \citenamefont {Howlett}, \citenamefont {Percival},
  \citenamefont {Burden},\ and\ \citenamefont {Manera}}]{Ross:2014qpa}%
  \BibitemOpen
  \bibfield  {author} {\bibinfo {author} {\bibfnamefont {A.~J.}\ \bibnamefont
  {Ross}}, \bibinfo {author} {\bibfnamefont {L.}~\bibnamefont {Samushia}},
  \bibinfo {author} {\bibfnamefont {C.}~\bibnamefont {Howlett}}, \bibinfo
  {author} {\bibfnamefont {W.~J.}\ \bibnamefont {Percival}}, \bibinfo {author}
  {\bibfnamefont {A.}~\bibnamefont {Burden}},\ and\ \bibinfo {author}
  {\bibfnamefont {M.}~\bibnamefont {Manera}},\ }\bibfield  {title} {\bibinfo
  {title} {{The clustering of the SDSS DR7 main Galaxy sample \textendash{} I.
  A 4 per cent distance measure at $z = 0.15$}},\ }\href
  {https://doi.org/10.1093/mnras/stv154} {\bibfield  {journal} {\bibinfo
  {journal} {Mon. Not. Roy. Astron. Soc.}\ }\textbf {\bibinfo {volume} {449}},\
  \bibinfo {pages} {835} (\bibinfo {year} {2015})},\ \Eprint
  {https://arxiv.org/abs/1409.3242} {arXiv:1409.3242 [astro-ph.CO]}
  \BibitemShut {NoStop}%
\bibitem [{\citenamefont {Beutler}\ \emph {et~al.}(2017)\citenamefont {Beutler}
  \emph {et~al.}}]{BOSS:2016hvq}%
  \BibitemOpen
  \bibfield  {author} {\bibinfo {author} {\bibfnamefont {F.}~\bibnamefont
  {Beutler}} \emph {et~al.} (\bibinfo {collaboration} {BOSS}),\ }\bibfield
  {title} {\bibinfo {title} {{The clustering of galaxies in the completed
  SDSS-III Baryon Oscillation Spectroscopic Survey: baryon acoustic
  oscillations in the Fourier space}},\ }\href
  {https://doi.org/10.1093/mnras/stw2373} {\bibfield  {journal} {\bibinfo
  {journal} {Mon. Not. Roy. Astron. Soc.}\ }\textbf {\bibinfo {volume} {464}},\
  \bibinfo {pages} {3409} (\bibinfo {year} {2017})},\ \Eprint
  {https://arxiv.org/abs/1607.03149} {arXiv:1607.03149 [astro-ph.CO]}
  \BibitemShut {NoStop}%
\bibitem [{\citenamefont {Beutler}\ \emph {et~al.}(2011)\citenamefont
  {Beutler}, \citenamefont {Blake}, \citenamefont {Colless}, \citenamefont
  {Jones}, \citenamefont {Staveley-Smith}, \citenamefont {Campbell},
  \citenamefont {Parker}, \citenamefont {Saunders},\ and\ \citenamefont
  {Watson}}]{Beutler:2011hx}%
  \BibitemOpen
  \bibfield  {author} {\bibinfo {author} {\bibfnamefont {F.}~\bibnamefont
  {Beutler}}, \bibinfo {author} {\bibfnamefont {C.}~\bibnamefont {Blake}},
  \bibinfo {author} {\bibfnamefont {M.}~\bibnamefont {Colless}}, \bibinfo
  {author} {\bibfnamefont {D.~H.}\ \bibnamefont {Jones}}, \bibinfo {author}
  {\bibfnamefont {L.}~\bibnamefont {Staveley-Smith}}, \bibinfo {author}
  {\bibfnamefont {L.}~\bibnamefont {Campbell}}, \bibinfo {author}
  {\bibfnamefont {Q.}~\bibnamefont {Parker}}, \bibinfo {author} {\bibfnamefont
  {W.}~\bibnamefont {Saunders}},\ and\ \bibinfo {author} {\bibfnamefont
  {F.}~\bibnamefont {Watson}},\ }\bibfield  {title} {\bibinfo {title} {{The 6dF
  Galaxy Survey: Baryon Acoustic Oscillations and the Local Hubble Constant}},\
  }\href {https://doi.org/10.1111/j.1365-2966.2011.19250.x} {\bibfield
  {journal} {\bibinfo  {journal} {Mon. Not. Roy. Astron. Soc.}\ }\textbf
  {\bibinfo {volume} {416}},\ \bibinfo {pages} {3017} (\bibinfo {year}
  {2011})},\ \Eprint {https://arxiv.org/abs/1106.3366} {arXiv:1106.3366
  [astro-ph.CO]} \BibitemShut {NoStop}%
\bibitem [{\citenamefont {Scolnic}\ \emph {et~al.}(2018)\citenamefont {Scolnic}
  \emph {et~al.}}]{Pan-STARRS1:2017jku}%
  \BibitemOpen
  \bibfield  {author} {\bibinfo {author} {\bibfnamefont {D.~M.}\ \bibnamefont
  {Scolnic}} \emph {et~al.} (\bibinfo {collaboration} {Pan-STARRS1}),\
  }\bibfield  {title} {\bibinfo {title} {{The Complete Light-curve Sample of
  Spectroscopically Confirmed SNe Ia from Pan-STARRS1 and Cosmological
  Constraints from the Combined Pantheon Sample}},\ }\href
  {https://doi.org/10.3847/1538-4357/aab9bb} {\bibfield  {journal} {\bibinfo
  {journal} {Astrophys. J.}\ }\textbf {\bibinfo {volume} {859}},\ \bibinfo
  {pages} {101} (\bibinfo {year} {2018})},\ \Eprint
  {https://arxiv.org/abs/1710.00845} {arXiv:1710.00845 [astro-ph.CO]}
  \BibitemShut {NoStop}%
\bibitem [{\citenamefont {Farrar}\ and\ \citenamefont
  {Peebles}(2004)}]{Farrar:2003uw}%
  \BibitemOpen
  \bibfield  {author} {\bibinfo {author} {\bibfnamefont {G.~R.}\ \bibnamefont
  {Farrar}}\ and\ \bibinfo {author} {\bibfnamefont {P.~J.~E.}\ \bibnamefont
  {Peebles}},\ }\bibfield  {title} {\bibinfo {title} {{Interacting dark matter
  and dark energy}},\ }\href {https://doi.org/10.1086/381728} {\bibfield
  {journal} {\bibinfo  {journal} {Astrophys. J.}\ }\textbf {\bibinfo {volume}
  {604}},\ \bibinfo {pages} {1} (\bibinfo {year} {2004})},\ \Eprint
  {https://arxiv.org/abs/astro-ph/0307316} {arXiv:astro-ph/0307316}
  \BibitemShut {NoStop}%
\bibitem [{\citenamefont {Amendola}(2004)}]{Amendola:2003wa}%
  \BibitemOpen
  \bibfield  {author} {\bibinfo {author} {\bibfnamefont {L.}~\bibnamefont
  {Amendola}},\ }\bibfield  {title} {\bibinfo {title} {{Linear and non-linear
  perturbations in dark energy models}},\ }\href
  {https://doi.org/10.1103/PhysRevD.69.103524} {\bibfield  {journal} {\bibinfo
  {journal} {Phys. Rev. D}\ }\textbf {\bibinfo {volume} {69}},\ \bibinfo
  {pages} {103524} (\bibinfo {year} {2004})},\ \Eprint
  {https://arxiv.org/abs/astro-ph/0311175} {arXiv:astro-ph/0311175}
  \BibitemShut {NoStop}%
\bibitem [{\citenamefont {van~de Bruck}\ \emph {et~al.}(2017)\citenamefont
  {van~de Bruck}, \citenamefont {Mifsud},\ and\ \citenamefont
  {Morrice}}]{vandeBruck:2016hpz}%
  \BibitemOpen
  \bibfield  {author} {\bibinfo {author} {\bibfnamefont {C.}~\bibnamefont
  {van~de Bruck}}, \bibinfo {author} {\bibfnamefont {J.}~\bibnamefont
  {Mifsud}},\ and\ \bibinfo {author} {\bibfnamefont {J.}~\bibnamefont
  {Morrice}},\ }\bibfield  {title} {\bibinfo {title} {{Testing coupled dark
  energy models with their cosmological background evolution}},\ }\href
  {https://doi.org/10.1103/PhysRevD.95.043513} {\bibfield  {journal} {\bibinfo
  {journal} {Phys. Rev. D}\ }\textbf {\bibinfo {volume} {95}},\ \bibinfo
  {pages} {043513} (\bibinfo {year} {2017})},\ \Eprint
  {https://arxiv.org/abs/1609.09855} {arXiv:1609.09855 [astro-ph.CO]}
  \BibitemShut {NoStop}%
\bibitem [{\citenamefont {Van De~Bruck}\ and\ \citenamefont
  {Mifsud}(2018)}]{VanDeBruck:2017mua}%
  \BibitemOpen
  \bibfield  {author} {\bibinfo {author} {\bibfnamefont {C.}~\bibnamefont {Van
  De~Bruck}}\ and\ \bibinfo {author} {\bibfnamefont {J.}~\bibnamefont
  {Mifsud}},\ }\bibfield  {title} {\bibinfo {title} {{Searching for dark matter
  - dark energy interactions: going beyond the conformal case}},\ }\href
  {https://doi.org/10.1103/PhysRevD.97.023506} {\bibfield  {journal} {\bibinfo
  {journal} {Phys. Rev. D}\ }\textbf {\bibinfo {volume} {97}},\ \bibinfo
  {pages} {023506} (\bibinfo {year} {2018})},\ \Eprint
  {https://arxiv.org/abs/1709.04882} {arXiv:1709.04882 [astro-ph.CO]}
  \BibitemShut {NoStop}%
\bibitem [{\citenamefont {G\'omez-Valent}\ \emph {et~al.}(2020)\citenamefont
  {G\'omez-Valent}, \citenamefont {Pettorino},\ and\ \citenamefont
  {Amendola}}]{Gomez-Valent:2020mqn}%
  \BibitemOpen
  \bibfield  {author} {\bibinfo {author} {\bibfnamefont {A.}~\bibnamefont
  {G\'omez-Valent}}, \bibinfo {author} {\bibfnamefont {V.}~\bibnamefont
  {Pettorino}},\ and\ \bibinfo {author} {\bibfnamefont {L.}~\bibnamefont
  {Amendola}},\ }\bibfield  {title} {\bibinfo {title} {{Update on coupled dark
  energy and the $H_0$ tension}},\ }\href
  {https://doi.org/10.1103/PhysRevD.101.123513} {\bibfield  {journal} {\bibinfo
   {journal} {Phys. Rev. D}\ }\textbf {\bibinfo {volume} {101}},\ \bibinfo
  {pages} {123513} (\bibinfo {year} {2020})},\ \Eprint
  {https://arxiv.org/abs/2004.00610} {arXiv:2004.00610 [astro-ph.CO]}
  \BibitemShut {NoStop}%
\bibitem [{\citenamefont {Barros}(2019)}]{Barros:2019rdv}%
  \BibitemOpen
  \bibfield  {author} {\bibinfo {author} {\bibfnamefont {B.~J.}\ \bibnamefont
  {Barros}},\ }\bibfield  {title} {\bibinfo {title} {{Kinetically coupled dark
  energy}},\ }\href {https://doi.org/10.1103/PhysRevD.99.064051} {\bibfield
  {journal} {\bibinfo  {journal} {Phys. Rev. D}\ }\textbf {\bibinfo {volume}
  {99}},\ \bibinfo {pages} {064051} (\bibinfo {year} {2019})},\ \Eprint
  {https://arxiv.org/abs/1901.03972} {arXiv:1901.03972 [gr-qc]} \BibitemShut
  {NoStop}%
\end{thebibliography}%

\end{document}